\documentclass[english,11pt,a4paper]{article}
\pdfoutput=1
\usepackage{epsf,amsmath,amssymb,graphicx,dcolumn}
\usepackage{scalefnt,ulem,cite,hyperref}
\usepackage{cleveref,etoolbox,color}
\apptocmd{\thebibliography}{\setlength{\itemsep}{0.02cm}}{}{}
\Crefname{figure}{Fig.}{Figs.}

\newcommand{\citere}[1]{Ref.\,\cite{#1}}
\newcommand{\citeres}[1]{Refs.\,\cite{#1}}
\newcommand{\code}{\tt}

\newcommand{\sushi}[1]{{\code SusHi#1}}
\newcommand{\sushimi}[1]{{\code SusHiMi#1}}
\newcommand{\abbrev}{\scalefont{.9}}
\newcommand{\eqn}[1]{Eq.\,(\ref{#1})}
\newcommand{\fig}[1]{Fig.\,\ref{#1}}
\newcommand{\sct}[1]{Section~\ref{#1}}
\newcommand{\appref}[1]{Appendix~\ref{#1}}

\newcommand{\lhc}{{\abbrev LHC}}
\newcommand{\qcd}{{\abbrev QCD}}
\newcommand{\eft}{{\abbrev EFT}}
\newcommand{\sm}{{\abbrev SM}}
\newcommand{\thdm}{{\abbrev 2HDM}}
\newcommand{\mssm}{{\abbrev MSSM}}
\newcommand{\nmssm}{{\abbrev NMSSM}}
\newcommand{\ckm}{{\abbrev CKM}}
\newcommand{\susy}{{\abbrev SUSY}}

\newcommand{\pdf}{{\abbrev PDF}}
\newcommand{\lo}{{\abbrev LO}}
\newcommand{\atlas}{{\abbrev ATLAS}}
\newcommand{\cms}{{\abbrev CMS}}
\newcommand{\fortran}{{\abbrev FORTRAN}}
\newcommand{\cp}{{\abbrev $\mathcal{CP}$}}
\newcommand{\nlo}{{\abbrev NLO}}
\newcommand{\nnlo}{{\abbrev NNLO}}
\newcommand{\nklo}[1]{{\abbrev N$^{#1}$LO}}

\newcommand{\muF}{\mu_\text{F}}
\newcommand{\muR}{\mu_\text{R}}
\newcommand{\mhmod}{m_h^{\mathrm{mod}+}}
\newcommand{\zhat}{\hat{\mathrm{\bf{Z}}}}
\newcommand{\sinb}{s_\beta}
\newcommand{\cosb}{c_\beta}

\newcommand{\td}[1]{\ensuremath{\tilde{#1}}}
\newcommand{\mc}[1]{\ensuremath{\mathcal{#1}}}

\newcommand{\tx}[1]{\ensuremath{\text{#1}}}
\newcommand{\h}[1]{\ensuremath{\hat{#1}}}
\newcommand{\nn}{\nonumber}
\newcommand{\al}{\alpha}
\newcommand{\bb}{\beta}
\newcommand{\D}{\Delta}

\newcommand{\G}{\Gamma}

\title{\vspace*{-4em}
  \begin{flushright}
    {\sf\small
      DESY 16-190\\
    }
  \end{flushright}
\vspace*{2em}
Phenomenology of on-shell Higgs production in the \mssm{} with complex parameters\vspace{-0.2cm}}
\author{Stefan Liebler, Shruti Patel, Georg Weiglein\\[1.2em]
{\it DESY, Notkestra{\ss}e 85, D-22607 Hamburg, Germany}\\[1.2em]
{\small\tt stefan.liebler@desy.de}\\[-.3em]
{\small\tt shruti.patel@desy.de}\\[-.3em]
{\small\tt georg.weiglein@desy.de}
}
\date{}

\begin{document}
\maketitle

\begin{abstract}
\noindent
A computation of inclusive cross sections for neutral Higgs boson production through gluon fusion and bottom-quark 
annihilation is presented in the \mssm{} with complex parameters. The predictions for the gluon-fusion process are based 
on an explicit calculation of the leading-order cross section for arbitrary complex parameters which is supplemented 
by higher-order corrections: massive top- and bottom-quark contributions at  \nlo{} \qcd{}, in the heavy top-quark effective 
theory the top-quark contribution up to \nklo{3} \qcd{} including a soft expansion for the \cp{}-even component of the light Higgs 
boson. For its \cp{}-odd component and the heavy Higgs bosons the contributions are incorporated up to \nnlo{} \qcd. Two-loop 
electroweak effects are also incorporated, and \susy{} \qcd{} corrections at \nlo{} are interpolated from the \mssm{} with real 
parameters. Finite wave function normalisation factors ensuring correct on-shell properties of the external Higgs bosons 
are incorporated from the code {\tt FeynHiggs}. For the typical case of a strong admixture of the two heavy Higgs bosons 
it is demonstrated that squark effects are strongly dependent on the phases of the complex parameters. The 
remaining theoretical uncertainties for cross sections are discussed. The results have been implemented into an 
extension of the numerical code  \sushi{} called \sushimi{}.
\end{abstract}

\section{Introduction}

In 2012 the experimental collaborations \atlas{} and \cms{}
announced the discovery of a Higgs-like boson~\cite{Aad:2012tfa,Chatrchyan:2012xdj}
produced in collisions of protons at the Large Hadron Collider (\lhc{}).
Apart from the precise measurement of its production and decay properties in order to
test whether there are deviations from the expectations for a Standard Model (\sm{}) Higgs boson, an essential
part of the programme of the \lhc{} experiments in the upcoming years will be
the search for additional
Higgs bosons. The observed state can be easily accommodated in extended Higgs sectors
like a Two-Higgs-Doublet Model (\thdm{}) or supersymmetric extensions, e.g.\,the
Minimal Supersymmetric Standard Model (\mssm{}). For the search for additional
Higgs bosons and the test of deviations from the 
\sm{} expectations for the \sm{}-like Higgs boson,
the precise knowledge of production cross sections through gluon fusion
and bottom-quark annihilation for these Higgs bosons is a key ingredient.
Current efforts in this direction are summarised in the reports of the \lhc{} Higgs Cross
Section Working Group, see \citeres{Dittmaier:2011ti,Dittmaier:2012vm,Heinemeyer:2013tqa,deFlorian:2016spz}.
So far, the searches for additional Higgs bosons have been interpreted in various scenarios
beyond the Standard Model, including several supersymmetric ones.
However, those analyses do not yet cover the most general
case where \cp{} is violated and leads to mixing between \cp{}-even and -odd
eigenstates.
The reason that an analysis for the general case taking account the
possibility of \cp-violation has not been possible so far has mainly been
the lack of appropriate theoretical predictions for the Higgs production
rates at the \lhc{} for complex parameters in the MSSM and of a practical
prescription for taking into account relevant interference effects in Higgs
production and decay. A discussion of the latter has recently been given in 
\citere{Fuchs:2014ola}. Therefore it is our goal in the present paper to provide
state-of-the-art cross-section predictions in the \mssm{}, taking into
account \cp{}-violating effects, for the two main Higgs production channels
at the \lhc{}, which can be used as input for future experimental 
analyses in \cp{}-violating Higgs scenarios. We present in this paper
precise predictions for neutral Higgs boson
production
through gluon fusion and bottom-quark annihilation in the \mssm{}
with complex parameters, in which \cp{}-even and \cp{}-odd Higgs states
form three admixed Higgs mass eigenstates ${h_a},a\in\lbrace 1,2,3\rbrace$.
Complex parameters in the \mssm{} give rise to additional sources of
\cp{} violation beyond the one induced by the mixing of the quarks of the \sm{},
described by the Cabibbo-Kobayashi-Maskawa (\ckm{}) matrix~\cite{Cabibbo:1963yz,Kobayashi:1973fv}.
In order to explain the baryon asymmetry of the universe, such additional
sources of \cp{} violation beyond the \ckm{} phase are actually needed,
see e.g.\ \citeres{Riotto:1998bt,Cline:2006ts,Davidson:2008bu} for reviews.
It is thus of interest to investigate the \mssm{} with complex parameters.
Its Higgs sector is influenced by the additional phases only 
beyond tree level. Still, these phases are of relevance
in the Higgs boson collider phenomenology as they can induce a large mixing
among the heavy Higgs bosons, and squark and gluino loop contributions
also directly affect Higgs boson production and decay.

For a brief summary of higher-order corrections to the most important
production processes -- gluon fusion and bottom-quark annihilation -- in the \sm{}
and the \mssm{} with real parameters we refer to \sct{sec:ggphi} and
focus here on studies performed for the Higgs sector of the \mssm{} with complex parameters.
Early investigations of Higgs production through gluon fusion at hadron colliders in the \mssm{} with
complex parameters were carried out in \citeres{Dedes:1999zh,Choi:1999aj,Choi:2001iu}.
A thorough analysis taking different production channels into account was presented in \citere{Carena:2002bb},
and results for Higgs\-strahlung can be found in \citere{Arhrib:2001pg}.
Large effects of stops on the cross section for a \cp{}-odd Higgs boson neglecting \cp{}-even
and -odd Higgs mixing were discussed in \citere{Cao:2005zk}.
\citeres{Hesselbach:2007en,Hesselbach:2009gw} discuss the production of a light Higgs through gluon
fusion including its decay into two photons in the \mssm{} with complex parameters.
It should be noted that the mentioned references were published before
the Higgs discovery in 2012 and mostly employ only the lowest order in perturbation theory for
 the production processes. It is therefore timely to improve these predictions
by including up-to-date higher-order corrections and to investigate the compatibility with the
experimental results obtained for the observed signal at $125$\,GeV.
For this purpose we incorporate the prediction within
the \mssm{} with complex parameters into the numerical
code \sushi{}~\cite{Harlander:2012pb,Harlander:2016hcx}, which calculates
Higgs production through gluon fusion and heavy-quark annihilation~\cite{Harlander:2015xur}
in the \sm{}, the \mssm{}, the Two-Higgs-Doublet-Model (\thdm{}) and the
Next-to-Minimal Supersymmetric Standard Model (\nmssm{})~\cite{Liebler:2015bka}.
However until now, \sushi{} did not support complex parameters in the \mssm{}
and thus did not provide predictions for \cp{}-admixed Higgs bosons.

For the calculation of the masses and the wave function normalisation
factors ensuring the correct on-shell properties of external Higgs bosons,
which involves the evaluation of
Higgs boson self-energies and their renormalisation, we use the code
{\tt FeynHiggs}~\cite{Heinemeyer:1998yj,Heinemeyer:1998np,Degrassi:2002fi,Frank:2006yh,Hahn:2013ria}.
It employs a Feynman-diagrammatic approach and includes the full
one-loop~\cite{Frank:2006yh} and the dominant two-loop corrections of
$\mathcal{O}(\alpha_t\alpha_s)$~\cite{Heinemeyer:2007aq}
and $\mathcal{O}(\alpha_t^2)$~\cite{Hollik:2014wea,Hollik:2014bua}
in the \mssm{} with complex parameters.\footnote{Another
approach to calculate the Higgs boson sector in the
\mssm{} with complex parameters is based on the renormalisation group improved effective potential approach
and implemented in e.g.\ the code {\tt CPsuperH}~\cite{Lee:2003nta,Lee:2007gn}.}
A detailed description of the prediction for the Higgs boson masses and the wave function normalisation factors
as implemented in {\tt FeynHiggs} can be found in
\citeres{Frank:2006yh,Williams:2007dc,Fowler:2009ay,Williams:2011bu,Fuchs:2014ola,Fuchs:2015jwa,Fuchs:2016swt}.
Whereas the Higgs sector at tree level remains \cp{}-conserving, at higher
orders an admixture of all three neutral Higgs bosons, i.e.\ 
the two \cp{}-even Higgs bosons $h$, $H$ and the \cp{}-odd Higgs boson~$A$,
is induced. The case where the light Higgs boson describes the \sm{}-like Higgs at $\sim 125$\,GeV
is typically accompanied with a strong admixture of the two heavy
Higgs bosons. For a proper prediction in such a case 
interference effects need to be taken into account in the full 
process involving production and decay of the Higgs bosons, which requires
going beyond the usual narrow-width approximation (see also
\citeres{Kauer:2007zc,Berdine:2007uv,Uhlemann:2008pm,Cacciapaglia:2009ic,Reuter:2007me}).
A convenient way to incorporate interference effects is a 
generalised narrow-width approximation for the production and decay of
on-shell particles
as described in \citeres{alison_thesis,Barducci:2013zaa,Fuchs:2014ola}, where
in \citere{Barducci:2013zaa} only lowest-order contributions have been
considered, while in \citeres{alison_thesis,Fuchs:2014ola} also the
inclusion of higher-order corrections has been addressed.
The results for the cross sections for on-shell Higgs boson production
obtained in the present paper are suitable for direct incorporation into the
framework of a generalised narrow-width approximation.

Our paper is organised as follows: We start by outlining the relevant
quantities in the Higgs, the gluino and the squark sector of the \mssm{} with complex
parameters in \sct{sec:mssmcomplex}. We move to the description
of the gluon-fusion cross section in \sct{sec:ggphi}, where
we discuss the calculation of the cross section at \lo{} and
the applicability of higher-order corrections. Next we introduce in \sct{sec:sushi}
the code \sushi{} and its extension \sushimi{}, which we use for our
phenomenological studies carried out in \sct{sec:numerics}.
We discuss the remaining theoretical uncertainties in \sct{sec:uncertainties}.
Lastly, we conclude in \sct{sec:conclusions} and list
Higgs-(s)quark couplings in \appref{sec:appendix}.

\section{The \texorpdfstring{\mssm{}}{MSSM} with complex parameters}
\label{sec:mssmcomplex}

In this section we discuss the relevant sectors of the \mssm{} with complex
parameters, namely the gluino, the squark as well as the Higgs sector.
While the discussion of the gluino and the squark sector at tree level
is sufficient for our purposes, we will briefly describe the inclusion of higher-order corrections in the Higgs sector. The \mssm{} with complex parameters
allows for $12$~physical, independent phases of the complex parameters, once the phases
of the wino soft-breaking parameter $M_2$ and the soft-breaking parameter $m_{12}^2$ are rotated away.
Those independent phases are the ones of the soft-breaking gaugino masses $M_1$ and $M_3$,
the Higgsino mass parameter $\mu$ and trilinear soft-breaking
couplings~$A_f, f\in\lbrace e,\mu,\tau,u,d,c,s,t,b\rbrace$. In the subsequent discussion
we focus on these phases and their effect on the gluino, the squark and the Higgs sector 
as well as the Higgs boson cross sections.

\subsection{Gluino and squark sector}
\label{sec:gluinosquark}

The gluino $\td{g}$ does not mix with other fields and enters the Lagrangian in the form
\begin{align}
 \mathcal{L}\supset -\frac{1}{2}\overline{\td{g}}m_{\td{g}}\td{g}\,,
\end{align}
where $m_{\td{g}}$ is the absolute value of the complex soft-breaking parameter
$M_3=m_{\td{g}}e^{i\phi_{M_3}}$.\footnote{The soft-breaking parameter $M_1$
associated with the bino can also be complex, but has a minor impact on
the Higgs sector, and we neglect its phase dependence in the following.}
In the Feynman diagrams for the Higgs boson self-energies and the 
Higgs boson production via gluon fusion,
the gluino only contributes beyond the one-loop level. 
However it affects
the bottom-quark Yukawa coupling already at the one-loop level, where it enters
the leading corrections to the relation between the bottom-quark mass and the
bottom-quark Yukawa coupling which can be resummed to all orders,
see below.

In the \mssm{} without flavour mixing in the squark sector, squarks $\td{q}_{L,R}$ of one generation
mix into mass eigenstates $\td{q}_{1,2}$. The term of the Lagrangian containing the
squark mass matrix of one generation is given by~\cite{Williams:2011bu}
\begin{align}
\mathcal{L}&\supset -(\td{q}_L^\dagger,
\td{q}_R^\dagger)M_{\td{q}}^2\begin{pmatrix}\td{q}_L\\\td{q}_R\end{pmatrix}\qquad\text{with}\nonumber \\
&M_{\td{q}}^2 = \begin{pmatrix}
M^2_{\td{q}_L} + m_q^2 + M^2_Z \cos 2 \beta (I^3_q - Q_q s_W^2) & m_q X^*_q \\
m_q X_q & M^2_{\td{q}_R} + m^2_q + M_Z^2 \cos 2 \beta Q_q s^2_W
\end{pmatrix} .
\label{eq:sfermion_mass_matrix}
\end{align}
Here $X_q := A_q - \mu^* \cdot \lbrace { \cot \beta, \tan \beta \rbrace}$,
where $\cot \beta$ and $\tan \beta$ apply
to up- and down-type quarks, respectively. 
The soft-breaking masses $M^2_{\td{q}_L}$ and $M^2_{\td{q}_R}$, the third component of the weak isospin~$I^3_q$, the electric charge $Q_q$ 
and the mass of the quark $m_q$ are real parameters. This also applies to
the $Z$-boson mass $M_Z$ and the sine of the weak mixing angle $s_W\equiv\sin\theta_W$.
Contrarily, in the \cp{}-violating \mssm{} the parameters $A_q = |A_q| e^{i \phi_{A_q}}$
and $\mu = |\mu| e^{i \phi_{\mu}}$, and hence $X_q$, can be complex. These complex parameters enter the Higgs sector
via the Higgs-sfermion couplings, see \appref{sec:appendix}, which are thus also of direct relevance for Higgs boson production.

The mass matrix is diagonalised through the unitary matrix~$U_{\td{q}}$ having real
diagonal elements and complex off-diagonal elements
\begin{align}\label{eq:sqark_mass_matrix_diag}
\begin{pmatrix}
\td{q}_1 \\ \td{q}_2
\end{pmatrix} = U_{\td{q}}
\begin{pmatrix}
\td{q}_L \\ \td{q}_R
\end{pmatrix}\,.
\end{align}

The squark masses (using the convention $m_{\td{q}_1} \leq
m_{\td{q}_2}$)
are calculated as the eigenvalues of \eqn{eq:sfermion_mass_matrix}.
The fact that the left-handed soft-breaking parameter $M^2_{\td{q}_L}$ is
the same for the fields in an SU(2) doublet gives rise to a tree-level
relation between the stop and the sbottom masses. At the loop level, the
corresponding relation between the physical squark masses receives a finite
shift, see \citeres{Djouadi:1998sq}, which we have incorporated 
as a shift in 
the left-handed soft-breaking parameter $M^2_{\td{q}_L}$ in the sbottom
sector, as obtained from {\tt FeynHiggs}.

In the $b / \tilde b$ sector we take into account higher-order corrections
to the relation between the bottom-quark mass and the bottom-Yukawa 
coupling~\cite{Banks:1987iu,Hall:1993gn,Hempfling:1993kv,Carena:1994bv,Carena:1999py,Carena:2000uj}. The couplings of the lowest-order mass eigenstates~$\phi$, 
where $\phi\in\lbrace h,H,A\rbrace$, see \sct{sec:higgssector} below, 
of the Higgs bosons to bottom quarks are given by the effective Lagrangian
\begin{align}
\mc{L}_{\text{eff}} = \frac{m_b}{v} \sum_{\phi^e  \in \lbrace
h,H \rbrace} \bar{b} \left[ g_{b_L}^{\phi^e} P_L + (g_{b_L}^{\phi^e})^* P_R
\right] b\phi^e
+ \frac{im_b}{v} \bar{b} \left[ g_{b_L}^{A} P_L - (g_{b_L}^{A})^* P_R
\right] bA 
\label{eq:efflag}
\end{align}
in terms of the left-handed and right-handed couplings 
$g_{b_L}^{\phi}$ and $g_{b_R}^{\phi}=(g_{b_L}^{\phi})^*$,
where $P_{L/R}=\frac{1}{2}(1\mp \gamma_5)$ are the left- and right-handed
projection operators, respectively.
The explicit form of the couplings is given by
\begin{align}
g_{b_L}^{h} &= \frac{f_{\alpha\beta}^h}{1+ \D_b} \left[ 1 - \frac{\cot \al}{\tan \bb} \D_b \right],\,
g_{b_L}^{H} = \frac{f_{\alpha\beta}^H}{1+ \D_b} \left[ 1 + \frac{\tan \al}{\tan \bb} \D_b \right],\,
g_{b_L}^{A} = \frac{f_{\alpha\beta}^A}{1+ \D_b} \left[  1-
\frac{\D_b}{\tan^2 \bb} \right] \,,
\label{eq:fulldeltab}
\end{align}
with
$f_{\alpha\beta}^h=\sin \al/\cos \bb$, $f_{\alpha\beta}^H=\cos \al/\cos \bb$ and
$f_{\alpha\beta}^A=\tan \bb$
(see also \citeres{Williams:2011bu,Baglio:2013iia}).
The effective Lagrangian provides a resummation of leading 
$\tan\beta$-enhanced contributions entering via the quantity
$\D_b$.
The leading \qcd{} contribution to $\D_b$ has the form
\begin{align}
 \D_b = \frac{2}{3}\frac{\alpha_s(\mu_d)}{\pi}M_3^*\mu^*\tan\beta \, I(m_{\td{b}_1}^2,m_{\td{b}_2}^2,m_{\td{g}}^2)\,,
\label{eq:deltablead}
\end{align}
where $\alpha_s$ is typically evaluated at an averaged \susy{} scale 
$\mu_d=(m_{\td{b}_1}+m_{\td{b}_2}+m_{\td{g}})/3$,
and the function $I(a,b,c)$ is given by $I(a,b,c)=\left(ab\log\left(\tfrac{a}{b}\right)+bc\log\left(\tfrac{b}{c}\right)+ca\log\left(\tfrac{c}{a}\right)\right)/\left((a-b)(b-c)(a-c))\right)$.
As one can see from \eqn{eq:deltablead}, the leading contribution 
to $\D_b$ has an explicit dependence on the complex parameters $M_3$ and $\mu$. 
In our numerical analysis below we use the value for $\D_b$ as obtained 
from {\tt FeynHiggs} (see \citere{Frank:2013hba}), 
which includes additional \qcd{} and electroweak
contributions~\cite{Hofer:2009xb,Noth:2010jy,Noth:2008tw,Bauer:2008bj}.
In our implementation in the program
\sushimi{}, see \sct{sec:sushi} below, both the $\D_b$ value from 
{\tt FeynHiggs} and the leading contribution from \eqn{eq:deltablead} can be
selected.

We will use the expression for the bottom-quark Yukawa coupling according to
the effective Lagrangian of \eqn{eq:efflag} and \eqn{eq:fulldeltab} in our
leading-order expressions for the (loop-induced) gluon-fusion process.
For bottom-quark annihilation and the implementation of higher-order corrections to the gluon-fusion process, see \sct{sec:ggphi}, 
we use as a simplified version~\cite{Williams:2011bu}
\begin{align}
g_{b}^\phi\equiv g_{b_L}^\phi=g_{b_R}^\phi=\frac{1}{|1+\D_b|}f^\phi_{\alpha\beta}\,,
\label{eq:simpledeltab}
\end{align}
in which the left- and right-handed couplings to bottom quarks are
identical to each other. 
We will compare the numerical
impact of the two implementations at 
\lo{} in \sct{sec:numerics}. 
The effective Yukawa coupling in \eqn{eq:efflag} is complex. The phase
of this coupling could be rotated away by an appropriate redefinition of the 
(s)quark fields, see e.g.\ \citere{Hofer:2009xb}. We prefer to use the
general expression for a complex Yukawa coupling. In our 
phenomenological discussion in \sct{sec:numerics} below we will compare the
effect of the complex Yukawa coupling of \eqn{eq:efflag} with the simplified
real coupling of \eqn{eq:simpledeltab} (which are not equivalent to each
other) and we will show that the numerical
differences are small.

\subsection{Higgs sector}
\label{sec:higgssector}

The \mssm{} contains two Higgs doublets with opposite hypercharges $Y_{\mathcal{H}_{1,2}} = \pm 1$
in order to introduce masses for both the up- and down-type fermions.
The neutral fields of the two Higgs doublets
can be decomposed in \cp{}-even ($\phi_1^0, \phi_2^0$) and \cp{}--odd ($\chi_1^0, \chi_2^0$)
components as follows\footnote{We note that the convention
differs from the convention employed by {\tt FeynHiggs} by a different sign of $\chi_1^0$ and $\phi_1^-$, which induces different signs in the corresponding elements of
the matrices in \eqn{eq:HiggsMixing_tree_neutral} and \eqn{eq:HiggsMixing_tree_charged} and the $\chi_1^0$ couplings to (s)quarks displayed in the Appendix.}
\begin{align}
	\mathcal{H}_1 =& \begin{pmatrix}
	                	h_d^0 \\ h_d^-
	                \end{pmatrix} = 
                        \begin{pmatrix}
				v_d + \frac{1}{\sqrt{2}}(\phi_1^0 + i \chi_1^0)\\
				\phi^-_1
                        \end{pmatrix}\label{eq:H_1} \\
      \mathcal{H}_2 =& \begin{pmatrix}
	                	h_u^+ \\ h_u^0
	                \end{pmatrix} = e^{i \xi}
                        \begin{pmatrix}
				\phi^+_2\\
				v_u + \frac{1}{\sqrt{2}}(\phi_2^0 + i \chi_2^0)	
                        \end{pmatrix} \label{eq:H_2}\,,
\end{align}
such that the Higgs potential $V_H$ in terms of the neutral Higgs states is given by
\begin{align} \label{eq:V_min}
	V_H^0 = &(|\mu|^2 + m^2_{\mathcal{H}_2})|h^0_u|^2 + (|\mu|^2 + m^2_{\mathcal{H}_1})|h^0_d|^2 \\\nonumber 
	&-[m_{12}^2 h_u^0h_d^0 +h.c.] + \frac{g_1^2+ g_2^2}{8}[|h_u^0|^2-|h_d^0|^2]^2\,.
\end{align}
The quadratic terms of $V_H$ contain the \susy{} parameter $|\mu|^2$ and the soft terms $m_{\mathcal{H}_1}$, $m_{\mathcal{H}_2}$.
The bilinear terms have the soft coefficient $m^2_{12}$, which is a complex parameter in general but whose
phase can be absorbed through a Peccei-Quinn transformation~\cite{Peccei:1977ur,Peccei:1977hh}.  
The relative phase $\xi$ between the Higgs doublets vanishes when the Higgs potential is minimised, making the Higgs sector of the \mssm{} \cp{}-invariant at lowest order. 

The tree-level neutral mass eigenstates $\lbrace h, H, A, G\rbrace$ are related to the tree-level
neutral fields $\lbrace\phi_1^0, \phi_2^0, \chi_1^0, \chi_2^0\rbrace$
through a unitary matrix as follows
\begin{align}\label{eq:HiggsMixing_tree_neutral}
	\begin{pmatrix} 
		h\\H\\A\\G
	\end{pmatrix} = 
        \begin{pmatrix}
        	-s_{\alpha} & c_{\alpha} & 0 & 0 \\
		 c_{\alpha} & s_{\alpha} & 0 & 0 \\
		 0 & 0 & s_{\beta_n} & c_{\beta_n}\\
		 0 & 0 & -c_{\beta_n} & s_{\beta_n}
        \end{pmatrix} 
        \begin{pmatrix}
        	\phi_1^0 \\ \phi_2^0 \\ \chi_1^0 \\ \chi_2^0
        \end{pmatrix}\,.
\end{align}
Similarly, for the charged Higgs states one obtains
\begin{align}\label{eq:HiggsMixing_tree_charged}
	\begin{pmatrix}
		H^{\pm} \\ G^{\pm}
	\end{pmatrix} = 
	\begin{pmatrix}
		s_{\beta_c} & c_{\beta_c} \\ -c_{\beta_c} & s_{\beta_c} 
	\end{pmatrix}
	\begin{pmatrix}
		\phi_1^{\pm} \\ \phi_2^{\pm}
	\end{pmatrix}\,,
\end{align}
where $s_x \equiv \sin x, c_x \equiv \cos x$.  
$\alpha$, $\beta_n$ and $\beta_c$ are the mixing angles for the \cp{}-even Higgs bosons ($h, H$),
the neutral \cp{}-odd states ($A$, $G$), and the charged states ($H^{\pm}$, $G^{\pm}$), respectively.
Minimising the Higgs potential leads to $\beta:=\beta_n = \beta_c$ at tree level.
The masses of the charged Higgs bosons and the neutral \cp{}-odd Higgs boson 
at tree level are given by
\begin{align}
	m^2_{H^{\pm}} = m_A^2 + M_W^2,\qquad
	m_A^2= \frac{2 m^2_{12}}{\sin (2 \beta)}\,.
\end{align}
The Higgs sector of the \mssm{} at lowest order is fully determined
(besides the gauge couplings) by two parameters, which are usually chosen as
$m_{H^{\pm}}$ ($m_A$) and $\tan \beta := \frac{v_u}{v_d}$ for the case of
the \mssm{} with complex (real) parameters.

\subsection{Higgs mixing at higher orders}
\label{sec:higgshigherorder}

\cp{}-violating mixing between the neutral Higgs bosons $\lbrace
h,H,A\rbrace$ arises as a consequence of radiative corrections
and results in the neutral mass eigenstates $\lbrace h_1, h_2, h_3\rbrace$,
where by convention $m_{h_1} \leq m_{h_2} \leq m_{h_3}$.
The full mixing in higher orders takes place not just between $\lbrace h, H, A\rbrace$,
but also with the Goldstone boson and the electroweak gauge bosons.
In general, $(6 \times 6)$-mixing contributions involving the fields 
$\left\lbrace h,H,A,G,Z,\gamma \right\rbrace$ need to be taken into account. 
For the calculation of the Higgs boson masses and wave function
normalisation factors at the considered order it is sufficient to restrict to 
a $(3 \times 3)$-mixing matrix among $\left\lbrace h,H,A \right\rbrace$,
since mixing effects with $\lbrace G,Z,\gamma\rbrace$ only 
appear at the sub-leading two-loop level and beyond. 
In processes with external Higgs bosons, on the other hand,
mixing contributions with $G$ and $Z$ already enter at the one-loop level,
but the numerical effect of these contributions has been found to be very
small, 
see e.g.\
\citeres{Williams:2007dc,Fowler:2009ay,Williams:2011bu,Bharucha:2012nx}.
In our numerical analysis of the Higgs production through gluon fusion and
bottom-quark annihilation
below we will neglect these kinds of (electroweak) mixing contributions 
of the external Higgs bosons with Goldstone and gauge bosons.
Concerning electroweak corrections, we only incorporate the potentially
numerically large contributions to the Higgs boson masses and wave function
normalisation factors as well as the electroweak contribution to the
correction affecting the relation between the bottom-Yukawa coupling and the
bottom quark mass (see above), while all other contributions considered here
like e.g.\ electroweak corrections to gluon fusion
involve at least one power of the strong coupling.
For the contribution of the $Z$ boson and the
Goldstone boson to the gluon-fusion process via
$gg\to \lbrace Z^*,G^*\rbrace \to h_i$ (the photon only enters at higher
orders) it should be noted that 
contributions from mass-degenerate quark weak-isodoublets vanish
and only top- and bottom-quark contributions proportional to their masses
are of relevance, see
the discussion of the 
Higgsstrahlung process in \citeres{Brein:2003wg,Harlander:2013mla}.
This is a consequence of the fact that only the 
axial component of the quark-quark-$Z$ boson coupling contributes to the
loop-induced coupling of the $Z$ boson to two gluons.
Similarly, squark contributions in $gg\to \lbrace Z^*, G^*\rbrace$ are completely absent at the
one-loop level,
even in case of \cp{} violation in the squark sector.
The one-loop contributions to $gg\to \lbrace Z^*, G^*\rbrace$ therefore have no dependence
on the phases of complex parameters.

Thus, we focus our discussion on the contributions to the $(3\times
3)$-mass matrix~$\mathbf{M}$, which contains
the tree level masses $m_i^2$ on the diagonal and has non-zero (off-)diagonal 
self-energies involving the Higgs states. It enters the Lagrangian,
with $\Phi=\left(h,H,A\right)$, as follows
\begin{align}\label{eq:HiggsMassMatrix}
\mathcal{L}\supset -\frac{1}{2}\Phi\mathbf{M}\Phi^T \quad\text{with}\quad \mathbf{M}=\begin{pmatrix}
	                  	m_h^2 - \hat{\Sigma}_{hh}(p^2) & -\hat{\Sigma}_{hH}(p^2) & - \hat{\Sigma}_{hA}(p^2) \\
                                - \hat{\Sigma}_{Hh}(p^2) & m_H^2 - \hat{\Sigma}_{HH}(p^2) &  - \hat{\Sigma}_{HA}(p^2) \\
                                 - \hat{\Sigma}_{Ah}(p^2) & - \hat{\Sigma}_{AH}(p^2) & m_A^2 - \hat{\Sigma}_{AA}(p^2)
 	                  \end{pmatrix}\,.
\end{align}
The propagator matrix is then given by
\begin{align}\label{eq:inv_gamma_matrix}
	[-\mathbf{\Delta}_{hHA} (p^2)]^{-1} =  \mathbf{\hat{\Gamma}}_{hHA}(p^2) \qquad\text{with}\qquad
	[\mathbf{\hat{\Gamma}}_{hHA}(p^2)]_{ij}=\hat{\Gamma}_{ij}=  i[(p^2
-m_i^2) \delta_{ij} + \hat{\Sigma}_{ij}(p^2)] ,
\end{align}
and the roots of the determinant of this matrix yield the
loop-corrected Higgs boson masses.
The non-diagonal and diagonal propagators can be written as follows
\begin{align}\label{eq:offdiag_selfenergy}
	&\Delta_{ij}  = \frac{\hat{\Gamma}_{ij}\hat{\Gamma}_{kk}-\hat{\Gamma}_{jk}\hat{\Gamma}_{ki}}{\hat{\Gamma}_{ii}\hat{\Gamma}_{jj}\hat{\Gamma}_{kk}
	+2\hat{\Gamma}_{ij}\hat{\Gamma}_{jk}\hat{\Gamma}_{ki}-\hat{\Gamma}_{ii}\hat{\Gamma}_{jk}^2-\hat{\Gamma}_{jj}\hat{\Gamma}_{ki}^2-\hat{\Gamma}_{kk}\hat{\Gamma}_{ij}^2} , \\
	&\Delta_{ii} = \frac{i}{p^2 - m_i^2 + \hat{\Sigma}_{ii}^{\text{eff}}} ,
\end{align}
with effective self-energies that contain mixed terms
\begin{align}\label{eq:eff_self_energy}
	\hat{\Sigma}_{ii}^{\text{eff}} = \hat{\Sigma}_{ii} - i \frac{2\hat{\Gamma}_{ij}\hat{\Gamma}_{jk}
	\hat{\Gamma}_{ki}-\hat{\Gamma}_{jj}\hat{\Gamma}_{ki}^2-\hat{\Gamma}_{kk}\hat{\Gamma}_{ij}^2}
	{\hat{\Gamma}_{jj}\hat{\Gamma}_{kk}-\hat{\Gamma}_{jk}^2}
	=\hat{\Sigma}_{ii} + \frac{\Delta_{ij}}{\Delta_{ii}} \hat{\Sigma}_{ij}
	+  \frac{\Delta_{ik}}{\Delta_{ii}} \hat{\Sigma}_{ik} \,,
\end{align}
where we suppressed the $p^2$ arguments of all terms and $i\neq j\neq k$.

\subsection{Wave function normalisation factors for external Higgs bosons}
\label{sec:zfactordefine}

For Higgs bosons that appear as external particles in a process
appropriate on-shell properties are required
for a correct normalisation of the S-matrix. 
Unless the field renormalisation constants have been chosen such that
all mixing contributions between the mass eigenstates 
$\lbrace h_1, h_2, h_3\rbrace$ vanish on-shell and 
the propagators of the external particles have unit residue, the correct
on-shell properties need to be ensured via the 
introduction of finite wave function normalisation factors, 
see e.g.\ 
\citeres{Chankowski:1992er,Dabelstein:1995js,Heinemeyer:2001iy},
as a consequence of the LSZ formalism~\cite{Lehmann:1954rq}.
The matrix of those so-called $\zhat$ factors contains the correction factors
for the external Higgs bosons $\lbrace h_1, h_2, h_3\rbrace$ relative to the
lowest-order mass eigenstates $\lbrace h, H, A\rbrace$.
The matrix elements $\h{\mathbf{Z}}_{aj}$~\cite{Fuchs:2016swt}
(see also \citeres{Williams:2007dc,Williams:2011bu})
are composed of the root of the external wave function
normalisation factor
\begin{align}
	\h{Z}_i^a := \text{Res}_{\mc{M}_a^2} \lbrace \Delta_{ii} (p^2) \rbrace
\end{align}
and the on-shell transition ratio
\begin{align}
	\h{Z}^a_{ij} = \frac{\Delta_{ij}(p^2)}{\Delta_{jj}(p^2)} \Bigg|_{p^2 = \mc{M}_a^2}\,,
\end{align}
which are evaluated at the complex pole $\mc{M}_a^2$.
Here the indices $\lbrace a, b, c \rbrace$ refer to the loop-corrected mass eigenstates, while 
$\lbrace i, j, k \rbrace$ label the lowest-order mass eigenstates.
With an appropriate assignment of the indices of the two types of states
(see \citere{Fuchs:2016swt}) the matrix elements can be written as
\begin{align}
	\h{\mathbf{Z}}_{aj} = \sqrt{\h{Z}}_a \h{Z}_{aj}\,,
\end{align}
corresponding to the (non-unitary) matrix
\begin{align} \label{eq:Zmix}
\zhat = \begin{pmatrix}
                	\sqrt{\h{Z}}_1 \h{Z}_{1h} & \sqrt{\h{Z}}_1 \h{Z}_{1H} & \sqrt{\h{Z}}_1 \h{Z}_{1A} \\
                        \sqrt{\h{Z}}_2 \h{Z}_{2h} & \sqrt{\h{Z}}_2 \h{Z}_{2H} & \sqrt{\h{Z}}_2 \h{Z}_{2A} \\
                        \sqrt{\h{Z}}_3 \h{Z}_{3h} & \sqrt{\h{Z}}_3 \h{Z}_{3H} & \sqrt{\h{Z}}_3 \h{Z}_{3A} 
                \end{pmatrix}\,.
\end{align}

As explained above, these $\zhat$ factors provide the correct normalisation 
of a matrix element with an external on-shell Higgs boson $h_a$, 
$a\in\lbrace 1,2,3\rbrace$, at $p^2 = \mc{M}_a^2$. 
The application of the $\zhat$ factors yields an expression of the
amplitude $\mathcal{A}_{h_a}$ for an external on-shell Higgs boson $h_a$
in terms of a linear combination of the amplitudes resulting from the 
one-particle irreducible diagrams for each of the lowest-order mass
eigenstates $\lbrace h, H, A\rbrace$ according to
\begin{align}
	\mathcal{A}_{h_a}= \zhat_{ah}\mathcal{A}_{h} + \zhat_{aH} \mathcal{A}_{H} + \zhat_{aA} \mathcal{A}_{A}+\ldots
                    = \sqrt{\h{Z}_a} \left( \h{Z}_{ah} \mathcal{A}_h + \h{Z}_{aH}\mathcal{A}_H + \h{Z}_{aA}\mathcal{A}_A \right)+\ldots\,. \label{eq:eigenstate_mixing}
\end{align}
The ellipsis indicate additional mixing effects with Goldstone bosons and 
gauge bosons, which we neglect in our numerical analysis, see the
discussion in \sct{sec:higgshigherorder}.

\section{The gluon-fusion cross section}
\label{sec:ggphi}

\begin{figure}[t]
\begin{center}
\begin{tabular}{cc}
\includegraphics[width=0.47\textwidth]{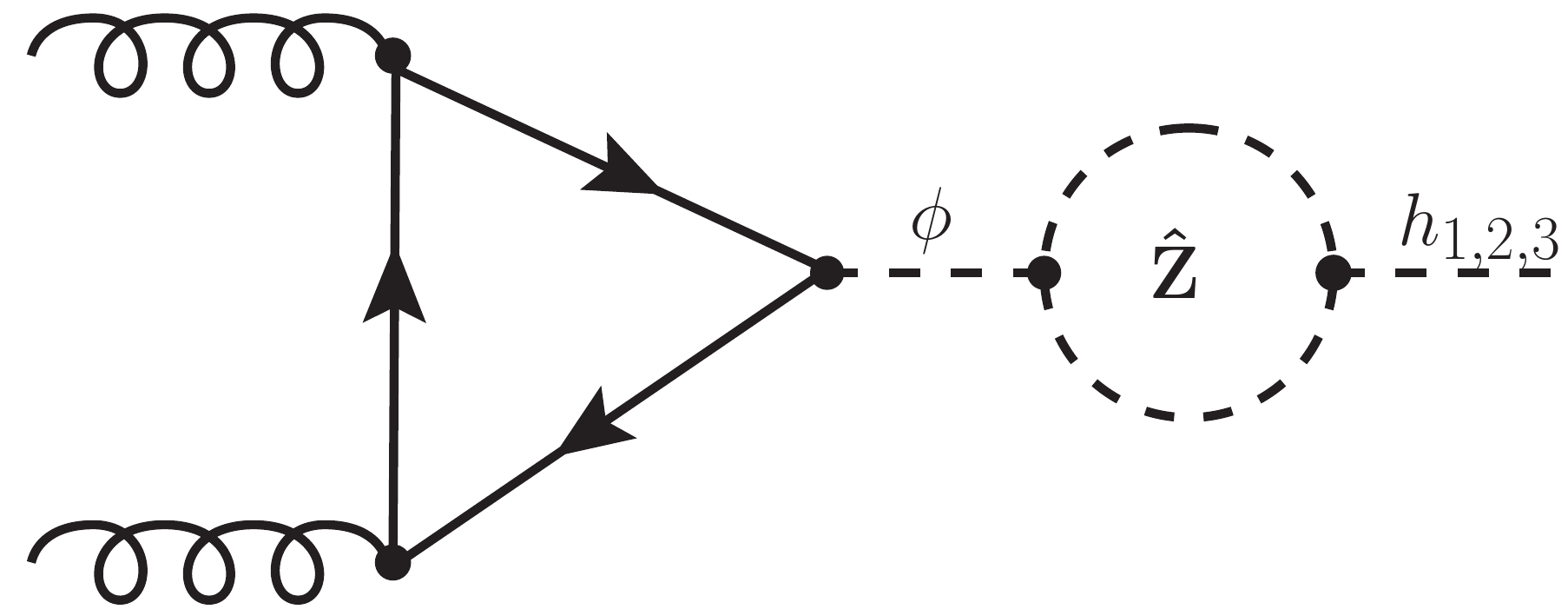}&
\includegraphics[width=0.47\textwidth]{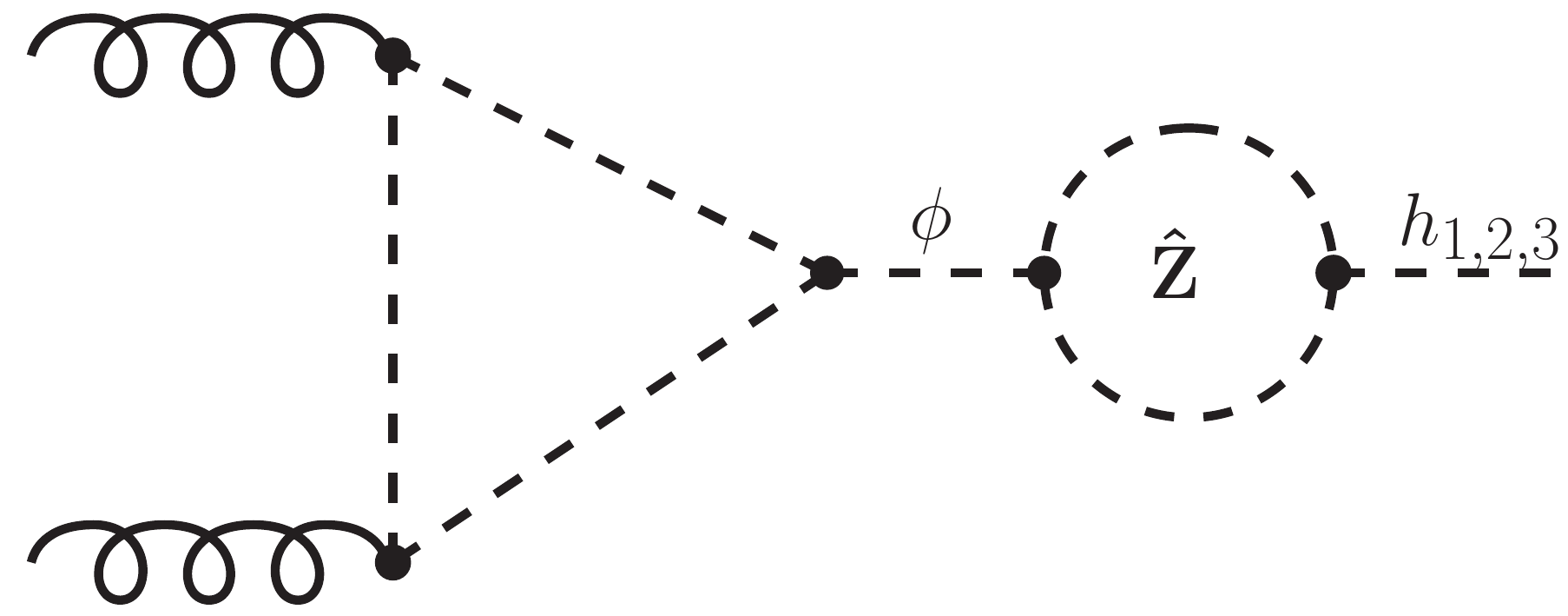}\\
 (a) & (b)
\end{tabular}
\end{center}
\vspace{-0.7cm}
\caption{Feynman diagrams for the
\lo{} cross section with (a) quark and (b) squark contributions.}
\label{fig: LOXS}
\end{figure}

In this section we discuss the calculation of the gluon-fusion cross section
with particular emphasis on the effects of complex parameters. We
first focus on individual ingredients
and then combine them in \sct{sec:sushi}. For this purpose our notation closely
follows \citere{Harlander:2012pb}. At leading order (\lo{}) the gluon-fusion
cross section is known since a long time~\cite{Georgi:1977gs}.
In addition to the quark-induced contributions,
the squark-induced contributions to the gluon-fusion process are also of relevance in supersymmetric extensions of
the \sm{}, even though they are suppressed by inverse powers of the supersymmetric
particle masses if those masses are heavy. Subsequently we present our calculation of the \lo{} cross section 
for the case of the \mssm{} with complex parameters for the three physical
Higgs bosons~$h_a, a\in \lbrace 1,2,3\rbrace$.
Differences with respect to the calculation in the \mssm{} with real parameters are
induced through\footnote{In the \mssm{} with real parameters only couplings
involving $\td{f}_i-\td{f}_j-A$ with $i\neq j$ are non-vanishing,
and left- and right-handed quark Yukawa couplings are identical,
$g^\phi_q\equiv g^\phi_{q_L}=g^\phi_{q_R}$.}
\begin{itemize}
\item $\zhat$ factors, which 
relate the amplitude for an external on-shell Higgs $h_a$ (in the mass
eigenstate basis) to the amplitudes of both the \cp{}-even lowest-order 
states $h$ and $H$ and the \cp{}-odd state~$A$,
see \sct{sec:zfactordefine}.
\item Non-vanishing couplings of squarks $g^A_{\td{f}ii}$ to the pseudoscalar
component~$A$.
\item Different left- and right-handed quark couplings $g^\phi_{q_L}$ and
$g^\phi_{q_R}$ with $\phi\in \lbrace h,H,A\rbrace$, see
\sct{sec:gluinosquark}.
\end{itemize}

\subsection{Lowest-order cross section}

The \lo{} production cross section of the mass eigenstates~$h_a$ can be written as follows
\begin{align}
\label{eq:xs}
 \sigma_{\text{\lo{}}}(pp\to h_a)=\sigma_0^{h_a}\tau_{h_a}\mathcal{L}^{gg}(\tau_{h_a})\quad \text{with}\quad
 \mathcal{L}^{gg}(\tau)=\int_{\tau}^1\frac{dx}{x}g(x)g(\tau/x) \,,
\end{align}
where $\tau_{h_a}=m_{h_a}^2/s$. The hadronic squared 
centre-of-mass energy is denoted by $s$, and the gluon-gluon luminosity by
$\mathcal{L}^{gg}$. 
Therein, the partonic \lo{} cross section for $gg\to h_a$ is given by
\begin{align}
\label{eq:partonicxs}
\sigma_0^{h_a} = \frac{G_F \al_s^2 (\mu_R)}{288 \sqrt{\pi}}&\left[ \left|\mc{A}^{h_a,\mathrm{e}}  \right|^2 +
\left| \mc{A}^{h_a,\mathrm{o}}  \right|^2\right]\\\nonumber
\text{with} \quad &\mc{A}^{h_a,\mathrm{e}}=\zhat_{ah}\mc{A}_+^{h} + \zhat_{aH} \mc{A}_+^{H} + \zhat_{aA}  \mc{A}_-^{A}\\\nonumber
\text{and}\quad &\mc{A}^{h_a,\mathrm{o}}=\zhat_{ah}\mc{A}_-^{h} + \zhat_{aH} \mc{A}_-^{H} + \zhat_{aA}  \mc{A}_+^{A}\,,
\end{align}
where $G_F$ denotes Fermi's constant, and $\zhat_{a\phi}$ are the elements of the $\zhat$ factor matrix.
$\muR$ is the renormalisation scale, which at \lo{} only enters through the scale dependence of the strong coupling constant~$\alpha_s$.
We denote the cross section $pp\to h_a$, which involves one-loop diagrams in
the production process $pp\to \phi$, as ``\lo{} cross section''
despite the fact that it contains higher-order effects through the application of the $\zhat$~factors (Fig. \ref{fig: LOXS}).
We note that in the effective field theory approach of heavy quark and \susy{} masses, where the gluon-gluon-Higgs interaction
is condensed into a single vertex, the amplitudes of the first term in \eqn{eq:partonicxs} can be identified with a contribution that
stems from $\mathcal{L}\supset G^{\mu\nu}G_{\mu\nu}\phi$ with the gluon field strength $G^{\mu\nu}$. The amplitudes of the second term stem from
$\mathcal{L}\supset \tilde{G}^{\mu\nu}G_{\mu\nu}\phi$, which involves the dual of the gluon field strength tensor $\tilde{G}^{\mu\nu}$, resulting in the cross section being
expressible as the sum of two non-interfering squared amplitudes.
This explains the naming of the first and the second term with $\mc{A}^{h_a,\mathrm{e}}$ and $\mc{A}^{h_a,\mathrm{o}}$, respectively.
Similarly, we can split $\sigma_{\text{\lo{}}}$ into $\sigma_{\text{\lo{}}}^{\mathrm{e}}$ and $\sigma_{\text{\lo{}}}^{\mathrm{o}}$.

For the two \cp{}-even lowest-order mass eigenstates $\phi^e\in \lbrace h,H\rbrace$ we obtain the amplitudes
\begin{align}
\mc{A}_+^{\phi^e} = \sum_{q \in \lbrace t,b \rbrace} \left(a^{\phi^e}_{q,+} + \td{a}_{q}^{\phi^e}\right)\,,\quad  \mc{A}_-^{\phi^e} =  \sum_{q \in \lbrace t,b \rbrace} a^{\phi^e}_{q,-}
\end{align}
with
\begin{align}\nonumber
a^{\phi^e}_{q,+} &= \frac{1}{2} \left( g_{q_L}^{\phi^e} + g_{q_R}^{\phi^e} \right) \frac{3}{2} \tau_q^{h_a} \left[ 1 + (1- \tau_q^{h_a}) f(\tau_q^{h_a}) \right]\,,\quad
a^{\phi^e}_{q,-} = \frac{i}{2} \left( g_{q_R}^{\phi^e} - g_{q_L}^{\phi^e}  \right) \frac{3}{2} \tau_q^{h_a}f( \tau_q^{h_a})\\
\td{a}_{q}^{\phi^e} &= -\frac{3}{8} \tau_q^{h_a} \sum_{i=1}^2 g^{\phi^e}_{\td{q}ii} \left[ 1- \tau_{\tilde{q}i}^{h_a} f(\tau_{\tilde{q}i}^{h_a}) \right]\,,
\end{align}
where $g_{q_L}^{\phi^e}$ and $g_{q_R}^{\phi^e}$ are the couplings of the
Higgs boson~$\phi^e$ to the left- and right-handed quarks, respectively.
They are normalised to the \sm{} Higgs-quark couplings.
$g^{\phi^e}_{\td{q}ij}$ are the couplings of the Higgs boson~$\phi^e$
to squarks $\td{q}_i$ and $\td{q}_j$. The explicit expressions for the
Higgs-squark and relative Higgs-quark couplings are listed
in Appendix~\ref{sec:appendix}.
Similarly, for the \cp{}-odd Higgs boson~$A$ we have
\begin{align}
\mc{A}_-^{A} =  \sum_{q \in \lbrace t,b \rbrace} \left(a^{A}_{q,-} + \td{a}_{q}^{A}\right)\,,\quad \mc{A}_+^{A} =  \sum_{q \in \lbrace t,b \rbrace} a^{A}_{q,+}
\end{align}
with
\begin{align}\nonumber
a^{A}_{q,+} &= \frac{1}{2} \left( g_{q_{L}}^A + g_{q_R}^A \right) \frac{3}{2} \tau_q^{h_a} f(\tau_q^{h_a})\,,\quad
a^{A}_{q,-} = \frac{i}{2} \left( g_{q_L}^A - g_{q_R}^A \right) \frac{3}{2}  \tau_q^{h_a} \left[ 1 + (1- \tau_q^{h_a}) f(\tau_q^{h_a}) \right]\\
\td{a}_{q}^{A} &= -\frac{3}{8}\tau_{q}^A \sum_{i=1}^2 g_{\td{q}ii}^A \left [  1- \tau_{\tilde{q}_i}^{h_a} f(\tau_{\tilde{q}_i}^{h_a})  \right]\,.
\end{align}
Within the previous formulas we use the notation
\begin{align}
\tau_q^{h_a} = \frac{4m^2_q}{m_{h_a}^2}, \hspace{5mm}\tau_{\tilde{q}_i}^{h_a}
= \frac{4m_{\tilde{q}_i}^2}{m_{h_a}^2} \,,
\end{align}

and $f(\tau)$ is given by
\begin{align}
f(\tau) = \Bigg\lbrace\begin{matrix} \tx{arcsin}^2 \frac{1}{\sqrt{\tau}} & \mathrm{for}\quad\tau \geq 1\\
-\frac{1}{4} \left( \log \frac{1+ \sqrt{1-\tau}}{1-\sqrt{1-\tau}} - i\pi \right)^2 & \mathrm{for}\quad\tau < 1 \end{matrix}\,.
\end{align}
Our result is consistent with \citere{Dedes:1999zh}, which however assumes $g_{q_L}^\phi=g_{q_R}^\phi$ 
(see our discussion of this issue in \sct{sec:gluinosquark})
and does not take into account
the mixing among the tree-level mass eigenstates~$\phi\in\lbrace h,H,A\rbrace$.
All squark contributions, i.e.\ $\td{a}_q^{\phi^e}$ and $\td{a}_q^{A}$,
enter the first term, $\mc{A}^{h_a,\mathrm{e}}$,
in \eqn{eq:partonicxs}. Quark contributions to 
$\mc{A}^{h_a,\mathrm{e}}$
which couple to the \cp{}-odd lowest-order mass eigenstate $A$
are proportional to the difference between
the left- and right-handed quark Yukawa couplings. The same holds for
the contributions to the second term $\mc{A}^{h_a,\mathrm{o}}$ in
\eqn{eq:partonicxs} which couple to the \cp{}-even lowest-order mass
eigenstates $\phi^e$.
All these terms are therefore denoted with the subscript $\mathcal{A}_-$.
It should be noted that the amplitudes $\mathcal{A}_-^{\phi_e, A}$
only arise due to the complex nature of the Yukawa couplings,
which is a consequence of the incorporation of higher-order
contributions entering via $\D_b$, see \eqn{eq:fulldeltab},
and our choice of working with a complex Yukawa coupling.
Accordingly, the amplitudes $\mathcal{A}_-^{\phi_e, A}$
are zero in the case of the \mssm{} with real parameters.

\subsection{Higher-order contributions}

Gluon fusion receives sizeable corrections at higher
orders in \qcd{}. The \nlo{} corrections for the \sm{} quark contributions are known for arbitrary quark
masses~\cite{Djouadi:1991tka,Dawson:1990zj,Spira:1995rr,Harlander:2005rq,Anastasiou:2006hc,Aglietti:2006tp}.
\nnlo{} (\sm{}-) \qcd{} contributions were calculated in the limit of a heavy top-quark
mass~\cite{Harlander:2002wh,Anastasiou:2002yz,Ravindran:2003um}, similar to
the recently published \nklo{3} contributions for a \cp{}-even Higgs
boson in an expansion around the threshold of Higgs production~\cite{Anastasiou:2014vaa,
Li:2014afw,Anastasiou:2014lda,Anastasiou:2015ema,Anastasiou:2016cez}.\footnote{Most
recently also \nklo{3} \qcd{} corrections for \cp{}-odd Higgs bosons became available~\cite{Ahmed:2015qda,Ahmed:2016otz}.
We neglect those corrections in our analysis.}
Finite top-quark mass effects at \nnlo{} are known in an expansion of inverse powers
of the top-quark mass~\cite{Harlander:2009my,Harlander:2009mq,Harlander:2009bw,
Pak:2009dg,Pak:2009bx,Pak:2011hs,Marzani:2008az,Neumann:2016dny}.
All of the previously mentioned corrections are implemented in
\sushi{}~\cite{Harlander:2012pb,Harlander:2016hcx} and can be added
in all supported models. We will later discuss in more detail
for which Higgs mass ranges
these corrections are applicable, which also explains
why the above mentioned \nklo{3} contributions are only employed for the
\cp{}-even component of the light Higgs boson.

As explained above a complex Yukawa coupling is only induced for the
bottom quark through the incorporation of $\D_b$ contributions.
According to this approach, 
for the top-quark Yukawa coupling~$g_t^\phi$ left- and right handed components
are identical also in the \mssm{} with complex parameters. Therefore we can directly adapt
the known higher-order \qcd{} corrections to the top-quark loop contribution
for the \mssm{} with complex parameters. They are incorporated in the extension 
\sushimi{}, see \sct{sec:sushi}.
For the incorporation of the bottom-quark contribution at \nlo{} (\sm{}-) \qcd{}, 
on the other hand, we have to rely on the simplified
version of the $\D_b$ corrections to the bottom-Yukawa coupling
as specified in \eqn{eq:simpledeltab}.
Electroweak two-loop corrections as discussed in \citeres{Actis:2008ug,Aglietti:2004nj,Bonciani:2010ms}
can be added as well. We take into account the contributions mediated
by light quarks, which can be reweighted to the \mssm{} with complex
parameters. We follow \citere{Bagnaschi:2011tu} and define the
correction factor
\begin{align}
\label{eq:ewfactor}
 \delta_{\text{EW}}^\text{lf} = \frac{\alpha_{\text{EM}}}{\pi}
 \frac{2\text{Re}\left(\mathcal{A}^{h_a,\mathrm{e}}\mathcal{A}^{h_a,\text{EW}*}\right)}{|\mathcal{A}^{{h_a},\mathrm{e}}|^2}\,,
\end{align}
where $\mathcal{A}^{h_a,\mathrm{e}}$,
which has been given in \eqn{eq:partonicxs}, denotes
the \cp{}-even part of the \lo{} amplitude
including quark and squark contributions.
Accordingly, this electroweak correction factor is only applied
to the \cp{}-even component of the \lo{} and \nlo{} cross section, see \sct{sec:sushi}.
The electroweak amplitude is given by~\cite{Bonciani:2010ms}
\begin{align}
\mathcal{A}^{h_a,\text{EW}} =
- \frac{3}{8} \frac{x_W}{s_W^2}
&\left[\frac{2}{c_W^4}\left(\frac{5}{4}-\frac{7}{3}s_W^2+
\frac{22}{9}s_W^4\right)A_1[x_Z] + 4A_1[x_W]\right]\\\nonumber
&\cdot\left(-\zhat_{ah}\sin\al\cos\bb + \zhat_{aH}\cos\al\sin\bb\right)\,,
\end{align}
with the abbreviation
\begin{equation}
\begin{split}
x_V = \frac{1}{m_{h_a}^2}\left(M_V-i\frac{\Gamma_V}{2}\right)^2\,,
\qquad V\in\{W,Z\}\,.
\end{split}
\end{equation}
In \eqn{eq:ewfactor} $\alpha_{\text{EM}}$
denotes the electro-magnetic coupling, and 
$s_W\equiv\sin\theta_W=(1-c_W^2)^{1/2}=(1-M_W^2/M_Z^2)^{1/2}$
is the sine of the weak mixing angle. $M_V$ and $\Gamma_V$ are the mass
and the width of the heavy gauge bosons $V\in\{W,Z\}$, and the function
$A_1[x]$ can be found in \citere{Bonciani:2010ms}.

In the \mssm{} with real parameters analytical \nlo{} virtual contributions
involving squarks, quarks and gluinos are either known in the
limit of a vanishing Higgs
mass~\cite{Harlander:2003bb,Harlander:2004tp,Harlander:2005if,Degrassi:2008zj}
or in an expansion of heavy \susy{}
masses~\cite{Degrassi:2010eu,Degrassi:2011vq,Degrassi:2012vt}\footnote{Exact numerical
and for certain contributions analytical results for \nlo{} virtual contributions
were presented in \citeres{Anastasiou:2008rm,Muhlleitner:2010nm,Anastasiou:2006hc,Aglietti:2006tp,Muhlleitner:2006wx}.}.
Even \nnlo{} corrections of stop-induced contributions to
gluon fusion are known~\cite{Pak:2010cu,Pak:2012xr}; \sushi{} can
approximate these \nnlo{} stop effects~\cite{Harlander:2003kf} in the
\cp{}-conserving \mssm{}.
We neglect those contributions in our analysis for the \mssm{} with
complex parameters.

At \nlo{} in the \mssm{} with complex parameters, supersymmetric contributions
are present both in virtual and real corrections. The real corrections
show a similar behaviour as observed for the \lo{} cross section, i.e.\ the squark induced
contributions of \cp{}-odd components proportional to $g_{\tilde{q}ii}^A$ 
are added as a complex component to the \cp{}-even couplings.
Since beyond \lo{} we employ the simplified version of the $\D_b$ resummation 
according to \eqn{eq:simpledeltab},
the higher-order quark contributions, both real and virtual,
are of the same structure as in the \cp{}-conserving \mssm{}.
The \nlo{} virtual contributions as described in the previous paragraph
are however not easily adjustable to the \mssm{} with complex parameters.
We therefore interpolate the \nlo{} virtual contributions between phases $0$ and $\pi$
of the various \mssm{} parameters
using a cosine interpolation, see \citeres{Heinemeyer:2006px,Hahn:2007fq}.
This interpolation makes use of on-shell stop- and sbottom-quark masses
defined at phases $0$ and $\pi$. Thus, within the interpolated result we have to ensure the correct
subtraction of the \nlo{} contributions that have already been
taken into account through $\D_b$ effects in the bottom-quark Yukawa coupling. This is done by expanding
the $\D_b$ correction to next-to-leading order in the subtraction term. 
For a certain value of the phase $\phi_z$ of a complex parameter $z$,
the virtual \nlo{} amplitude $\mc{A}_{\mathrm\nlo{}}^{\phi} (\phi_z)$
can be approximated using
\begin{align}\label{eq:cos_interpol}
\mc{A}_{\mathrm\nlo{}}^{\phi} (\phi_z) = \frac{1 + \cos \phi_z}{2}\mc{A}_{\mathrm\nlo{}}^{\phi} (0)
+ \frac{1 - \cos \phi_z}{2}\mc{A}_{\mathrm\nlo{}}^{\phi} ( \pi)
\end{align}
for each of the lowest-order mass eigenstates $\phi\in\lbrace h,H,A\rbrace$.
Here $\mc{A}_{\mathrm\nlo{}}^{\phi}(0)$ is the
analytical result for the \mssm{} with real parameters, and
$\mc{A}_{\mathrm\nlo{}}^{\phi}(\pi)$
is the analytical result with $z \rightarrow -z$. Using the factors $\cos \phi_z$ ensures
a smooth interpolation such that the known results for a vanishing phase are recovered.
Whereas a dependence on the phases of $A_q$ and $\mu$ is already apparent in the lowest-order
diagrams of $gg\to\phi$, the phase of $M_3$ only enters through the \nlo{}
virtual corrections. Besides the $\D_b$ contributions, where the full
phase dependence is incorporated, the treatment of the phase of $M_3$
therefore relies on the performed interpolation. While the implemented routines 
for the \mssm{} with real
parameters are expressed in terms of the gluino mass, they can also be 
used for a negative soft-breaking parameter~$M_3$,
such that we can obtain interpolated results for a
complex-valued parameter~$M_3$.
We note that the \nlo{} virtual amplitudes
with a negative $M_3$ are identical to the
virtual amplitudes for positive $M_3$ with opposite signs of the parameters
$A_t, A_b$ and $\mu$.
This can be understood from the structure of
the \nlo{} diagrams involving the squark--quark--gluino couplings.
It should however be noted in this context that 
due to the generation of Higgs-squark
couplings~$g_{\tilde{q}ii}^A$ for non-vanishing phases
a new class of \nlo{} virtual diagrams arises which is not 
present in the \mssm{} with real parameters. Since the interpolation is
based on the result for the \mssm{} with real parameters as input for the
predictions at the phases 0 and $\pi$, the additional set of diagrams may
not be adequately approximated in this way.

Despite this fact, we expect that the interpolation of the
virtual two-loop contributions involving squarks and gluinos to the
gluon-fusion amplitude provides a reasonable approximation, for the
following reasons (we discuss the theoretical uncertainty associated with the
interpolation in \sct{sec:uncertainties} and assign a conservative estimate
of the uncertainty in our numerical analysis). We focus here on the gluon
fusion amplitude without $\zhat$ factors, since in the $\zhat$ factors the
full phase dependence is incorporated without approximations. Gluino
contributions are generally suppressed for gluino masses that are
sufficiently heavy to be in accordance with the present bounds from LHC
searches, 
while gluon-exchange contributions do not add an additional phase dependence
compared to the dependence on the phases of $A_q$ and $\mu$ in the \lo{} 
cross section, which is fully taken into account. The dependence of the
\nlo\ amplitude on the phases of $A_q$ and $\mu$ is therefore expected to
follow a similar pattern as the \lo{} amplitude, which is also what we find
in the application of the interpolation method.

One can also compare the higher-order corrections to the gluon-fusion
process with the ones to the Higgs boson masses and $\zhat$ factors.
In fact, a similar interpolation was probed in the prediction for
Higgs boson masses in the \mssm{} with complex parameters, see e.g.\
\citeres{Frank:2006yh,Heinemeyer:2007aq,Hahn:2007fq,Hahn:2009zz},
where the phase dependence of sub-leading two-loop contributions beyond 
${\cal O}(\alpha_t \alpha_s)$ were approximated with an interpolation
before the full phase dependence of the corresponding two-loop corrections 
at ${\cal O}(\alpha_t^2)$ was calculated~\cite{Hollik:2014wea,Hollik:2014bua}.
Generally good agreement was found between the full result and the
approximation~\cite{Hollik:2014wea,Hollik:2014bua}.
In order to investigate the interpolation of the phase of $M_3$ that is
associated with the gluino we performed a similar
check concerning the phase dependence of two-loop squark and
gluino loop contributions. We numerically compared the full result for the Higgs mass
prediction at this order from {\tt FeynHiggs} with an approximation where the phases at the
two-loop level are interpolated. Despite the fact that also for the Higgs
mass calculation new diagrams proportional to $g_{\tilde{q}ii}^A$ arise
away from phases $0$ and $\pi$,
the phase dependence of the interpolated results generically follows
the behaviour of the full results very well.

Based on the \nlo{} amplitude that has been obtained as described
above, we can construct the \nlo{} cross sections
$\sigma_{\text{\nlo{}}}^{\mathrm{e}}$
and $\sigma_{\text{\nlo{}}}^{\mathrm{o}}$ 
individually, following \citere{Harlander:2012pb},
by defining the \nlo{} correction factors $C^{\mathrm{e}}$ and~$C^{\mathrm{o}}$:
\begin{align}
\label{eq:cfactors}
 C^{\mathrm{e/o}}=2\text{Re}\left[\frac{\mc{A}_{\text{\nlo}}^{h_a,\mathrm{e/o}}}{\mc{A}^{h_a,\mathrm{e/o}}}\right]+\pi^2+\beta_0\log\left(\frac{\muR^2}{\muF^2}\right)\,.
\end{align}
The amplitudes are given by $\mc{A}_{\text{\nlo}}^{h_a,\mathrm{e}}=\zhat_{ah}\mc{A}_{\text{\nlo}}^h+\zhat_{aH}\mc{A}_{\text{\nlo}}^H$
and
$\mc{A}_{\text{\nlo}}^{h_a,\mathrm{o}}=\zhat_{aA}\mc{A}_{\text{\nlo}}^A$,
$\muF$ denotes the factorisation scale,
and $\beta_0=11/2-n_f/3$ with $n_f=5$. Note that the \lo{} amplitudes $\mc{A}^{h_a,\mathrm{e/o}}$ entering
\eqn{eq:cfactors} are taken in the limit of large stop and sbottom masses, see \citere{Harlander:2012pb}.
The correction factors enter the \nlo{} cross section as follows
\begin{align}
\label{eq:xsnlo}
 \sigma_{\text{\nlo{}}}^{\mathrm{e/o}}(pp\to h_a+X)=\sigma_0^{h_a,\mathrm{e/o}}\tau_{h_a}\mathcal{L}^{gg}(\tau_{h_a})\left[1+C^{\mathrm{e/o}}\frac{\alpha_s}{\pi}\right]
 +\Delta \sigma_{gg}^{\mathrm{e/o}}+\Delta\sigma_{gq}^{\mathrm{e/o}}+\Delta\sigma_{q\bar q}^{\mathrm{e/o}}\,.
\end{align}
The terms $\Delta\sigma$ denote the real corrections, which are not fully
displayed here. We emphasise again that at \nlo{} we work
with \eqn{eq:simpledeltab}, such that in the real corrections the only new
ingredients are Higgs-squark couplings $g_{\td{q}ii}^A$, which are added
to the \cp{}-even components~$\Delta\sigma^{\mathrm{e}}$. The real corrections can be split in
$\Delta\sigma^{\mathrm{e}}$ and $\Delta\sigma^{\mathrm{o}}$ since no interference terms arise.

\section{The program \sushi{} and the extension \sushimi{}}
\label{sec:sushi}

\begin{figure}[ht]
\includegraphics[width=\textwidth]{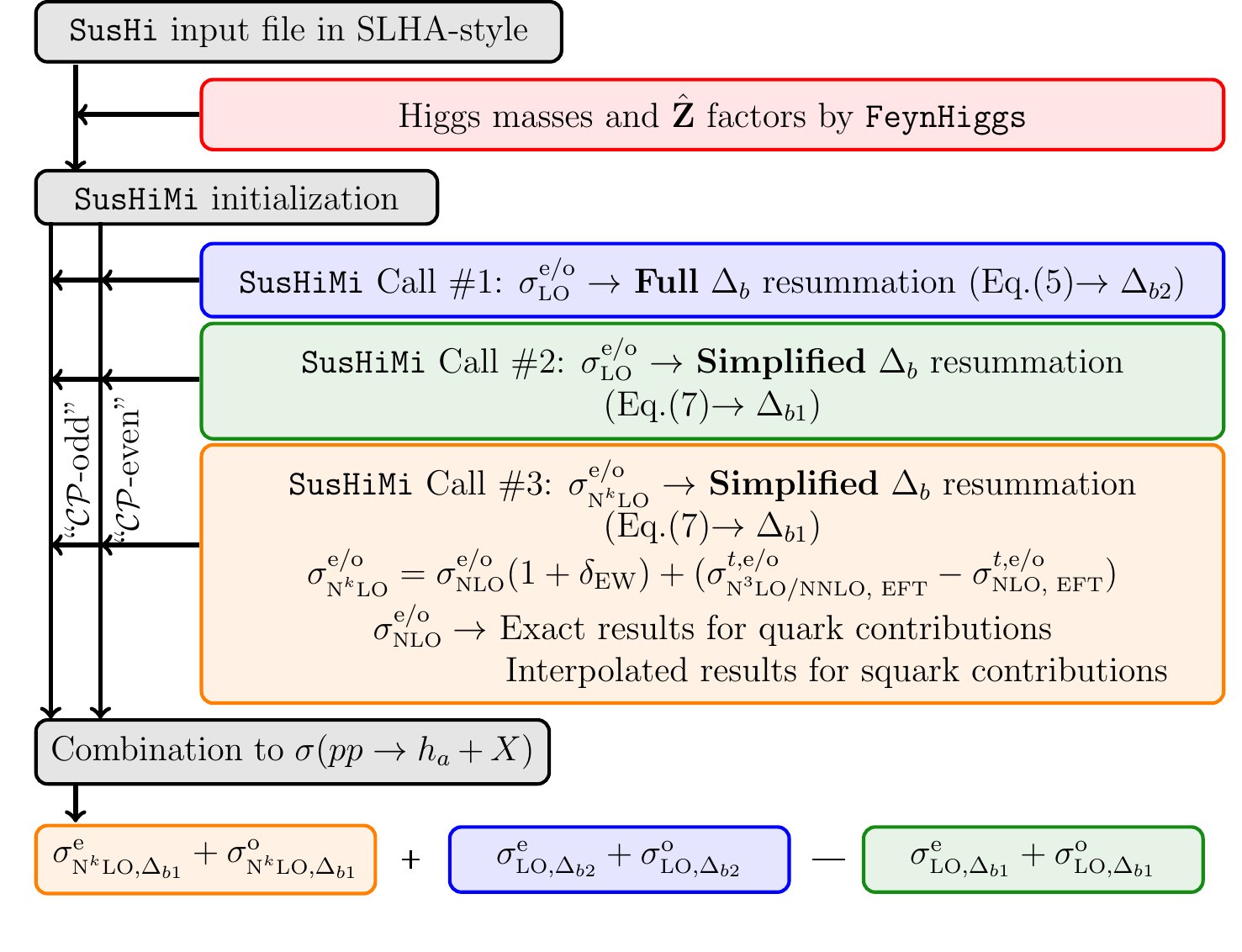}
\caption{Pictorial view of the gluon-fusion cross section calculation.
}
\label{fig:workflow}
\end{figure}

\sushi{} is a numerical \fortran{} code~\cite{Harlander:2012pb,Harlander:2016hcx} which combines analytical results
for the calculation of Higgs boson cross sections through gluon fusion and heavy-quark annihilation
in models beyond the Standard Model up to the highest known orders
in perturbation theory.
However, the current release does not allow for \cp{} violation in the Higgs sector.
Following our discussion in \sct{sec:ggphi} we present the calculation of
Higgs boson production in the context of the \mssm{} with complex parameters,
which we included in an extension of \sushi{} named \sushimi{}\footnote{The
name is inspired by the mixing of the Higgs bosons. \sushimi{} can be obtained upon request.}.
For this purpose we proceed along the lines of \fig{fig:workflow} and calculate the Higgs boson
production cross section through gluon fusion as follows:
\sushimi{} calls \sushi{} twice and in these two calls performs
a ``\cp{}-even'' calculation for $\sigma_{\text{\nlo{}}}^{\mathrm{e}}$
and a ``\cp{}-odd'' calculation for
$\sigma_{\text{\nlo{}}}^{\mathrm{o}}$ according to \eqn{eq:xsnlo}.
Thus, the total gluon-fusion cross section is the sum of the two parts
\begin{align}
\label{eq:ggphimaster}
\sigma_{\text{\nklo{k}}}(pp\to
h_a+X)=\sigma_{\text{\nklo{k}}}^{\mathrm{e}}(pp\to
h_a+X)+\sigma_{\text{\nklo{k}}}^{\mathrm{o}}(pp\to h_a+X)\,.
\end{align}
We obtain the result beyond \lo{} \qcd{} through\footnote{These formulas equal the
master formulas employed in previous \sushi{} releases~\cite{Harlander:2012pb,Harlander:2016hcx}.}
\begin{align}
\label{eq:ggphimasternlo}
&\sigma^{\mathrm{e}}_{\text{\nklo{k}}} =
\sigma_{\text{\nlo{}}}^{\mathrm{e}}(1+\delta_{\mathrm{EW}}^{\mathrm{lf}})
+\left(\sigma_{\text{\nklo{k}, \text{\eft{}}}}^{t,\mathrm{e}} - \sigma_{\text{\nlo{}, \text{\eft{}}}}^{t,\mathrm{e}}\right)\\
&\sigma^{\mathrm{o}}_{\text{\nklo{k}}} =
\sigma_{\text{\nlo{}}}^{\mathrm{o}}
+\left(\sigma_{\text{\nklo{k}, \text{\eft{}}}}^{t,\mathrm{o}} -
\sigma_{\text{\nlo{}, \text{\eft{}}}}^{t,\mathrm{o}}\right) \,,
\end{align}
whereas $\sigma^{\mathrm{e/o}}_{\text{\lo{}}}$ was specified in
\sct{sec:ggphi}, and $k\in \lbrace 1,2,3\rbrace$.
The \cp{}-odd component $\sigma_{\text{\nklo{k}, \text{\eft{}}}}^{t,\mathrm{o}}$ is only implemented up to $k=2$ (see below).
In the previous formulas $\sigma_{\text{\nlo{}}}^{\mathrm{e/o}}$ are the
\nlo{} cross sections including real contributions and the interpolated \nlo{} virtual corrections as discussed in \sct{sec:ggphi}.
They employ the simplified $\Delta_b$ resummation according to \eqn{eq:simpledeltab}, named $\Delta_{b1}$.
$\sigma_{\text{\nklo{k}, \text{\eft{}}}}^{t,\mathrm{e}}$ and $\sigma_{\text{\nklo{k}, \text{\eft{}}}}^{t,\mathrm{o}}$
are cross sections including the top-quark contribution only.
They are based on a $K$-factor calculated in the \eft{} approach of an infinitely heavy top-quark
obtained for a \sm{} Higgs boson~$H$ and a pseudoscalar~$A$ (in a \thdm{} with $\tan\beta=1$) with mass $m_{h_a}$, respectively.
This $K$-factor is subsequently reweighted with the exact \lo{} cross section.
For this purpose the employed \lo{} cross sections $\sigma_{\text{\lo}}^{t,\mathrm{e}}$
and $\sigma_{\text{\lo}}^{t,\mathrm{o}}$ are again evaluated as discussed in \sct{sec:ggphi}
with full $\zhat$ factors, but include only the top-quark contribution. They are multiplied with the $K$-factors in
$\sigma_{\text{\nklo{k}, \text{\eft{}}}}^{t,\mathrm{e}}$ and $\sigma_{\text{\nklo{k}, \text{\eft{}}}}^{t,\mathrm{o}}$, respectively.
Due to their small numerical impact in $\sigma_{\text{\nklo{k}, \text{\eft{}}}}^{t,\mathrm{e/o}}$
we do not take into account top-quark mass effects beyond \nlo{} even though they are implemented in \sushi{}.
An alternative approach, which is not discussed in this paper but can be implemented in \sushimi{}, is to
include the relative couplings $g_t^\phi$ and the $\zhat$~factors into the complex-valued Wilson coefficients of the \eft{} directly.

As already mentioned \nklo{3} \qcd{} corrections are only taken into account for the \cp{}-even component
of the light Higgs boson, which allows us to match the precision of the light Higgs
boson cross section in the \sm{} employed in up-to-date predictions. This
is motivated by the fact that the light Higgs boson that is identified with the observed signal at 
$125\,\text{GeV}$ is usually assumed to have a dominant \cp{}-even
component, which is also the case in the scenarios which are considered in our numerical discussion.
For the \cp{}-odd component of the light Higgs and the heavy Higgs bosons we employ
the \nnlo{} corrections for the top-quark induced contributions to gluon fusion
in the effective theory of a heavy top-quark, i.e.\ we do not take into account
top-quark mass effects beyond \nlo{}, but only factor out the \lo{} \qcd{} cross
sections~$\sigma_{\text{\lo}}^{t,\mathrm{e}}$ and $\sigma_{\text{\lo}}^{t,\mathrm{o}}$.
The strategy to employ the \eft{} result at \nnlo{} beyond the top-quark mass threshold can be justified from the comparison of
\nlo{} corrections, which are known in the \eft{} approach and exactly with full quark-mass dependence
and agree also beyond the top-quark mass threshold. 
On the other hand, the \nklo{3} \qcd{} corrections that were obtained
for the top-quark contribution
are only known in the \eft{} approach and for an expansion
around the threshold of Higgs production at $x=m_{h_a}^2/s\to 1$, which we
can take into account up to $\mathcal{O}(1-x)^{16}$. Since the combination
of the \eft\ approach and the threshold expansion
becomes questionable above the top-quark mass threshold, we apply \nklo{3} \qcd{} corrections
only for the \cp{}-even component of the light Higgs boson
and thus match the precision of the \sm{} prediction.
The electroweak correction factor $\delta_{\mathrm{EW}}^{\mathrm{lf}}$
multiplied in the ``\cp{}-even'' run is obtained from \eqn{eq:ewfactor}.

As shown in \fig{fig:workflow} we call \sushimi{} three times 
in order
to take into account the different possibilities of the
resummation of $\tan\beta$ enhanced sbottom effects in the
\lo{} \qcd{} contributions. We add the results as follows
\begin{align}
\label{eq:ggphisum}
\sigma(pp\to h_a+X)=\sigma_{\text{\nklo{k}}}^{\Delta_{b1}}
+\sigma_{\text{\lo{}}}^{\Delta_{b2}}
-\sigma_{\text{\lo{}}}^{\Delta_{b1}}\,,
\end{align}
where in the \nklo{k} \qcd{} cross section following \eqn{eq:ggphimaster}
the simplified resummation according to \eqn{eq:simpledeltab} is employed, indicated
through the index $\Delta_{b1}$. We add and subtract the \lo{} \qcd{}
cross section using the full resummation according to \eqn{eq:fulldeltab}, named $\Delta_{b2}$,
and the simplified resummation, respectively. As we will demonstrate
the differences between the two versions of resummation are small, which
can partially be understood from a possible rephasing of complex Yukawa couplings
by a redefinition of all (s)quark fields (see the discussion in 
\sct{sec:gluinosquark}).

\sushi{} also allows one to obtain differential cross sections as a function
of the transverse momentum or the (pseudo-)rapidity of the Higgs boson.
These effects can be studied also in the \mssm{} with complex parameters.
In the case of non-vanishing transverse momentum, which
is only possible through additional radiation, i.e.\ real corrections,
the precision for massive quark contributions
in extended Higgs sectors is currently limited to the \lo{} 
prediction~\cite{Bagnaschi:2015bop,Liebler:2016dpn}.
The predictions of the $p_T$ distributions in \sushimi{} have been obtained
from the \lo\ contributions
with arbitrary complex parameters, and in contrast to the
total cross sections are therefore 
not affected by additional interpolation uncertainties from higher orders in
comparison to the case of the MSSM with real parameters.

Higgs production through bottom-quark annihilation is calculated in \sushi{}
for a \sm{} Higgs boson at \nnlo{} \qcd{}. In the employed five-flavour scheme,
where the bottom quarks are understood as partons,
the result equals the cross section of a pseudoscalar~$A$ (in a \thdm{} with $\tan\beta=1$).
For the production of the Higgs boson~$h_a$ in the \mssm{} with complex
parameters, as implemented in \sushimi{}, the results for the \sm{}
Higgs boson are reweighted to the \mssm{} with $|\zhat_{ah}g_b^h+\zhat_{aH}g_b^H|^2+|\zhat_{aA}g_b^A|^2$,
which includes $\tan\beta$-enhanced squark
effects through $\Delta_b$ according to \eqn{eq:simpledeltab}. This procedure equals
the application of a $K$-factor on the full \lo{} cross section including $\zhat$ factors.
In case of non-equal left- and right-handed
couplings $g_{bL}$ and $g_{bR}$ due to the application of the full resummation in \eqn{eq:fulldeltab},
the \sm{} cross section has to be multiplied with
\begin{align}\nonumber
&|\zhat_{ah}(g_{bL}^h+g_{bR}^h)+\zhat_{aH}(g_{bL}^H+g_{bR}^H)+i\zhat_{aA}(g_{bL}^A-g_{bR}^A)|^2\\
&+|i\zhat_{ah}(g_{bR}^h-g_{bL}^h)+i\zhat_{aH}(g_{bR}^H-g_{bL}^H)+\zhat_{aA}(g_{bL}^A+g_{bR}^A)|^2\,.
\end{align}
Though, due to the similarity of both approaches we only discuss Higgs production
through bottom-quark annihilation with simplified $\Delta_b$ resummation.

\section{Numerical results}
\label{sec:numerics}

For our numerical analysis we slightly modify two standard \mssm{} scenarios introduced
in \citere{Carena:2013ytb}, namely the $\mhmod$
and the light-stop scenario. 
The scenarios have been chosen for illustration, featuring relatively
large squark and gluino contributions to the gluon
fusion process. The corresponding effects will be relevant in our 
discussion of the associated theoretical uncertainties.

The light-stop inspired scenario that we
use for our numerical analysis is defined as follows
\begin{align}\nonumber
 &M_1=340\,\text{GeV},\quad M_2= \mu =400\,\text{GeV},\quad M_3=1.5\,\text{TeV}\\
 &X_t=X_b=X_{\tau}=1.0\,\text{TeV},\quad A_q=A_l=0
\label{eq:lightstopscen}
\\\nonumber
 &\tilde{m}_{Q_2}=\tilde{m}_L=1\,\text{TeV}, \quad \tilde{m}_{Q_3}=0.5\,\text{TeV}\,,
\end{align}
where the modified values of $M_1$ and $M_2$ have been chosen to avoid direct bounds
from stop searches obtained in \lhc{}~Run~I (assuming $R$-parity conservation).
\footnote{Indirect bounds from the effects of stops on the measured Higgs rates are much weaker, see e.g. \citere{Liebler:2015ddv}}
For the $\mhmod$-inspired scenario we choose for vanishing phases of the complex parameters:
\begin{align}\nonumber
 &M_1=250\,\text{GeV},\quad M_2=500\,\text{GeV},\quad M_3=1.5\,\text{TeV}\\
 &X_t=X_b=X_{\tau}=1.5\,\text{TeV},\quad A_q=A_l=0\\\nonumber
 &\mu=\tilde{m}_{Q}=\tilde{m}_L=1\,\text{TeV}\,.
\end{align}
We use for the \sm{} parameters the values $m_t^{\textrm{OS}} = 173.20$\,GeV,
$m_b^{\overline{\textrm{MS}}}(m_b)=4.16\,$GeV,
$m_b^{\textrm{OS}}=4.75\,$GeV
and $\alpha_s (M_Z) = 0.119$.
The depicted on-shell bottom-quark mass is used as internal mass for
propagators and for the bottom-quark Yukawa coupling in the gluon-fusion process.
The depicted value of $\alpha_s$ is only used for the evaluations of {\tt FeynHiggs},
for the cross sections the value of $\alpha_s$ associated with the employed
\pdf{} set is taken. We employ the {\tt MMHT2014} \pdf{} sets
at \lo{}, \nlo{} and \nnlo{} \qcd{}~\cite{Harland-Lang:2014zoa}.
The central choice for the renormalisation and factorisation scales $\muR^0$ and $\muF^0$, respectively,
is $(\muR^0,\muF^0)=(m_{h_a}/2,m_{h_a}/2)$ for gluon fusion and
$(\muR^0,\muF^0)=(m_{h_a},m_{h_a}/4)$ for bottom-quark annihilation.
More details are described in \sct{sec:uncertainties}.

Whereas for the $\mhmod$-inspired scenario
we pick heavy Higgs bosons through $m_{H^\pm}=900$\,GeV 
with $\tan\beta=10$ and $40$ for the study of $\Delta_b$ effects,
we choose $m_{H^\pm}=500$\,GeV with $\tan\beta=16$ for
the light-stop inspired scenario. A detailed discussion of squark effects
for the Higgs boson cross sections in the light-stop scenario
can also be found in \citere{Bagnaschi:2014zla}. For the chosen
parameter point the squark effects are sizeable,  both for the light Higgs boson
and in particular also for the heavy \cp{}-even Higgs boson, where they
reduce the gluon-fusion cross section by about $\sim 90$\%.
The Higgs boson masses and the $\zhat$ factors are obtained
from {\tt FeynHiggs 2.11.2}. The cross sections are evaluated
with \sushimi{}, which is based on the latest release of \sushi{}, version {\tt 1.6.1}.
We will mostly focus on the gluon-fusion cross section and present the bottom-quark
annihilation cross section only for the scenario with $\tan\beta=40$.

For the parameter points associated with the mentioned scenarios in the \mssm{} with real parameters
we vary the phases of $A_t=|A_t|e^{i\phi_{A_t}}$ and $M_3=m_{\td{g}}e^{i\phi_{M_3}}$
leaving the absolute values constant in order
to address various aspects in the phenomenology of Higgs boson production.
The phases of $A_b$ and $\mu$ do not introduce new phenomenological
features, and we do not display results for the variation of those phases.
A variation of the phase of $X_t$ leads to very similar 
cross sections for all Higgs bosons
as observed for the variation of the phase of $A_t$. This can be understood from the fact
that we choose not too large values of $\mu$ and $\tan\beta\geq 10$, and so $X_t \approx A_t$.
Note that the stop masses are constant as a function of the phase of $X_t$,
if the absolute value of $X_t$ is fixed. Before we proceed we want to 
briefly discuss experimental constraints on
the phases: The most restrictive constraints on the phases arise from bounds on
the electric dipole moments (EDMs) of the electron and the neutron,
see \citeres{Demir:2003js,Chang:1998uc,Pilaftsis:1999td} and references therein.
EDMs from heavy quarks~\cite{Hollik:1997vb,Hollik:1997ph} and the deuteron~\cite{Lebedev:2004va} also have an impact.
\mssm{} contributions to these EDMs already contribute at the one-loop level
and primarily involve the first two generations
of sleptons and squarks. Thus, EDMs lead to severe constraints on
the phases of $A_{q}$ for $q\in\{ u,d,s,c\}$ and $A_l$ for $l\in\{e,\mu\}$.
Using the convention that the phase of 
the wino soft-breaking mass $M_2$ is rotated away, one finds tight
constraints on the phase of~$\mu$~\cite{Barger:2001nu}.
On the other hand constraints on the phases of the third-generation
trilinear couplings are significantly weaker.
We refer the reader to \citere{Li:2010ax} for a review.
While recent constraints from EDMs~\cite{Nakai:2016atk}
taking into account two-loop contributions~\cite{Giudice:2005rz}
have the potential to rule out the largest values of the 
phase of $A_t$, there is still significant room for variation of the phases
of $A_t$ and $M_3$. We therefore display the full range of the phases of
$A_t$ and $M_3$ in our considered scenarios without explicitly imposing EDM
constraints, following the common approach in benchmark scenarios for 
Higgs phenomenology (see e.g.\ \citere{Carena:2015uoe} for a recent
discussion).
It should be noted in this context that in particular the variation of $A_t$
affects the value of the stop masses.
Additionally, the Higgs boson masses are a function of the phases of the complex parameters.
The impact is particularly pronounced for the mass of the light Higgs boson. In order to factor out
the impact of phase space effects, we normalise 
the prediction for the cross section of the light Higgs boson in the \mssm{} to the
cross section of a \sm{} Higgs boson with identical mass
as the light Higgs mass eigenstate $m_{h_1}$. In case of the heavy Higgs bosons
for which the phase space effects are much less severe,
we stick to the inclusive cross sections without such a normalisation.
The predicted value for the Higgs boson mass $m_{h_1}$ deviates from $125$\,GeV
by up to a few GeV in our illustrative studies. 
Deviations from the experimental value in this ballpark are still
commensurate with the remaining theoretical uncertainties from unknown
higher-order corrections of current state-of-the-art calculations of the light Higgs boson mass in 
the \mssm{}~\cite{deFlorian:2016spz}.

Subsequently we discuss three aspects: We start with a discussion
of squark effects for the Higgs boson production cross sections.
They are of relevance both for the heavy Higgs bosons and the light Higgs boson.
Secondly, we focus on the admixture of the two heavy Higgs bosons
(described through $\zhat$ factors) and its effect on production cross sections.
Lastly we discuss $\Delta_b$ corrections in the context of
the $\mhmod$-inspired scenario with large $\tan\beta$, for
which the bottom-quark annihilation process
for the heavy Higgs bosons is relevant as well.

Note that given the large admixture of the two heavy Higgs bosons in the \mssm{}
with complex parameters, interference effects in the full processes
of production and decay can be large.
However, we restrict our discussion in the present paper to Higgs boson production.
The results for the cross sections obtained in our paper can be employed in
a generalised narrow-width approximation as described in \citere{Fuchs:2014ola} 
in order to incorporate interference effects.
We will address this issue elsewhere.

The prediction for Higgs boson cross sections is affected by various
theoretical uncertainties, which we discuss in detail in
\sct{sec:uncertainties}. In order to demonstrate the improvement
in precision through the inclusion of higher-order corrections, all
subsequent figures which show
the \lo{} cross section and our best prediction cross section
according to \eqn{eq:ggphimaster} include renormalisation
and factorisation scale uncertainties.
The procedure for obtaining these scale uncertainties is outlined in \sct{sec:uncertainties}.

\subsection{Squark contributions in the light-stop inspired scenario}

\begin{figure}[h]
\begin{center}
\begin{tabular}{cc}
\includegraphics[width=0.47\textwidth]{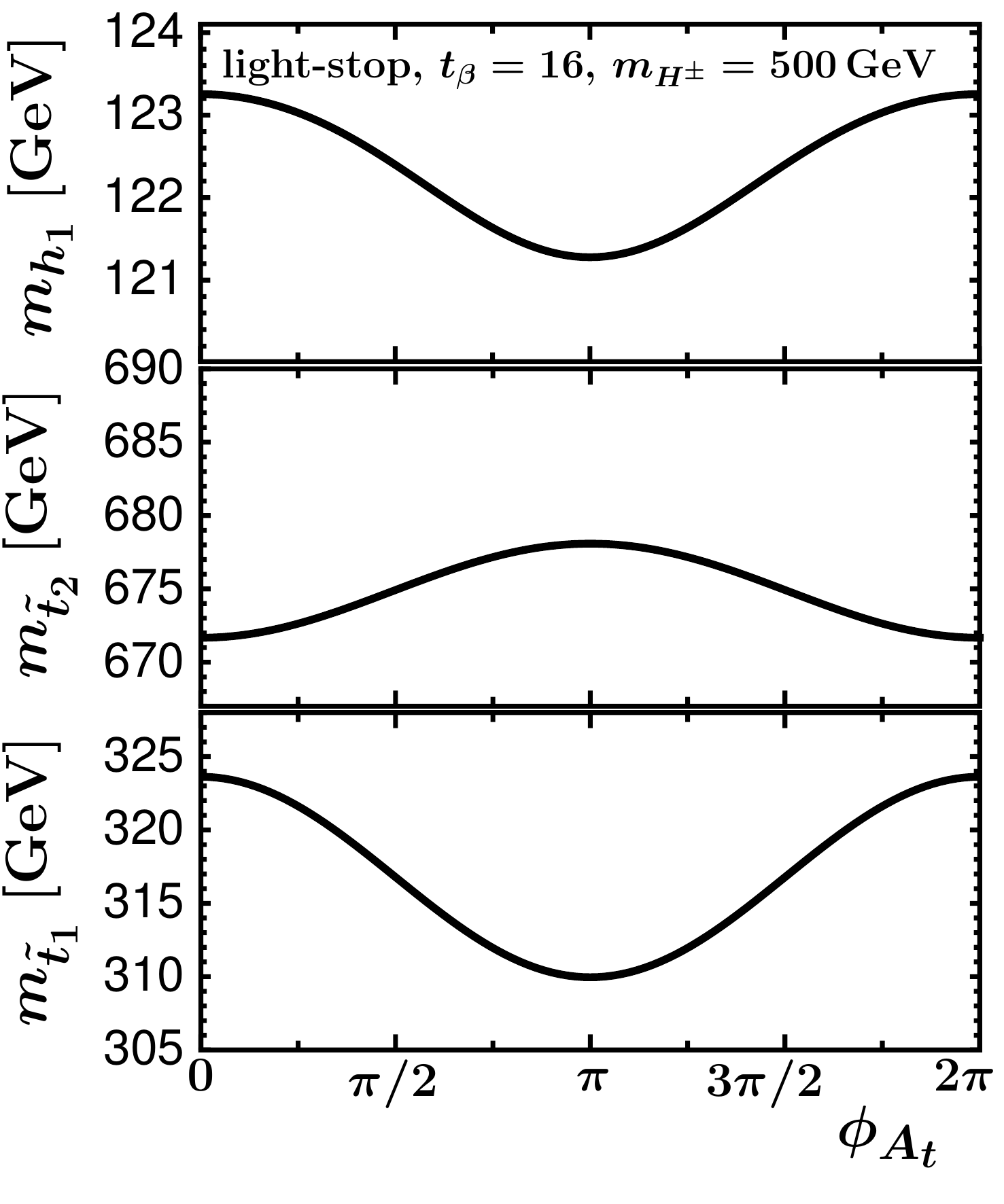} &
\includegraphics[width=0.47\textwidth]{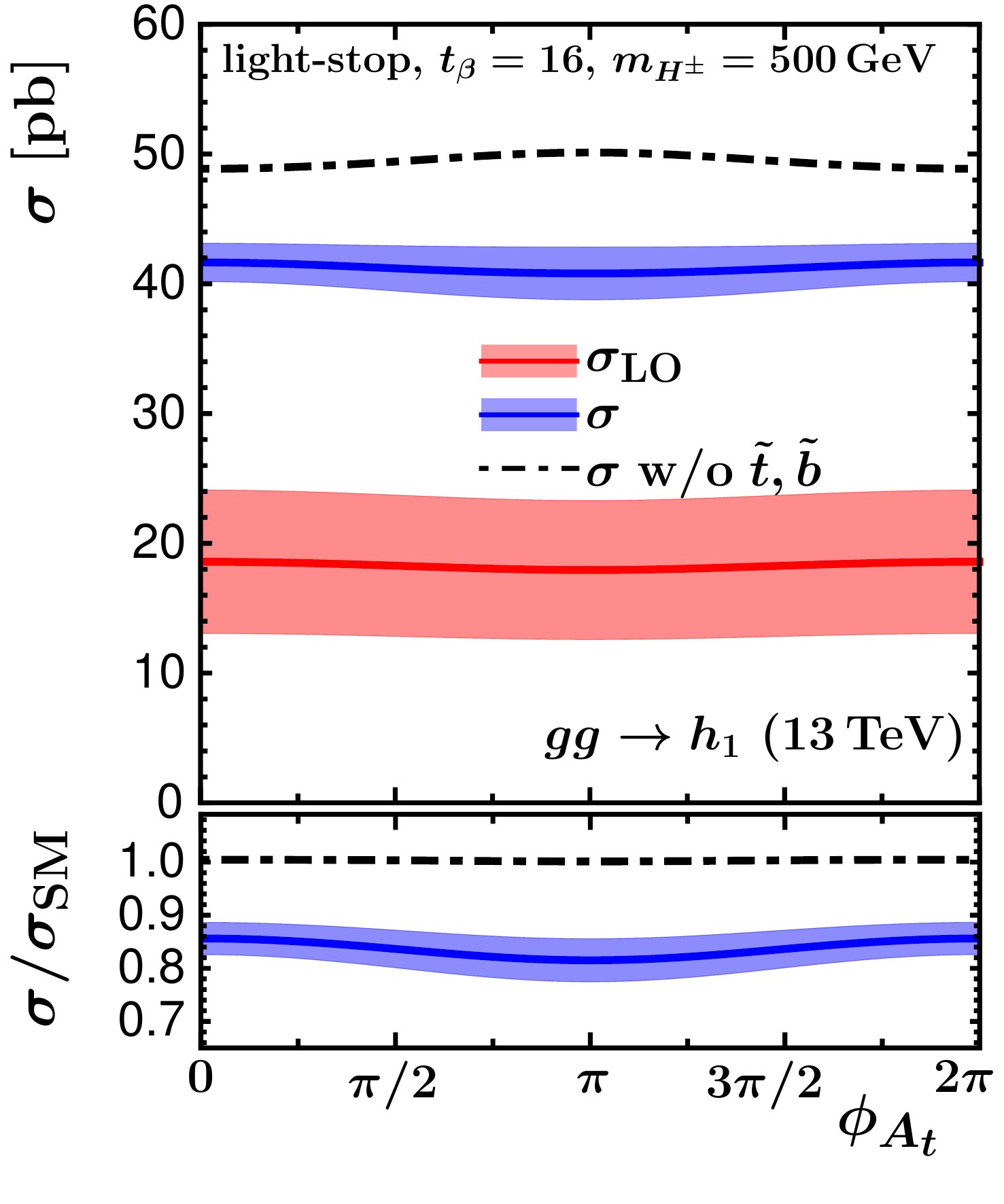}  \\[-.5cm]
 (a) & (b) \\
\end{tabular}
\end{center}
\vspace{-0.2cm}
\caption{(a) Mass of $h_1$ and stop masses in GeV as a function of $\phi_{A_t}$;
(b) \lo{} (red) and best prediction for the gluon-fusion cross section (blue) for the light Higgs~$h_1$ in pb
as a function of $\phi_{A_t}$. 
The results are shown for the light-stop inspired scenario as specified
in \eqn{eq:lightstopscen}.
The black, dot-dashed curve depicts the best prediction cross section
without squark contributions (except through $\zhat$ factors).
The depicted uncertainties are scale uncertainties.
In the lower panel we normalise to the cross section of a \sm{} Higgs boson with the same mass $m_{h_1}$.}
\label{fig:lhlightstop}
\end{figure}

We start with a discussion of squark effects to the Higgs boson cross section
$\sigma(gg\to h_i)$ for all three Higgs bosons $h_i$ in the context of the
light-stop inspired scenario with $m_{H^\pm}=500$\,GeV and $\tan\beta=16$.
The variation of the light Higgs boson mass~$m_{h_1}$ as well as the
stop masses is depicted in \fig{fig:lhlightstop}~(a) as a function of $\phi_{A_t}$.
The light Higgs mass in this scenario may appear to be too light 
to be compatible
with the signal observed at the \lhc{}, however we regard it as
sufficiently close in view of the facts that our 
discussion should demonstrate phenomenological effects only and that on the
other hand there are still sizeable theoretical uncertainties in the
\mssm{} prediction for the light Higgs boson mass.
The lightest stop mass has its
minimum at around $308$\,GeV. For the heavy Higgs bosons with
masses between $492$ and $494$\,GeV, which are not shown in the figures,
the \nlo{} squark-gluino contributions~\cite{Degrassi:2010eu,Degrassi:2011vq,Degrassi:2012vt} which assume
heavy squarks and gluinos are thus well applicable.
The variation of the heavy Higgs boson masses as a function of the phases
of $\phi_{A_t}$ and $\phi_{M_3}$ turns out to be
small, namely within $0.6$\,GeV.
Due to the strong admixture of the left- and right-handed stops through a large value of $A_t$,
also a phase dependence of the stop masses is observed. We checked that if we instead choose a phase
for $X_t$ keeping $|X_t|$ constant, we obtain constant stop masses if
the phase of $X_t$ is varied.

\fig{fig:lhlightstop}~(b) shows the production cross section through gluon fusion
for the light Higgs boson~$h_1$. The black, dot-dashed curve depicts the cross
section with top-quark and bottom-quark contributions and electroweak
corrections in the production amplitudes only, i.e.\ in the formulas of \sct{sec:sushi} we omit all squark
contributions which enter either directly or through $\D_b$. Note however that
squark contributions are always part of the $\zhat$ factors.
Due to the decoupling with large values of $m_{H^\pm}$ in our scenarios
the light Higgs~$h_1$ has mostly \sm{}-like couplings to quarks and gauge bosons.
Thus, thanks to the inclusion of \nklo{3} \qcd{} contributions
for the top-quark induced contribution, our prediction of the gluon fusion 
cross section omitting the squark contributions (black, dot-dashed curve 
in \fig{fig:lhlightstop}~(b))
is very close to the one for the \sm{} Higgs boson with the same mass
as provided by the \lhc{} Higgs Cross Section Working 
Group~\cite{deFlorian:2016spz,Forte:2194224}.
The inclusion of squark contributions explicitly and through $\Delta_b$ resummation lowers
the gluon-fusion cross section by about $20$\%, as can be inferred from the blue, solid curve,
which is shown together with its renormalisation and factorisation scale uncertainty,
see \sct{sec:uncertainties}.
For completeness we also show the \lo{} cross section calculated according to \eqn{eq:xs} including squark effects as the red curve.
It is apparent that the scale uncertainties are significantly
reduced from \lo{} \qcd{} to our best
prediction cross section calculated according to \eqn{eq:ggphimaster}. \fig{fig:lhlightstop}~(b)
also includes the cross section for $h_1$ normalised to the cross section of a \sm{} Higgs boson with
the same mass. Here the $\sim 20$\% reduction due to squark effects is apparent once again,
whereas the quark-induced cross section shows the well-known decoupling behaviour.
Not shown in the figures are the following effects, which we state
here for completeness:
The variation of $\phi_{M_3}$ leads to a very similar picture, even though
the light Higgs mass variation is not as pronounced and the stop masses
are unaffected. Moreover, in the comparison of the simplified
and the full resummation of $\D_b$ contributions in
the \lo{} gluon-fusion cross section of $h_1$ we observe a well-known behaviour,
namely the simplified resummation of \eqn{eq:simpledeltab} does not yield a decoupled bottom-quark Yukawa coupling,
whereas the full resummation of \eqn{eq:fulldeltab} does.

\begin{figure}[h]
\begin{center}
\begin{tabular}{cc}
\includegraphics[width=0.47\textwidth]{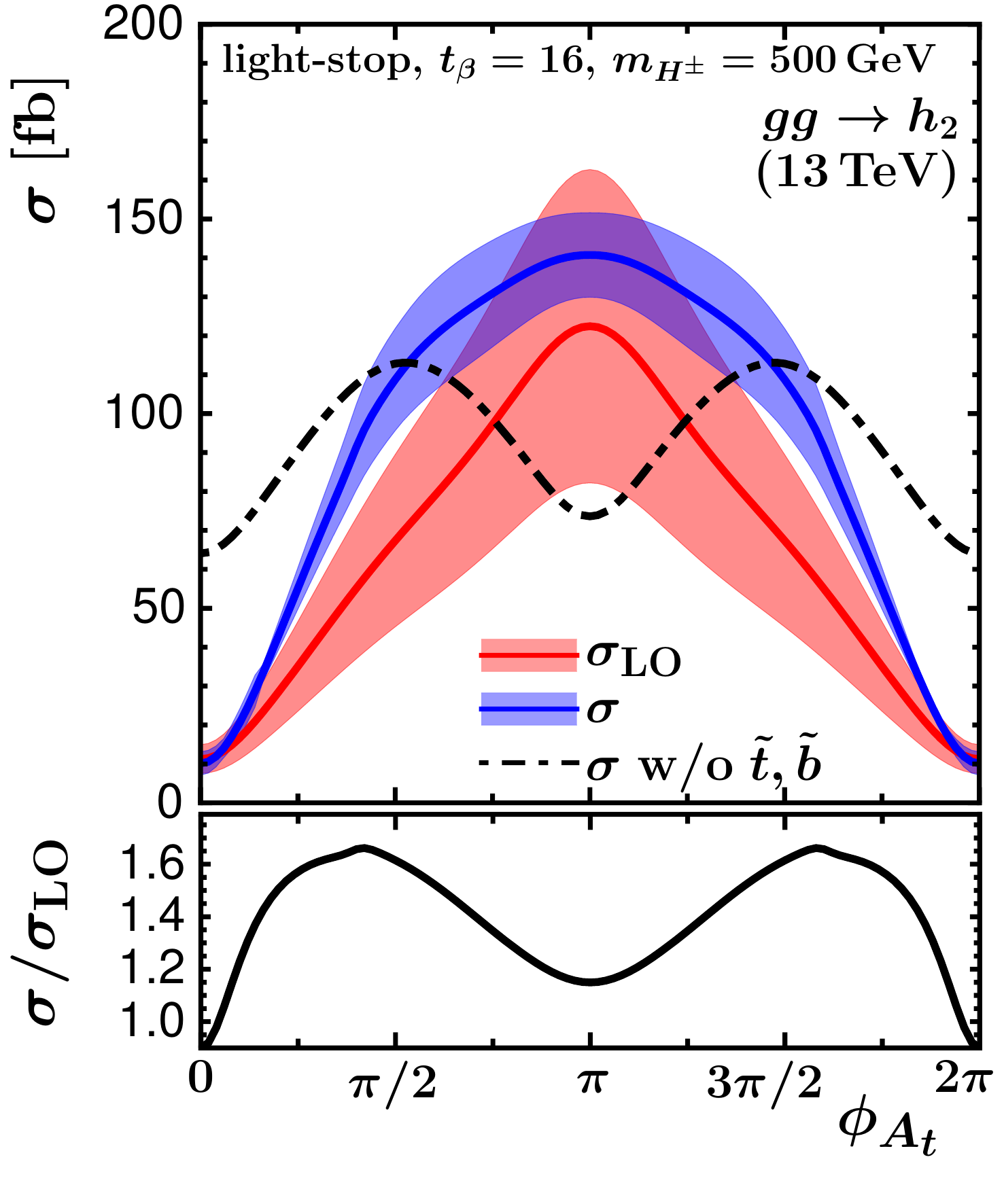} &
\includegraphics[width=0.47\textwidth]{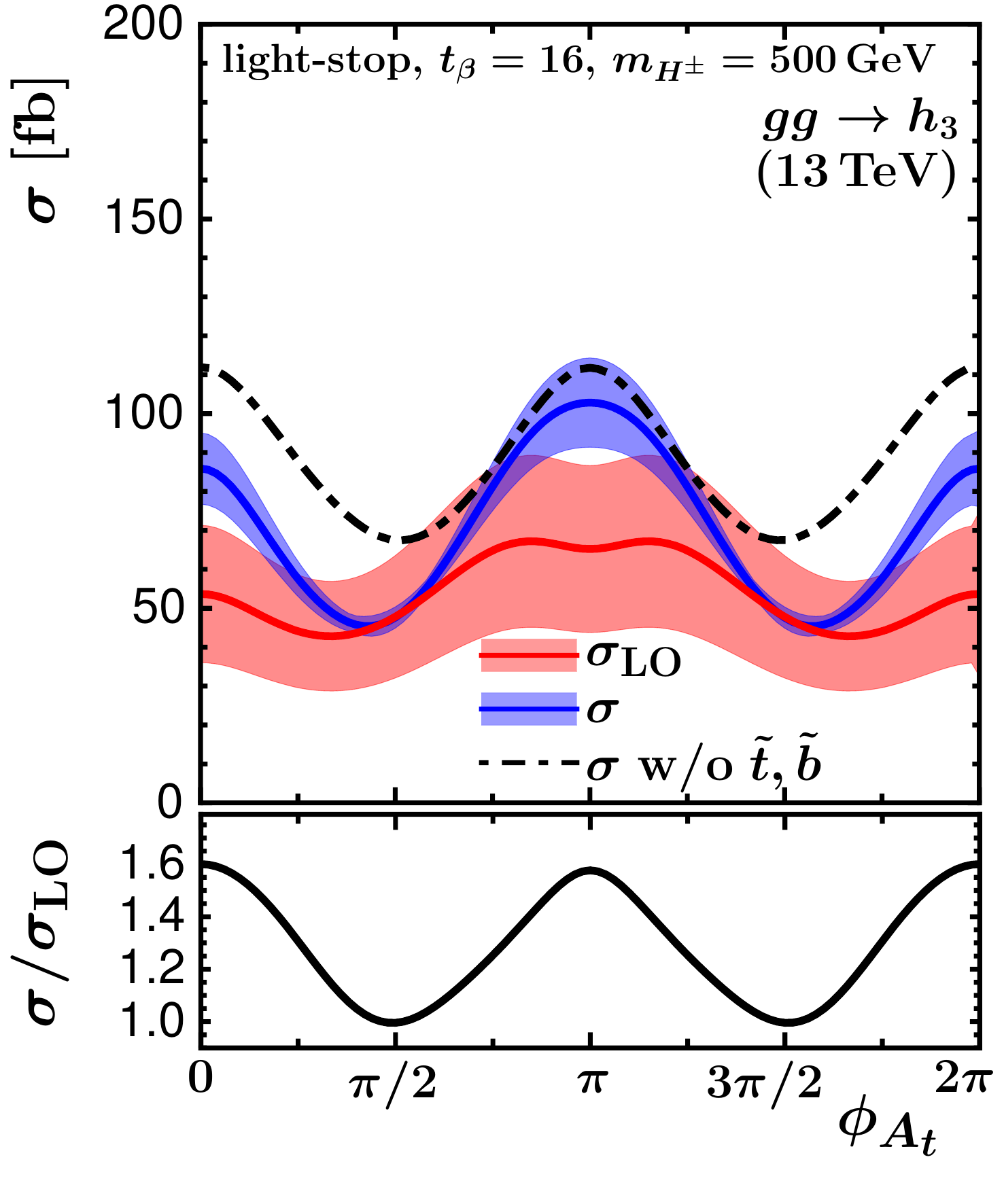}  \\[-.5cm]
 (a) & (b)
\end{tabular}
\end{center}
\vspace{-0.2cm}
\caption{\lo{} (red) and best prediction for the gluon-fusion cross section (blue) for (a) $h_2$ and (b) $h_3$ in fb
as a function of $\phi_{A_t}$. 
The results are shown for the light-stop inspired scenario as specified
in \eqn{eq:lightstopscen}.
The black, dot-dashed curve depicts the best prediction cross section
without squark contributions (except through $\zhat$ factors).
The depicted uncertainties are scale uncertainties.
In the lower panel we show the $K$-factor $\sigma/\sigma_{\lo}$.}
\label{fig:lightstop}
\end{figure}

In \fig{fig:lightstop} (a) and (b) we show the gluon-fusion cross sections of the heavy Higgs bosons $h_2$
and $h_3$, respectively, as a function of $\phi_{A_t}$. The colour coding is identical to \fig{fig:lhlightstop}
except for the fact that we show the $K$-factor of our best prediction for the cross section
with respect to the \lo{} cross section, $\sigma/\sigma_{\lo{}}$, rather than a cross section normalised to the \sm{}
Higgs boson cross section. In fact, the heavy Higgs masses
change only slightly 
as a function of the phase $\phi_{A_t}$, and therefore the associated 
phase space effect
is small. For vanishing phase $\phi_{A_t}=0$
it is known that squark effects are huge and reduce the cross section by $\sim 89$\% ($h_2$)
and $\sim 22$\% ($h_3$)~\cite{Bagnaschi:2014zla}. These squark effects are strongly dependent
on the phase $\phi_{A_t}$ and induce a large positive correction at phase $\phi_{A_t}=\pi$
in case of $h_2$. For $h_3$ the effects are not as pronounced, but still sizeable.
The $K$-factor for both processes $gg\to h_2$ and $gg\to h_3$ remains within
$[1,1.6]$, i.e.\ higher-order corrections mainly
follow the phase dependence of the \lo{} cross section. The dependence
of the $K$-factor on $\phi_{A_t}$ follows
the black, dot-dashed curve, which shows the cross section with quark contributions only.
The significant dependence of the cross section where only quark
contributions are included on the phase $\phi_{A_t}$
is induced by the admixture of the two Higgs bosons through $\zhat$ factors.
We will discuss this feature in detail for the $\mhmod$-inspired scenario
in \sct{sec:admix}.

\begin{figure}[h]
\begin{center}
\begin{tabular}{cc}
\includegraphics[width=0.47\textwidth]{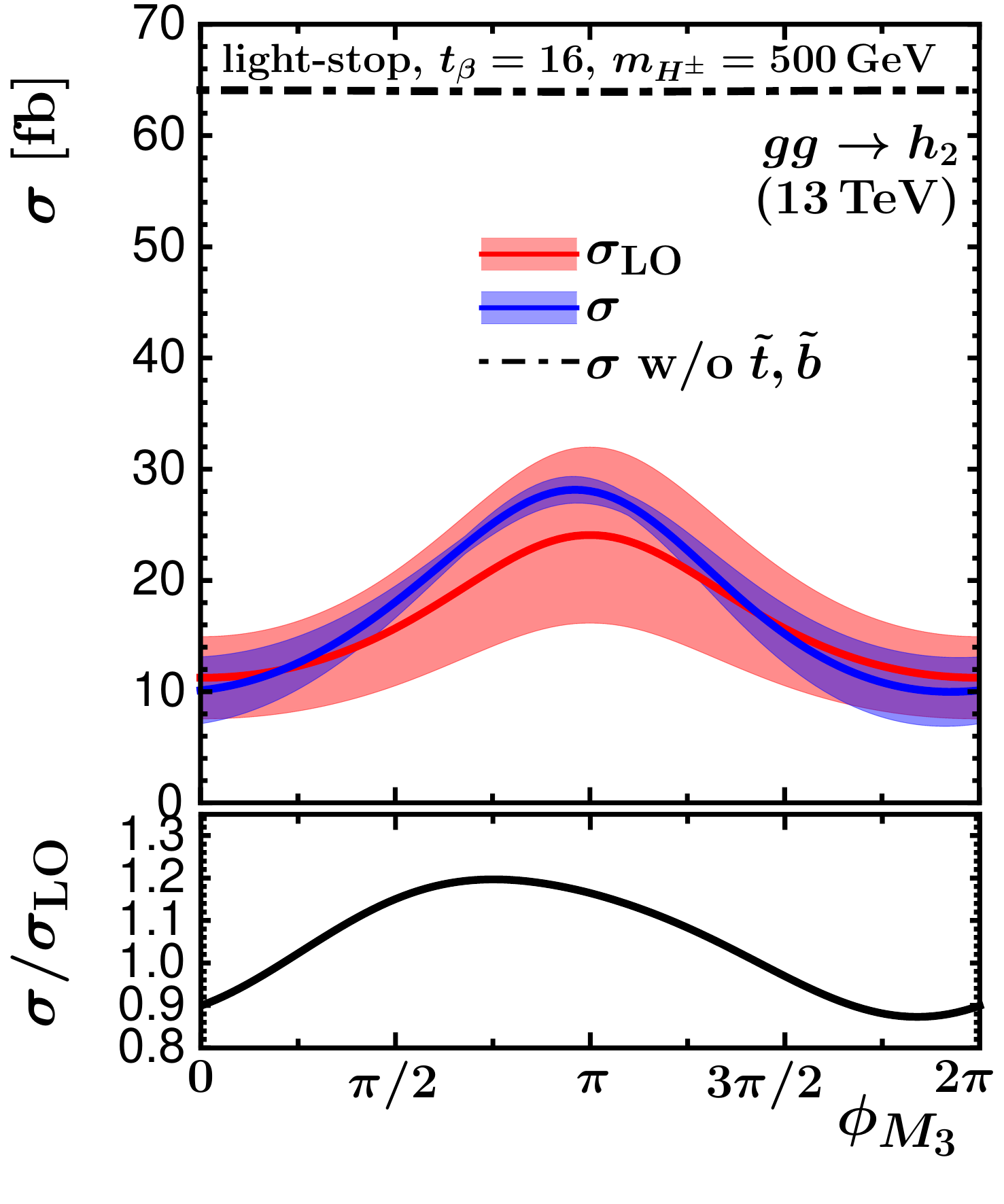} &
\includegraphics[width=0.47\textwidth]{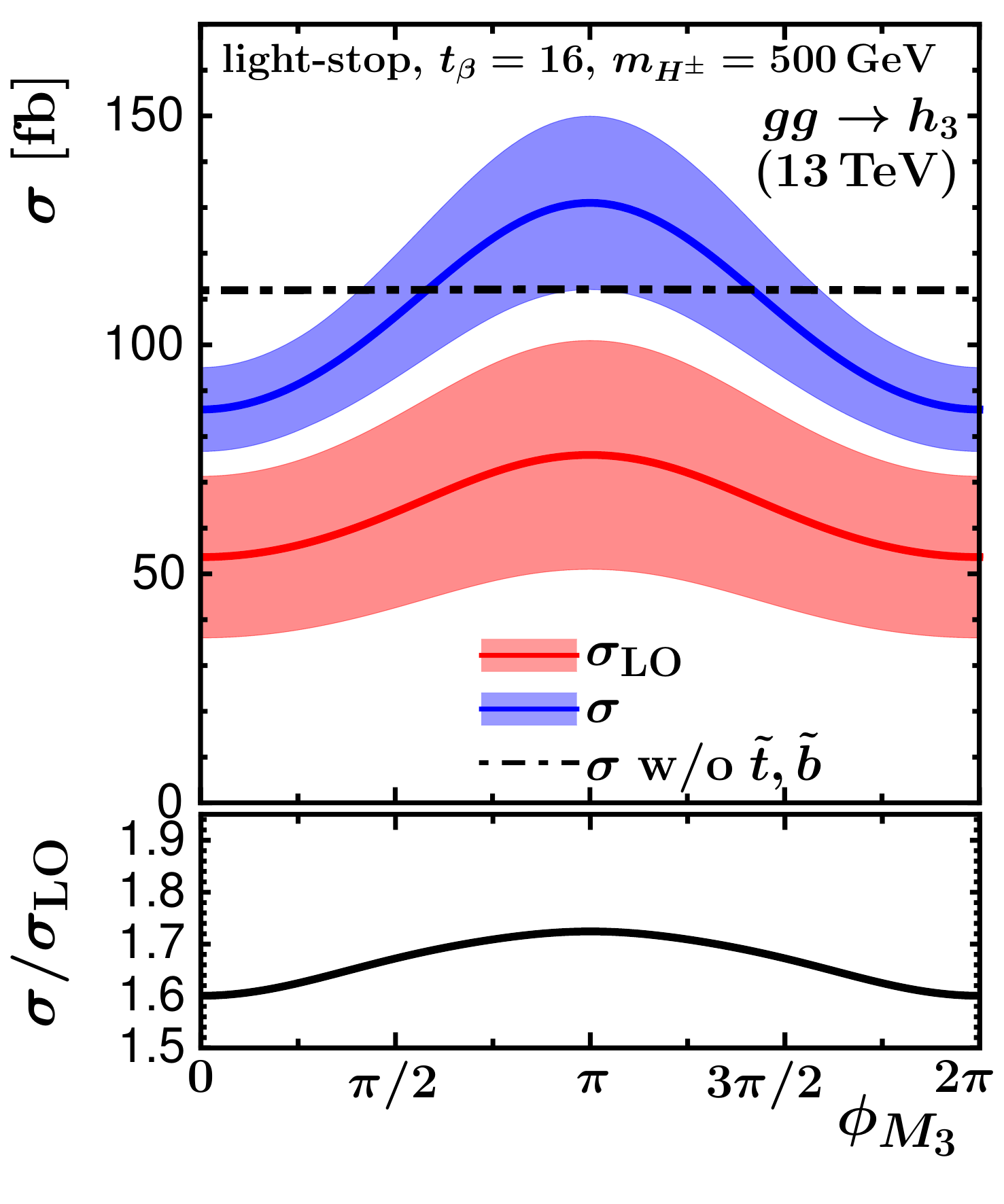}  \\[-.5cm]
 (a) & (b)
\end{tabular}
\end{center}
\vspace{-0.2cm}
\caption{\lo{} (red) and best prediction for the gluon-fusion cross section (blue) for (a) $h_2$ and (b) $h_3$ in fb
as a function of $\phi_{M_3}$. 
The results are shown for the light-stop inspired scenario as specified
in \eqn{eq:lightstopscen}.
The black, dot-dashed curve depicts the best prediction cross section
without squark contributions (except through $\zhat$ factors).
The depicted uncertainties are scale uncertainties.
In the lower panel we show the $K$-factor $\sigma/\sigma_{\lo}$.}
\label{fig:lightstopM3}
\end{figure}

The phase dependence on $\phi_{M_3}$ is less pronounced. We show the corresponding
cross sections for the two heavy Higgs bosons~$h_2$ and $h_3$ in \fig{fig:lightstopM3}.
As in previous figures we observe a significant
reduction in the scale dependence from \lo{} \qcd{} to our best prediction
for the cross section.
The inclusion of squark and gluino contributions through 
the $\zhat$ factors and through
$\D_b$ induces a dependence on the
gluino phase already for the \lo{} cross section.
The almost flat black dot-dashed curves show the cross section
with quark contributions only, and any variation with $\phi_{M_3}$ is an
effect of the $\zhat$ factors, which in this case is negligible since
$\phi_{M_3}$ only enters at the two-loop level.
The $K$-factor, which takes into account our interpolated \nlo{} virtual corrections,
only shows a relatively mild dependence on the phase.
We will discuss the interpolation uncertainty for this scenario in \sct{sec:uncertainties},
since we obtain the largest relative interpolation uncertainty in the cross section variation with phases
for the interpolation of the gluino phase $\phi_{M_3}$.

\subsection{Admixture of Higgs bosons in the $\mhmod$-inspired scenario}
\label{sec:admix}

\begin{figure}[htb]
\begin{center}
\begin{tabular}{ccc}
\includegraphics[width=0.31\textwidth]{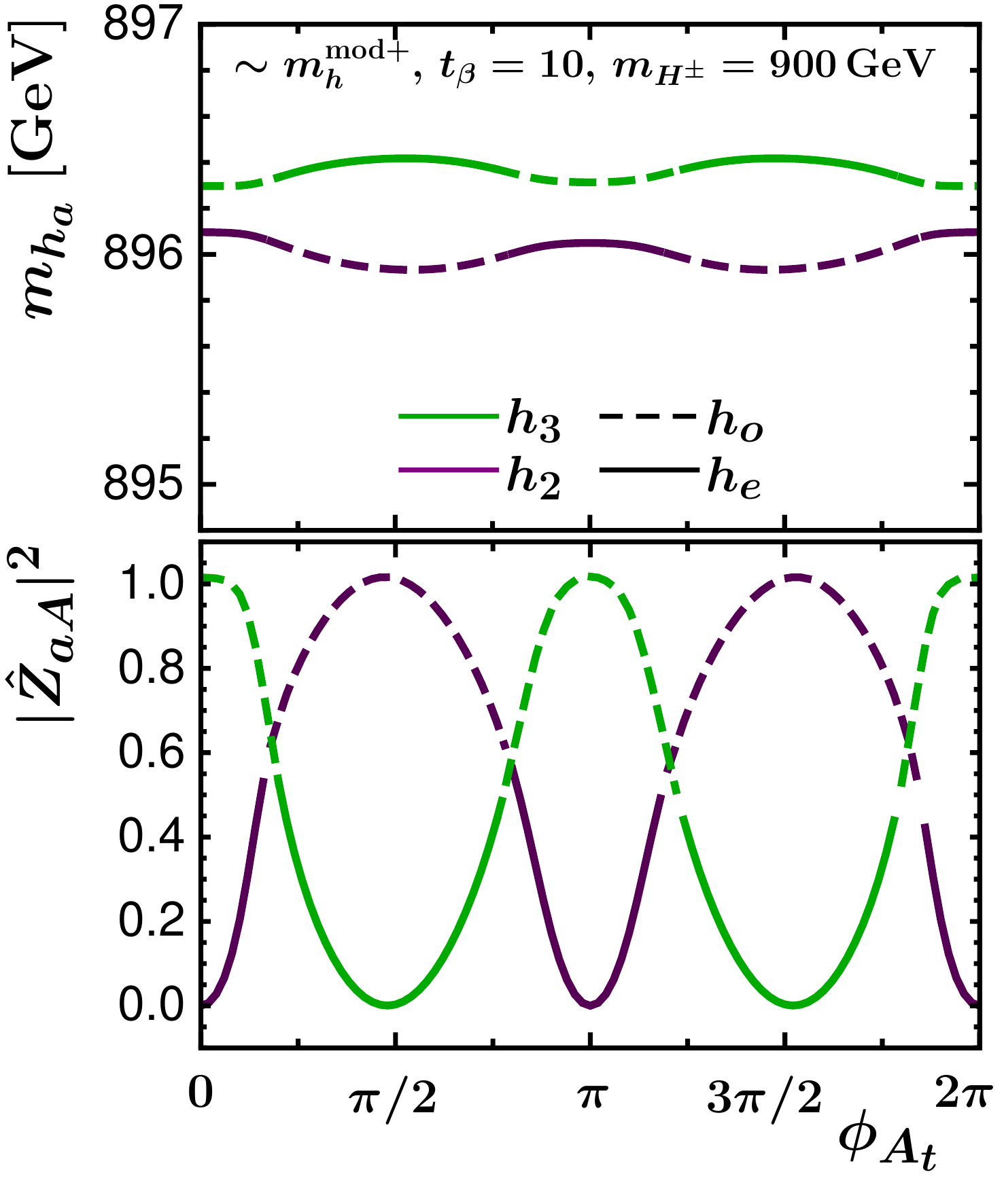} &
\includegraphics[width=0.31\textwidth]{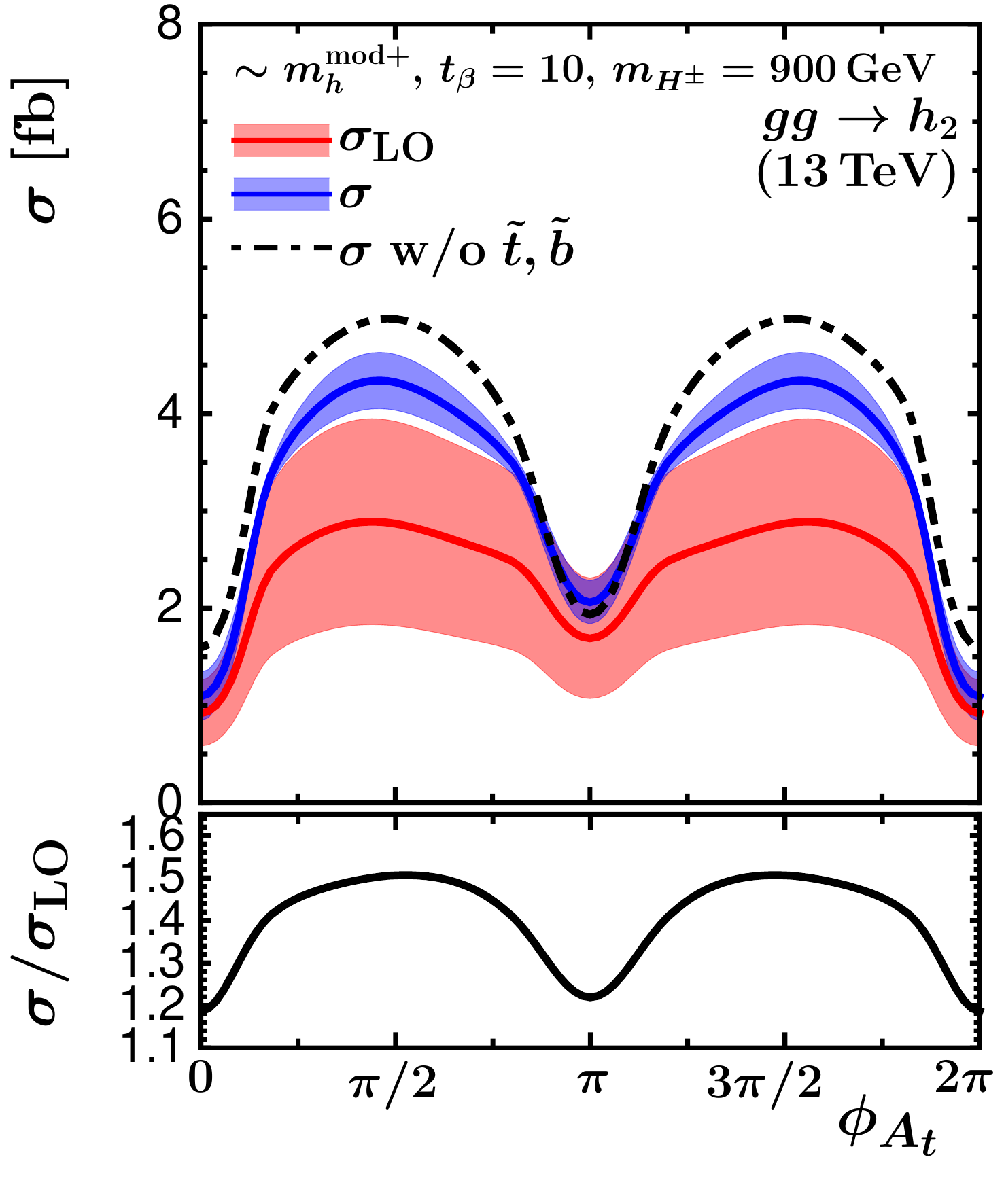} &
\includegraphics[width=0.31\textwidth]{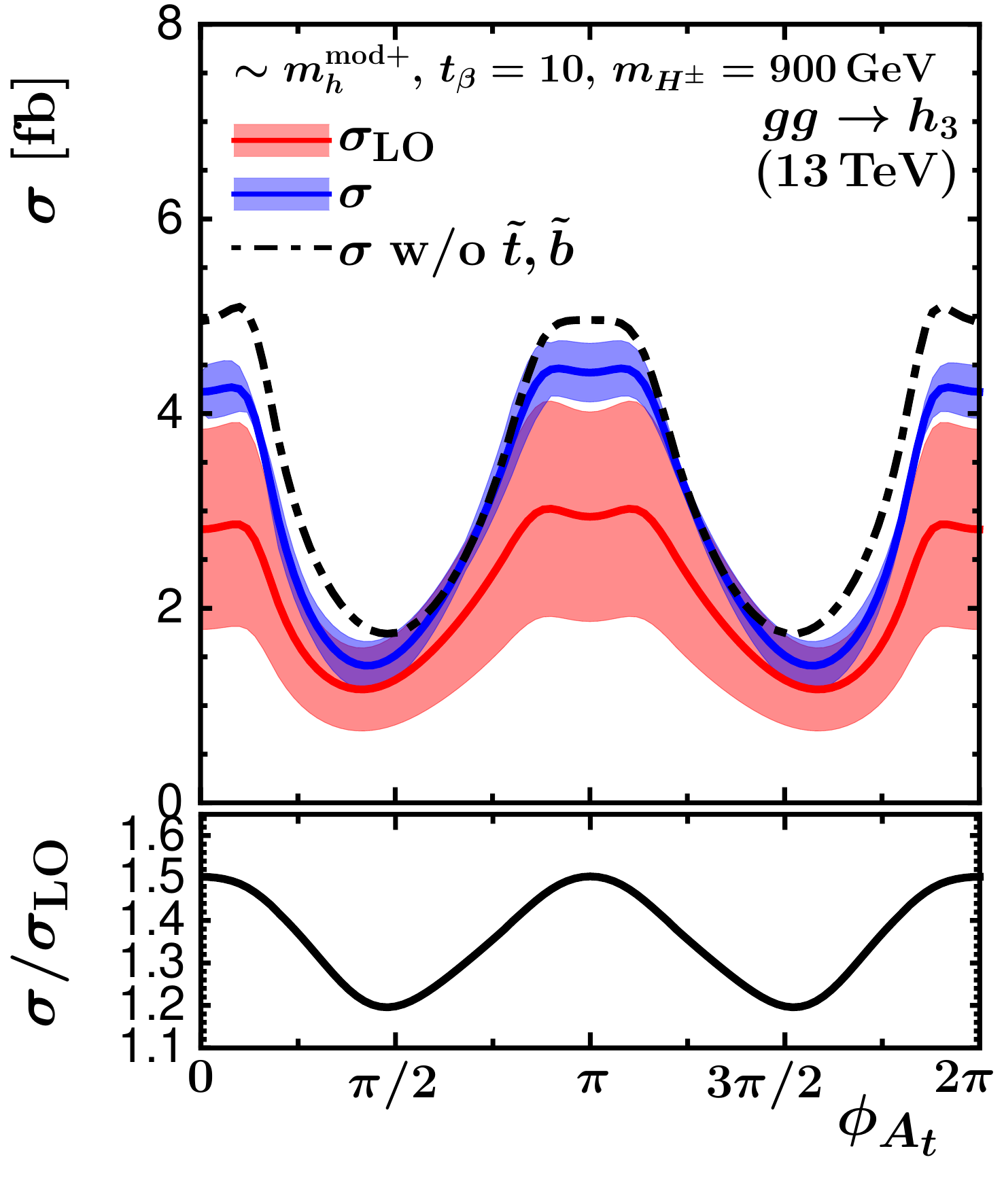} \\[-0.5cm]
 (a) & (b) & (c)\\
\end{tabular}
\end{center}
\vspace{-0.2cm}
\caption{(a) Masses of $h_2$ and $h_3$ in GeV as well as \cp{}-odd character $|\zhat_{aA}|^2$ as a function of $\phi_{A_t}$ in the $\mhmod$-inspired
scenario with $\tan\beta=10$. The solid and dashed curves depict
regions in $\phi_{A_t}$ where $h_2$ and $h_3$ are predominantly
\cp{}-even ($h_e$) or odd ($h_o$), respectively, corresponding to
$|\zhat_{aA}|^2$ being below or above $0.5$ as shown in the lower panel.
(b,c) \lo{} (red) and best prediction for the gluon-fusion cross section (blue) for (b) $h_2$ and (c) $h_3$ in fb
as a function of $\phi_{A_t}$ in the same scenario.
The black, dot-dashed curve depicts the best prediction for the
cross section
without squark contributions (except through $\zhat$ factors).
In the lower panel we show the $K$-factor $\sigma/\sigma_{\lo}$.
The depicted uncertainties are scale uncertainties.
}
\label{fig:mhmod10}
\end{figure}

In this subsection we discuss the $\mhmod$-inspired scenario with $\tan\beta=10$
and $m_{H^{\pm}}=900$\,GeV. Since the squark masses are at the TeV
level in this scenario, the numerical effect of the squark loops in the
gluon fusion vertex contributions is rather small
for the production cross section of the light Higgs boson~$h_1$.
We do not discuss the results for $h_1$ in this section. 
The results for the two heavy Higgs bosons are displayed in
\fig{fig:mhmod10}.
The effects from squark loops are at the level of about 
$\pm 20$\% in this case. The considered scenario is typical for the
decoupling region of supersymmetric theories, where a light \sm-like Higgs
boson (that is interpreted as the signal observed at about 125~GeV) is
accompanied by additional heavy Higgs bosons that are nearly
mass-degenerate. In the general case where the possibility of \cp-violating
interactions is taken into account, there can be a large mixing between the
\cp-even and \cp-odd neutral Higgs states. This feature is clearly visible
in \fig{fig:mhmod10}. The dependence on the phase $\phi_{A_t}$ is seen to be
closely correlated to the mixing character of the two neutral heavy Higgs
bosons.

\fig{fig:mhmod10}~(a) depicts the masses of the two heavy Higgs bosons
$h_2$ and $h_3$ as a function of $\phi_{A_t}$ together with
the \cp{}-odd character of $h_2$ and $h_3$,
being defined as $|\zhat_{aA}|^2$.
For illustration here and in the following we call the mass eigenstates
$h_2$ and $h_3$ either $h_e$ or $h_o$, depending on their mixing character:
if $|\zhat_{aA}|^2\gtrsim 1/2$ the mass eigenstate $h_a$ is called $h_o$,
otherwise it is called $h_e$.
It can be seen in \fig{fig:mhmod10}~(b) and (c) that
the behaviour of the cross sections as a function of $\phi_{A_t}$ closely
follows the variation in the \cp-even and \cp-odd 
character of the Higgs states.
A similar effect was already apparent in the top- and bottom-quark induced cross
sections depicted in the light-stop inspired scenario, see \fig{fig:lightstop}, however
there the effects of squark contributions are dominant.
Also in this case the scale uncertainty of our best prediction
for the cross section is significantly reduced in comparison to the uncertainty for the 
prediction in \lo{} \qcd.
The variation of the $K$-factors between about 1.2 and 1.5 with the phase 
$\phi_{A_t}$ also follows the modification of the 
mixing character of the two neutral heavy Higgs
bosons.

Since the two heavy Higgs bosons are nearly mass degenerate, it 
may not be possible in such a case to experimentally resolve the two Higgs
bosons as separate signals. Rather than the individual cross sections times
their respective branching ratios, the experimentally measurable quantity
then consists of the sum of the cross sections of the two Higgs states times
their respective branching ratios together with the interference
contribution involving the two Higgs states. The latter can be particularly
important if the mass difference between the two Higgs states is smaller
than the sum of their total widths~\cite{Fuchs:2014ola}. While we 
defer the incorporation of such interference effects into the prediction
for the production and decay process to a forthcoming publication, one can
already infer from the plots of \fig{fig:mhmod10}~(b) and (c) 
that in the overall contribution there will be sizeable
cancellations between the phase dependencies of the separate contributions.

\subsection{\texorpdfstring{$\Delta_b$}{Delta\_b} corrections in the $\mhmod$-inspired scenario}\label{sec:deltab}

\begin{figure}[htb]
\begin{center}
\begin{tabular}{cc}
\includegraphics[width=0.47\textwidth]{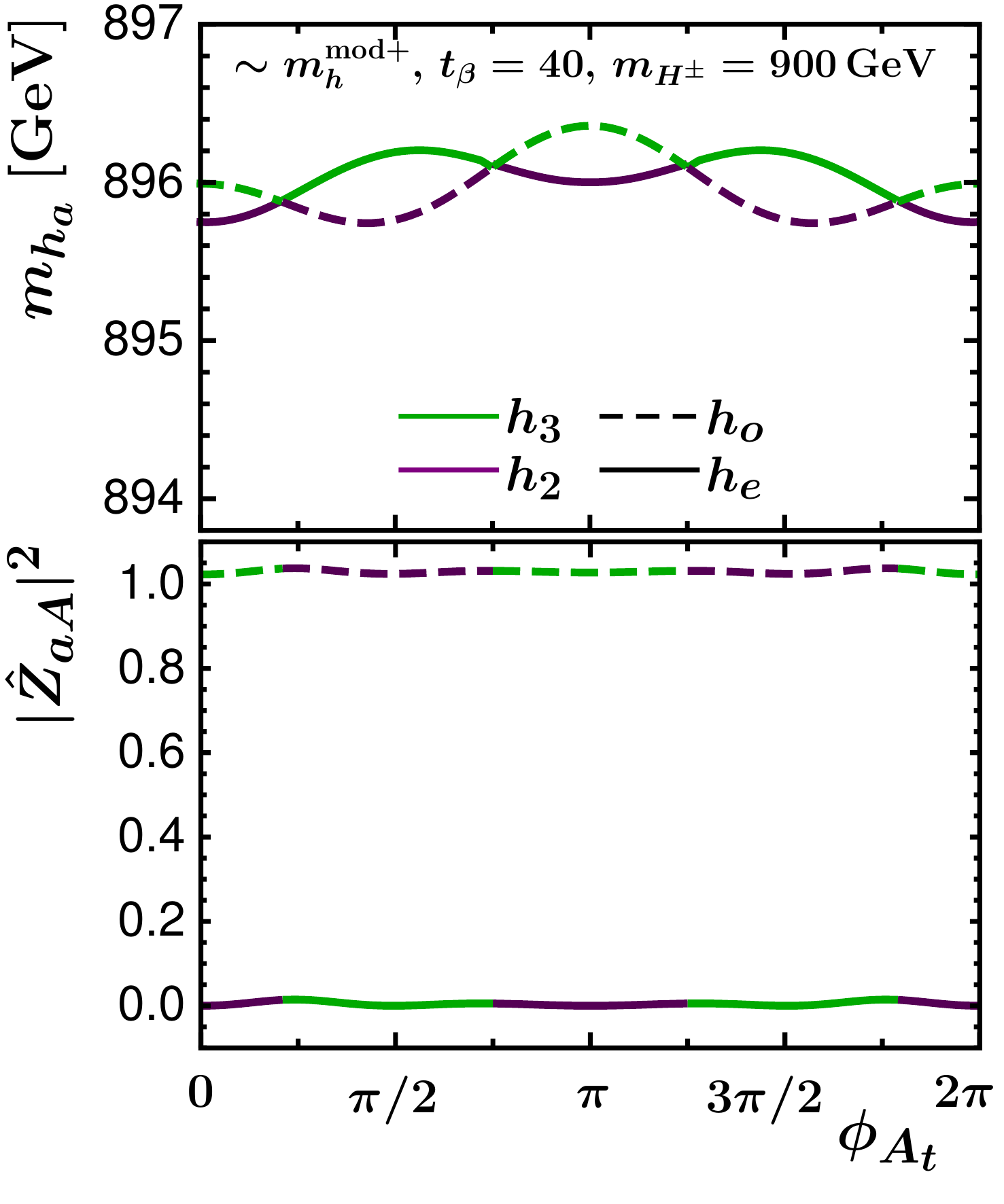} &
\includegraphics[width=0.47\textwidth]{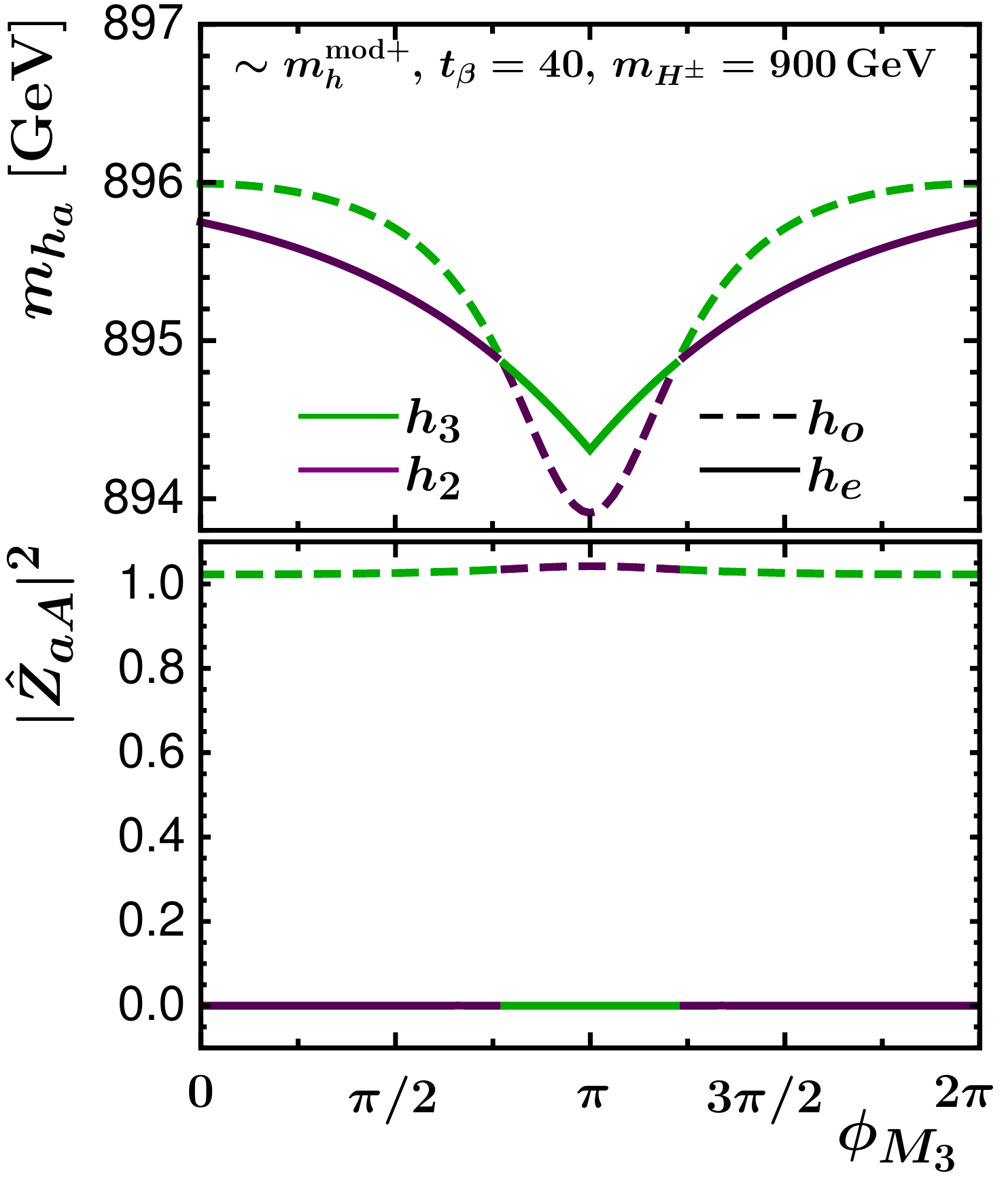}
\\[-.5cm]
 (a) & (b) \\
\end{tabular}
\end{center}
\vspace{-0.2cm}
\caption{Masses of $h_2$ and $h_3$ in GeV and \cp{}-odd character
as a function of (a) $\phi_{A_t}$ and (b) $\phi_{M_3}$ in the
$\mhmod$-inspired scenario with $\tan\beta=40$. 
As in \fig{fig:mhmod10}~(a),
the solid and dashed curves refer to $h_e$ and $h_o$, respectively.
}
\label{fig:mhmodmasses}
\end{figure}

We finally discuss the impact of $\Delta_b$ effects, which we investigate for the two heavy Higgs bosons in the $\mhmod$-inspired
scenario with $\tan\beta=40$. In this scenario the admixture between the two heavy Higgs bosons is
again sizeable both as a function of $\phi_{A_t}$ and as a function of
$\phi_{M_3}$. This even leads to mass crossings as seen in \fig{fig:mhmodmasses}. 
It is therefore convenient to discuss the results in terms of the
predominantly \cp-even mass eigenstate $h_e$ and the 
predominantly \cp-odd mass eigenstate $h_o$, as defined in 
\sct{sec:admix}, as for those states a smooth behaviour of the cross section 
as function of the phases is obtained.
The masses of the two heavy Higgs bosons and their \cp-character 
(defining $h_o$ and $h_e$) are shown 
in \fig{fig:mhmodmasses} as a function of $\phi_{A_t}$ and $\phi_{M_3}$.
One can see that the states $h_2$ and $h_3$ drastically change their \cp\
character upon variation of the phases $\phi_{A_t}$ and $\phi_{M_3}$, while
on the other hand the state $h_e$ is almost purely \cp-even and $h_o$ is
almost purely \cp-odd for the whole range of phase values. 
It should be kept in mind in this context 
that $|\zhat_{aA}|^2$ arises from a non-unitary
matrix and can therefore have values above 1.
For vanishing phases the mass eigenstate $h_2$ corresponds to 
$h_e$ and $h_3$ to $h_o$.

\begin{figure}[htb]
\begin{center}
\begin{tabular}{cc}
\includegraphics[width=0.47\textwidth]{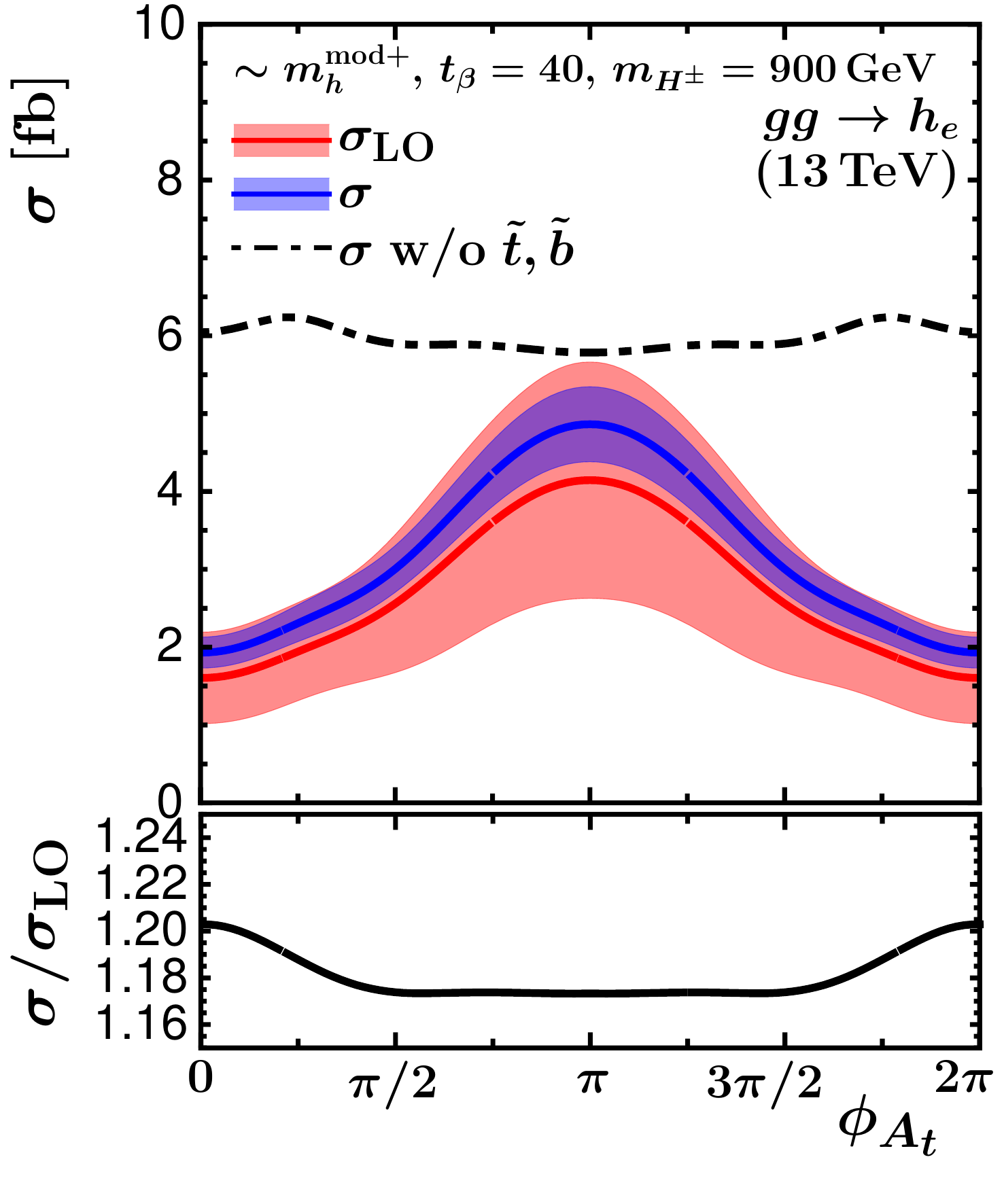} &
\includegraphics[width=0.47\textwidth]{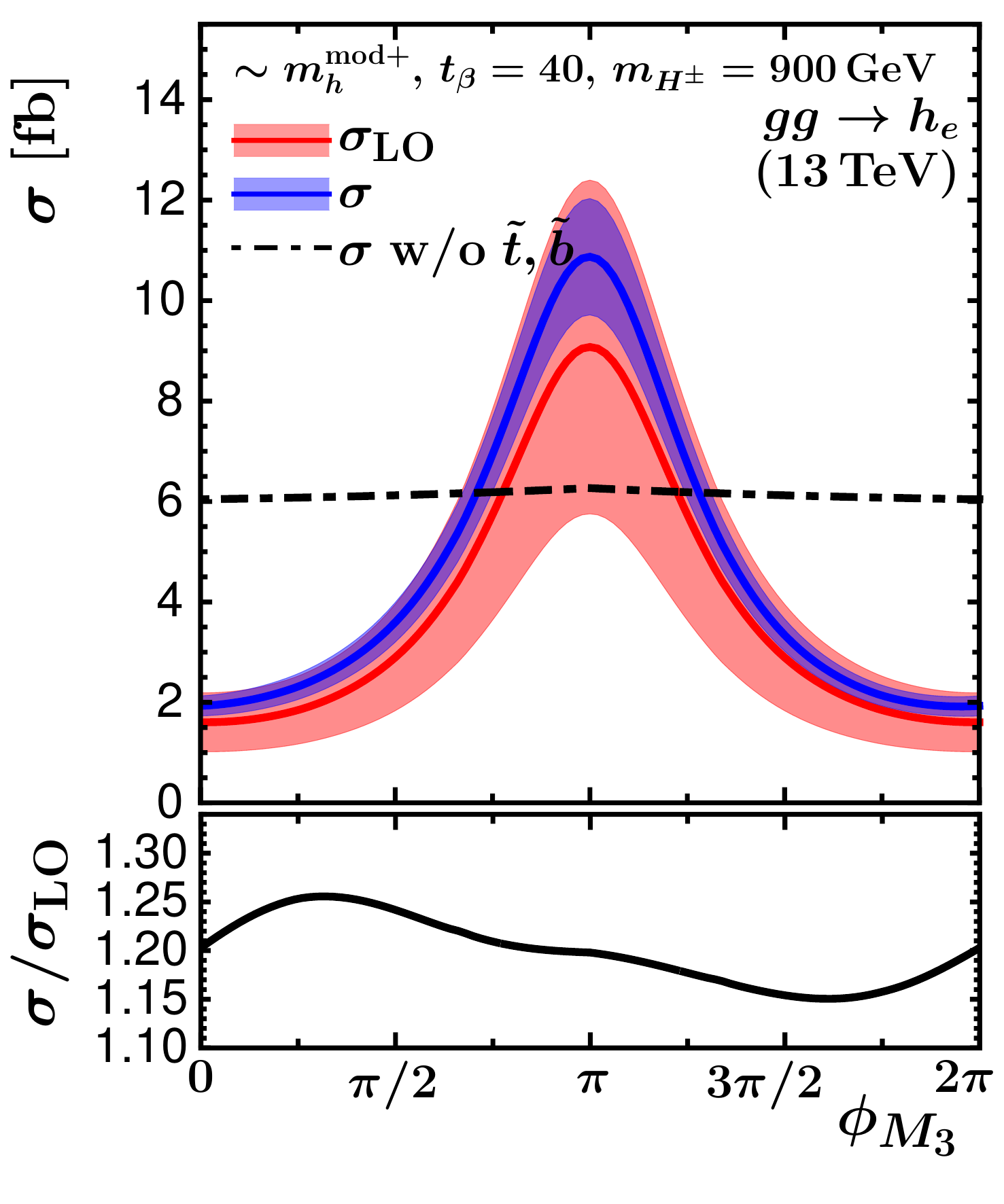}  \\[-.5cm]
 (a) & (b)
\end{tabular}
\end{center}
\vspace{-0.2cm}
\caption{\lo{} (red) and best prediction gluon-fusion cross section (blue) for $h_e$ in fb
as a function of (a) $\phi_{A_t}$ and (b) $\phi_{M_3}$ in the $\mhmod$-inspired scenario with $\tan\beta=40$.
The black dot-dashed curves depict the best prediction cross section
without squark contributions (except through $\zhat$ factors).
In the lower panel we show the $K$-factor $\sigma/\sigma_{\lo}$.
The depicted uncertainties are scale uncertainties.
}
\label{fig:mhmod}
\end{figure}

In the following we show results for the 
predominantly \cp-even mass eigenstate $h_e$.
The observations for $h_o$ are very similar and are not shown here, 
we will only add comments where appropriate.
In \fig{fig:mhmod} we show the gluon-fusion cross section as a function
of the phases $\phi_{A_t}$ and $\phi_{M_3}$. In both cases the behaviour 
for the full prediction, including the squark contributions, 
is dominated by $\Delta_b$ corrections. For vanishing phases those
corrections significantly reduce the cross sections compared to the case
where only quark contributions are taken into account. For phase values
around $\pi$, however, the $\Delta_b$ corrections 
can also give rise to a significant enhancement of the cross section.
In particular, for $\phi_{M_3}$ the quantity $\Delta_b$ changes sign 
between $\phi_{M_3}=0$ and $\phi_{M_3}=\pi$,
such that the bottom-Yukawa coupling is suppressed for small values of 
$\phi_{M_3}$ and enhanced for $\phi_{M_3}$ values close to $\pi$ as a
consequence of the resummation of the $\Delta_b$ corrections.
The reduction of the scale uncertainties from \lo{} \qcd{} to our 
best prediction for the cross section is similar as in the previous plots.
The $K$-factors in the lower panel show that the dependence of the \nlo{} cross sections
on the phases $\phi_{A_t}$ and $\phi_{M_3}$ follows a similar trend as the \lo{} cross section.
In the plot on the right, the asymmetric $K$-factor dependence on $\phi_{M_3}$ is related to the direct dependence of $\D_b$ on the phase $\phi_{M_3}$.

\begin{figure}
\begin{center}
\begin{tabular}{cc}
\includegraphics[width=0.47\textwidth]{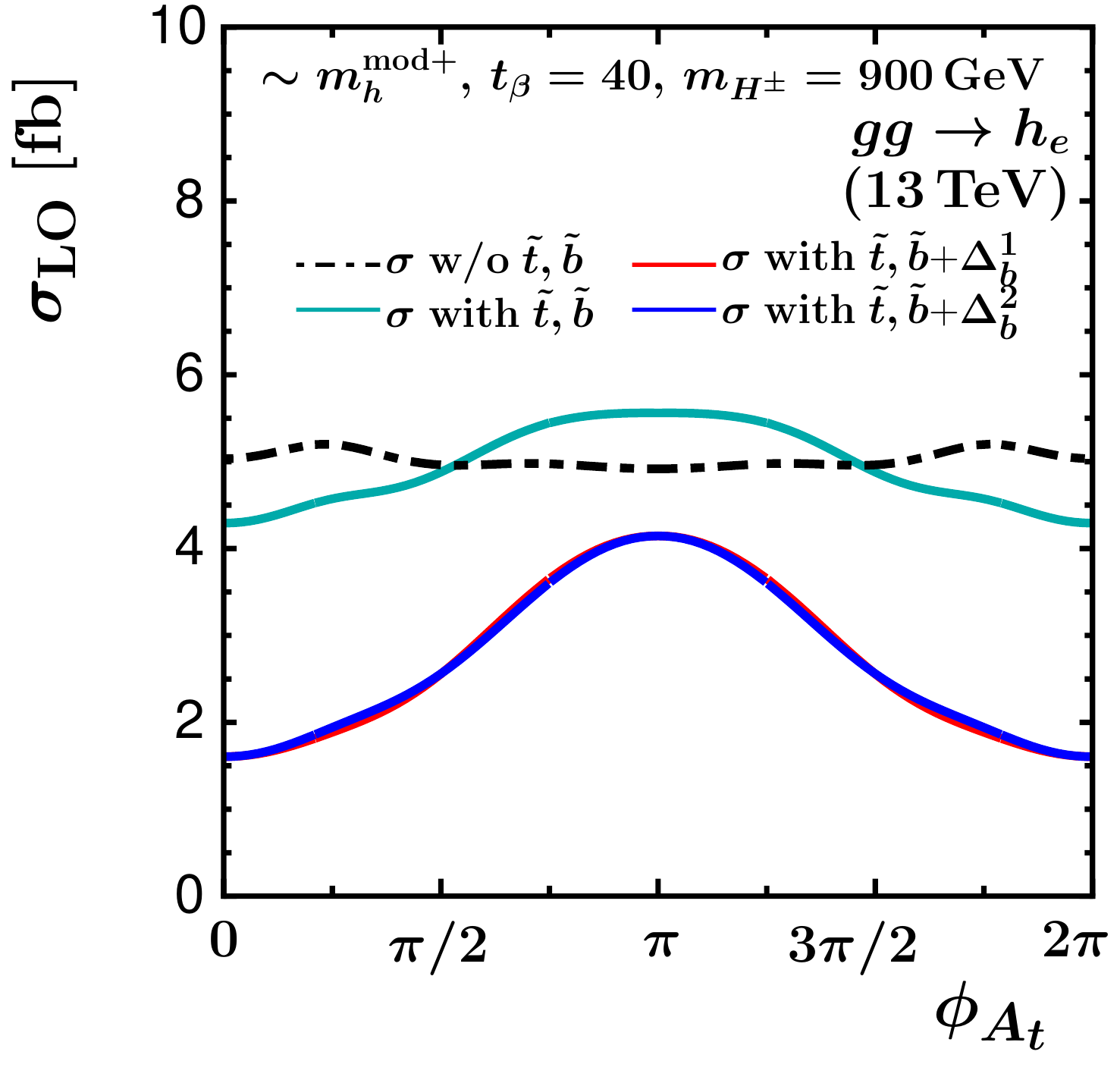} &
\includegraphics[width=0.47\textwidth]{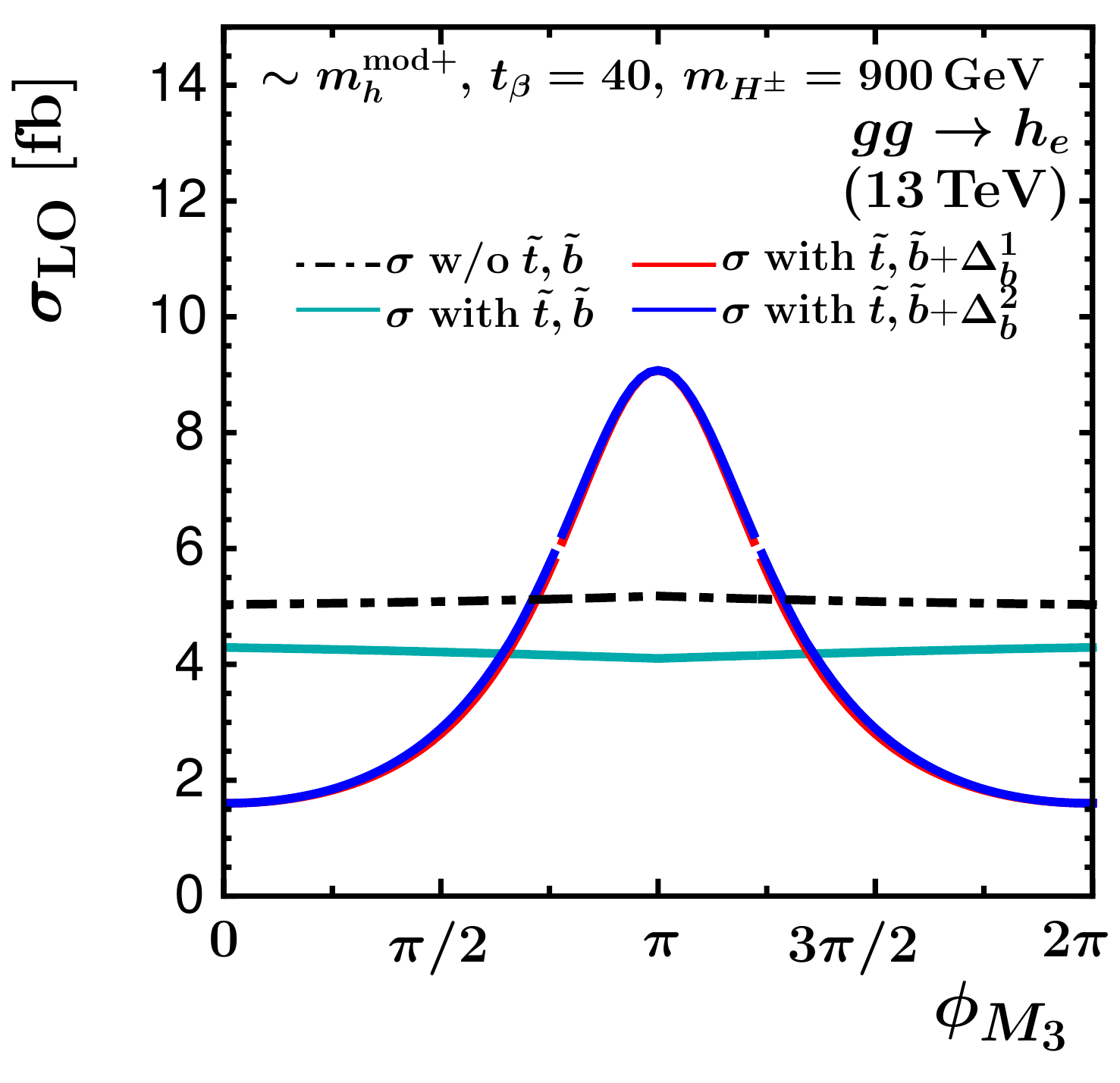}
\\[-.5cm]
 (a) & (b)
\end{tabular}
\end{center}
\vspace{-0.2cm}
\caption{Effect of $\Delta_b$ contributions on the \lo{} cross sections of $h_e$
as a function of (a) $\phi_{A_t}$ and (b) $\phi_{M_3}$ in the
$\mhmod$-inspired scenario with $\tan\beta=40$. 
The black dot-dashed curves depict the prediction 
without squark contributions (except through $\zhat$ factors), 
while the cyan lines correspond to the prediction where the squark
loop contributions at the one-loop level are included. In the red (blue) curves
furthermore the simplified (full) resummation of the 
$\Delta_b$ contributions is included.
}
\label{fig:mhmodsquarks}
\end{figure}

In \fig{fig:mhmodsquarks} we separately analyse the squark
contributions for the \lo{} cross section, 
i.e.\ 
the prediction omitting the squark loop contributions 
(black dot-dashed curves) is compared with the ones where first
the pure \lo{} squark contributions are added (depicted in cyan), 
and then the resummation of the 
$\Delta_b$ contributions to the bottom-quark Yukawa coupling is taken into
account. For the latter both the results for the full 
($\Delta_{b2}$, blue) and the simplified ($\Delta_{b1}$, red) 
resummation are shown. While the 
the pure \lo{} squark contributions are seen to have a moderate effect, it
can be seen that the incorporation of the resummation of the 
$\Delta_b$ contribution leads to a significant enhancement of the squark
loop effects.
We furthermore confirm
that for the heavy neutral Higgs bosons considered here 
the simplified resummation approximates the full resummation of the 
$\Delta_b$ contribution very well.
The curves corresponding to 
$\Delta_{b2}$ and $\Delta_{b1}$ 
hardly differ from each other both for the variation of $\phi_{A_t}$ and
$\phi_{M_3}$.
As before all curves include the same $\zhat$ factors obtained from {\tt FeynHiggs}.
The results for $h_o$, which are not shown here, are qualitatively very
similar.
The \lo{} squark contributions are less relevant for the $h_o$ cross section,
since those contributions are absent in the \mssm{} with real parameters. 
We also note that the curves for $h_o$ 
follow a similar behaviour as the ones for $h_e$, which implies that 
there are no large cancellations expected in the sum of the cross sections
for the two heavy Higgs bosons times their respective branching ratios.
Thus, the phases entering $\Delta_b$ could potentially lead to observable
effects in the production of the two heavy Higgs bosons even if the two
states cannot be experimentally resolved 
as separate signals.

\begin{figure}
\begin{center}
\begin{tabular}{cc}
\includegraphics[width=0.47\textwidth]{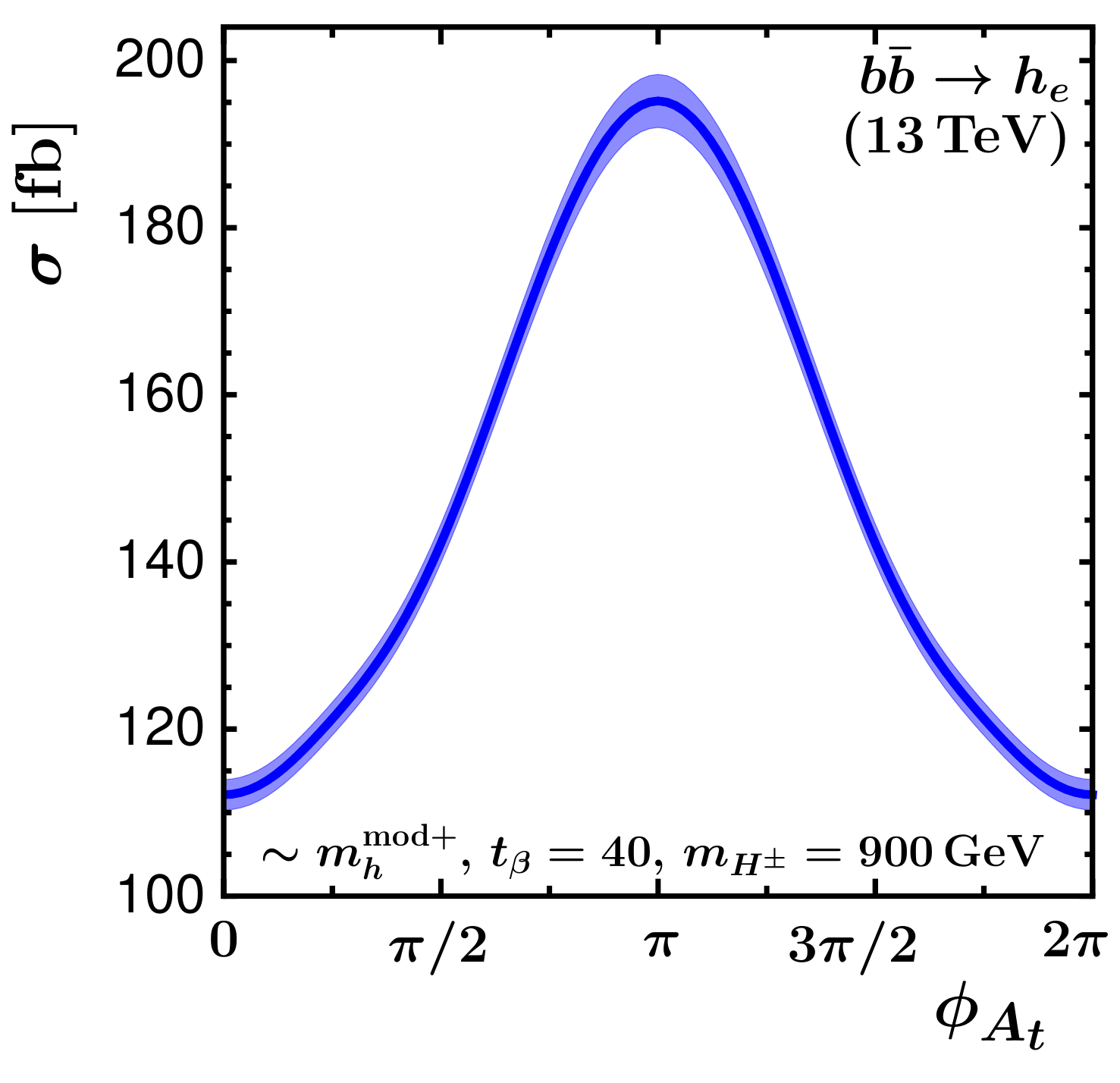} &
\includegraphics[width=0.47\textwidth]{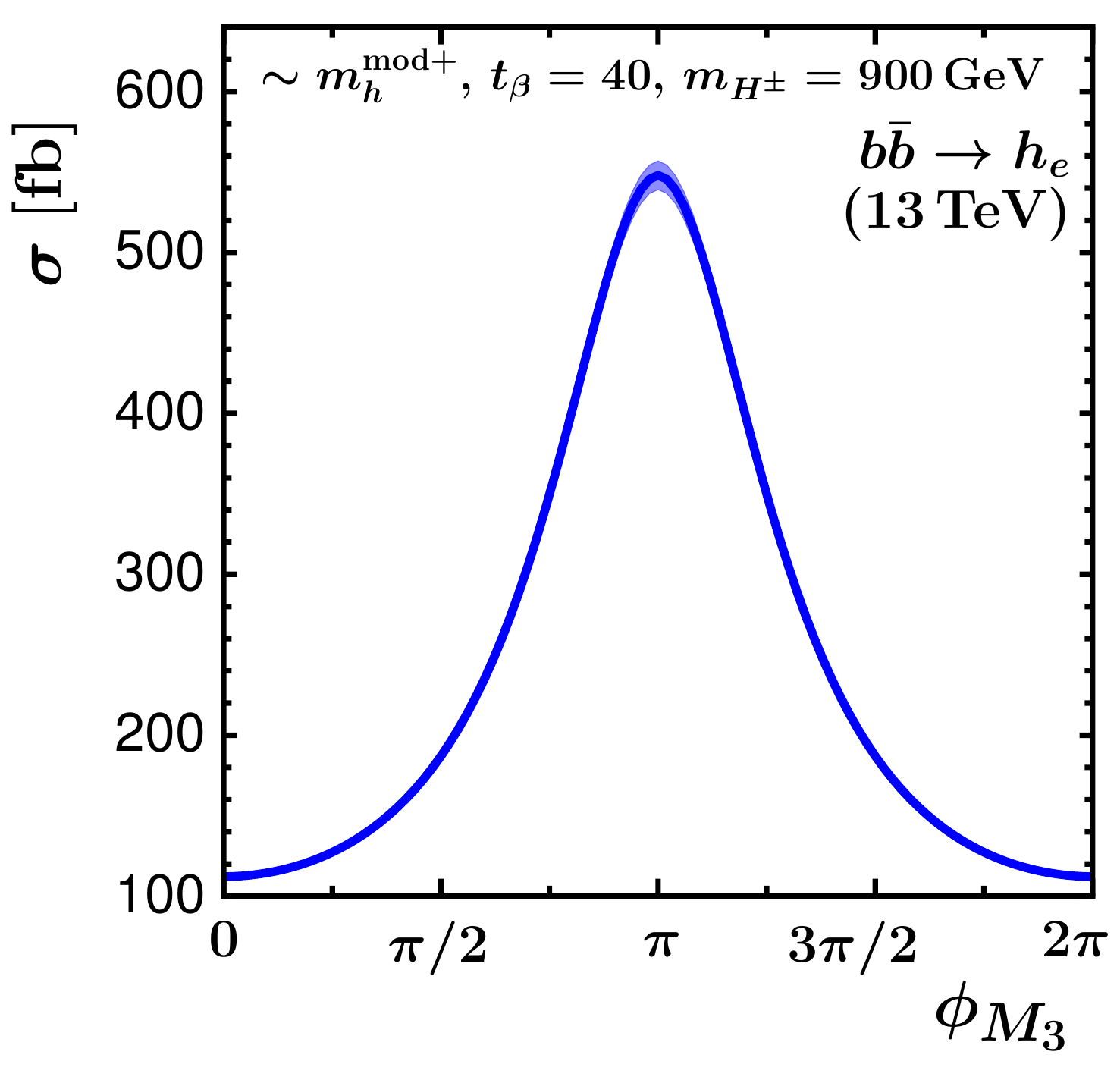}  \\[-.5cm]
 (a) & (b) \\
\end{tabular}
\end{center}
\vspace{-0.2cm}
\caption{Bottom-quark annihilation cross section for $h_e$ in fb
as a function of (a) $\phi_{A_t}$ and (b) $\phi_{M_3}$ in the $\mhmod$-inspired scenario with $\tan\beta=40$.
The depicted uncertainties are scale uncertainties.}
\label{fig:mhmodbbh}
\end{figure}

Having discussed the three different sources for \cp-violating effects
relevant for Higgs boson production through gluon fusion in the MSSM 
--- squark loop contributions, admixtures through $\zhat$ factors and 
resummation of $\Delta_b$ contributions ---
for completeness we also briefly discuss the
bottom-quark annihilation cross section for the $\mhmod$-inspired scenario
with $\tan\beta=40$. 
The corresponding cross section is shown in \fig{fig:mhmodbbh}
as a function of $\phi_{A_t}$ and $\phi_{M_3}$.
For such a large value of $\tan\beta$ this cross section
exceeds the gluon-fusion cross section by far.
It shows a very significant dependence on the phases 
$\phi_{A_t}$ and $\phi_{M_3}$, which is mainly induced by the 
$\Delta_b$~contribution.

\section{Remaining theoretical uncertainties}
\label{sec:uncertainties}

In the previous section we analysed our 
cross section predictions regarding \cp-violating effects entering via
squark loop contributions, $\zhat$ factors and $\Delta_b$ contributions.
Therein, we included renormalisation and factorisation scale uncertainties, 
which as expected are reduced upon inclusion of higher-order corrections.
However, the cross section predictions are also affected by other relevant 
theoretical uncertainties,
which we want to discuss in detail in this section.

Some of the theoretical uncertainties of cross sections in the \mssm{}
with complex parameters
are very similar to the ones in the \mssm{} with real parameters as
discussed in \citere{Bagnaschi:2014zla}. Therefore, we can directly transfer the discussion of \pdf{}$+\alpha_s$
uncertainties as well as the uncertainty associated with
the renormalisation prescription for the bottom-quark
Yukawa coupling from the case of the MSSM with real parameters:

\begin{itemize}
\item \pdf{}$+\alpha_s$ uncertainties: 
The fitted parton distribution functions (\pdf{})
 and the associated value of $\alpha_s$ induce an uncertainty in the prediction of the gluon-fusion cross section and, in particular, also the bottom-quark annihilation cross section.
 In our calculation we employ the {\tt MMHT2014} \pdf{} sets at \lo{}, \nlo{} and \nnlo{}~\cite{Harland-Lang:2014zoa},
 which can be used for both gluon fusion and bottom-quark annihilation. In \citeres{Bagnaschi:2014zla,Liebler:2015bka}
 it was observed that despite the effects of squarks in supersymmetric models, the \pdf{}$+\alpha_s$ uncertainties
 are mostly a function of the Higgs boson mass $m_{h_a}$. We will therefore not discuss them in more detail, since -- similar to the prescription for \mssm{} Higgs boson cross sections by the \lhc{} Higgs Cross
 Section Working Group~\cite{deFlorian:2016spz} -- relative uncertainties can be taken over from tabulated relative uncertainties
 obtained for the \sm{} Higgs boson or a pseudoscalar (in a \thdm{} with $\tan\beta=1$) as a function of its mass.
For Higgs masses in the range between $50$\,GeV and $1$\,TeV the typical size
of \pdf{}$+\alpha_s$ uncertainties for gluon fusion is $\pm (3-5)$\% following the prescription of \citere{Butterworth:2015oua}.
They increase up to $\pm 10$\% for Higgs masses up to $2$\,TeV. For bottom-quark annihilation they are
in the range $\pm(3-8)$\% for Higgs masses between $50$\,GeV and $1$\,TeV and up to $\pm 16$\%
for Higgs masses below $2$\,TeV.
\item Renormalisation of the bottom-quark mass and definition of the bottom
Yukawa coupling: 
In our calculation the bottom-quark mass is renormalised on-shell, and the 
bottom-Yukawa coupling is obtained from the bottom-quark mass as described
in \sct{sec:gluinosquark}.
The renormalisation of the bottom-quark mass and the freedom in the
definition of the bottom-Yukawa coupling are known to have a sizeable
numerical impact on the cross section predictions. This is in particular the 
case for large values of $\tan\beta$ where the bottom-Yukawa
coupling of the heavy Higgs bosons
is significantly enhanced and the top-quark Yukawa coupling is suppressed.
On the other hand,
in these regions of parameter space bottom-quark annihilation is the 
dominant process, for which there is less ambiguity regarding an appropriate
choice for the renormalisation scale.
The described uncertainties in the \mssm{} with complex parameters are
analogous to the case of real parameters. We therefore refer to the 
discussion in
\citere{Bagnaschi:2014zla} and references therein for further details.
\end{itemize}
We neglect approximate \nnlo{} stop-quark contributions and accordingly
the uncertainty associated with the approximation of the
involved Wilson coefficients, which was
discussed in \citere{Bagnaschi:2014zla}.
The impact of the 
\nnlo{} stop-quark contributions for the case of the MSSM with real
parameters can be compared with our estimate for 
the renormalisation and factorisation scale uncertainty of
our calculation. 
As an example, the \nnlo{} stop-quark contributions
lower the inclusive cross section 
for the light Higgs boson
by about $2$\,pb 
for zero phases
in the light-stop inspired scenario, 
which is at the lower edge of the scale uncertainty
depicted in \fig{fig:lhlightstop}~(b).
Other uncertainties discussed in \citere{Bagnaschi:2014zla} are renormalisation and factorisation
scale uncertainties and an uncertainty related to higher-order
contributions to $\Delta_b$.
Moreover, we add another uncertainty related to the performed
interpolation of supersymmetric \nlo{} \qcd{} contributions.
We discuss in the following our estimates for
the three previously mentioned uncertainties:
\begin{itemize}
\item We obtain the renormalisation and factorisation scale uncertainty as follows:
The central scale choice is $(\muR^0,\muF^0)=(m_{h_a}/2,m_{h_a}/2)$ for gluon fusion and
$(\muR^0,\muF^0)=(m_{h_a},m_{h_a}/4)$ for bottom-quark annihilation. We obtain the scale uncertainty
by taking the maximal deviation from the central scale choice $\Delta\sigma$ obtained from the additional scale choices
$(\muR,\muF)\in \lbrace{(2\muR^0,2\muF^0),(2\muR^0,\muF^0),(\muR^0,2\muF^0),(\muR^0,\muF^0/2),(\muR^0/2,\muF^0),(\muR^0/2,\muF^0/2)\rbrace}$.
We perform this procedure individually for all three cross sections in \eqn{eq:ggphisum} and then
obtain the overall absolute uncertainty through
\begin{align} 
\label{eq:overallunc}
\Delta\sigma^{\mathrm{scale}}=\sqrt{\left(\Delta\sigma_{\text{\nklo{k}}}^{\Delta_{b1}}\right)^2
+\left(\Delta\sigma_{\text{\lo{}}}^{\Delta_{b2}}
-\Delta\sigma_{\text{\lo{}}}^{\Delta_{b1}}\right)^2}\,,
\end{align}
where we assume the two \lo{} cross sections to be fully correlated. 
The uncertainty bands that we have displayed in the plots shown above
correspond to the cross section range covered by $\sigma\pm \Delta\sigma^{\mathrm{scale}}$.
\item In order to display the propagation of an uncertainty arising
from higher-order contributions to $\Delta_b$ to our cross section calculation,
we vary the value of $\Delta_b$ obtained from {\tt FeynHiggs} by $\pm 10\%$. 
This variation by $\pm 10\%$ roughly corresponds to the effect of a
variation of the renormalisation scales, see the discussion in
\citere{Bagnaschi:2014zla}.
We label the obtained uncertainty as $\Delta\sigma^{\mathrm{resum}}$
and assign an uncertainty band of $\sigma\pm \Delta\sigma^{\mathrm{resum}}$.
\item The employed interpolation for the two-loop virtual squark-gluino contributions following \eqn{eq:cos_interpol} leads
to a further uncertainty. A conservative estimate for it can be
obtained as follows: We determine the cross section $\sigma(\phi_z)$ following \eqn{eq:ggphimaster} not only
for the correct phase $\phi_z$ in \eqn{eq:cos_interpol}, but also leave the phase within \eqn{eq:cos_interpol} constant,
i.e.\,fixed to $0$ and $\pi$. We call the obtained cross sections $\sigma(0)$ and $\sigma(\pi)$. For each value of $\phi_z$ we take the difference
$\Delta\sigma^{\mathrm{int}}=\sin^2(\phi_z)|\sigma(0)-\sigma(\pi)|/2$. It is reweighted with $\sin^2(\phi_z)$, since we know that our result is
correct at phases $0$ and $\pi$. The obtained uncertainty band is given by $\sigma\pm \Delta\sigma^{\mathrm{int}}$.
\end{itemize}

In the following we display the effects of the estimated uncertainties
for certain scenarios, where we choose the displayed scenarios and the
displayed cross sections such that the effect of the uncertainties is
largest.
While the scale uncertainties were included in all previous figures for the \lo{} prediction as well as for
our best prediction already, we will discuss the interpolation uncertainty
for the light-stop inspired scenario with $\tan\beta=16$
and the resummation uncertainty for the $\mhmod$-inspired scenario with 
$\tan\beta=40$.

\begin{figure}[t!]
\begin{center}
\begin{tabular}{cc}
\includegraphics[width=0.47\textwidth]{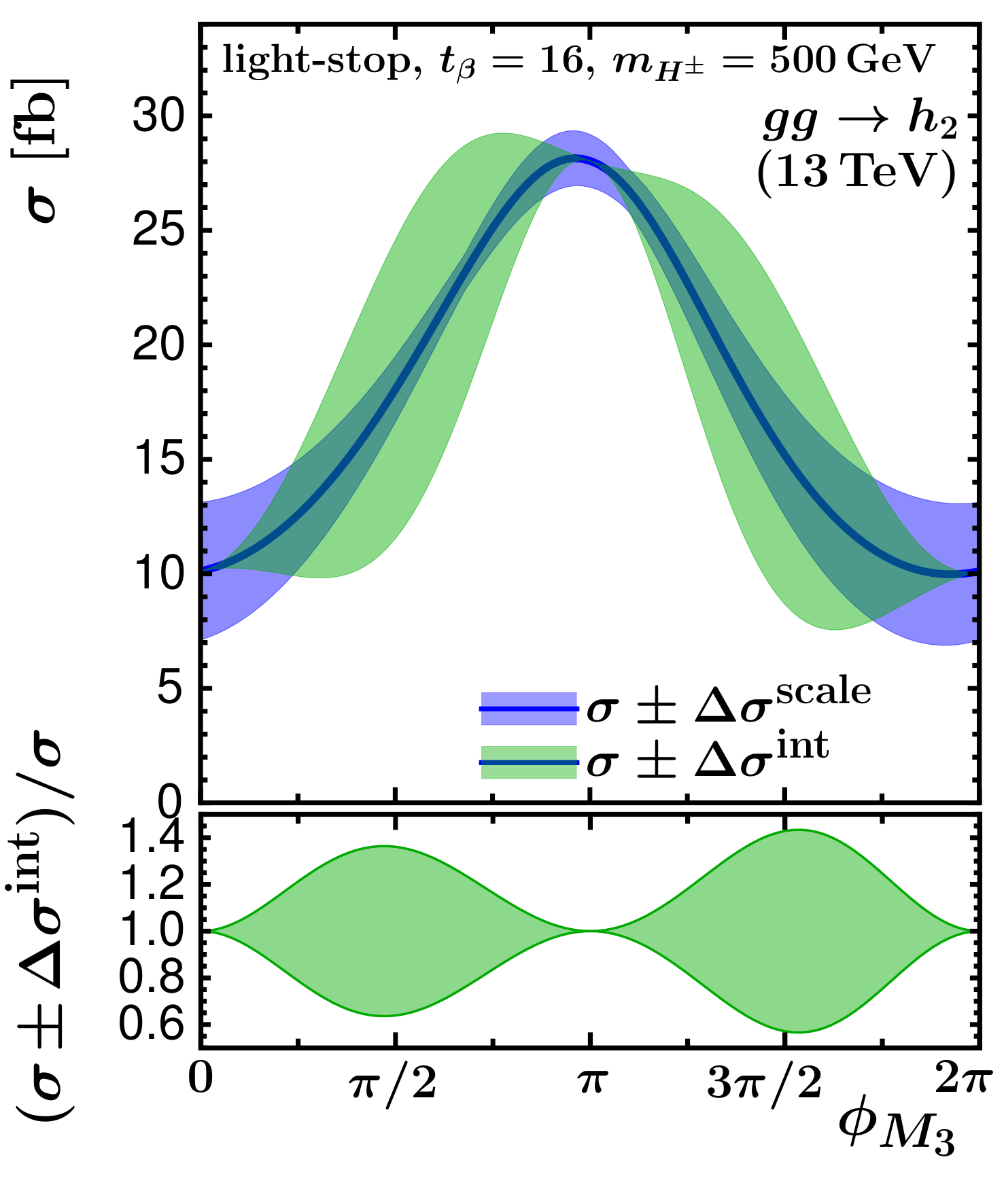} &
\includegraphics[width=0.47\textwidth]{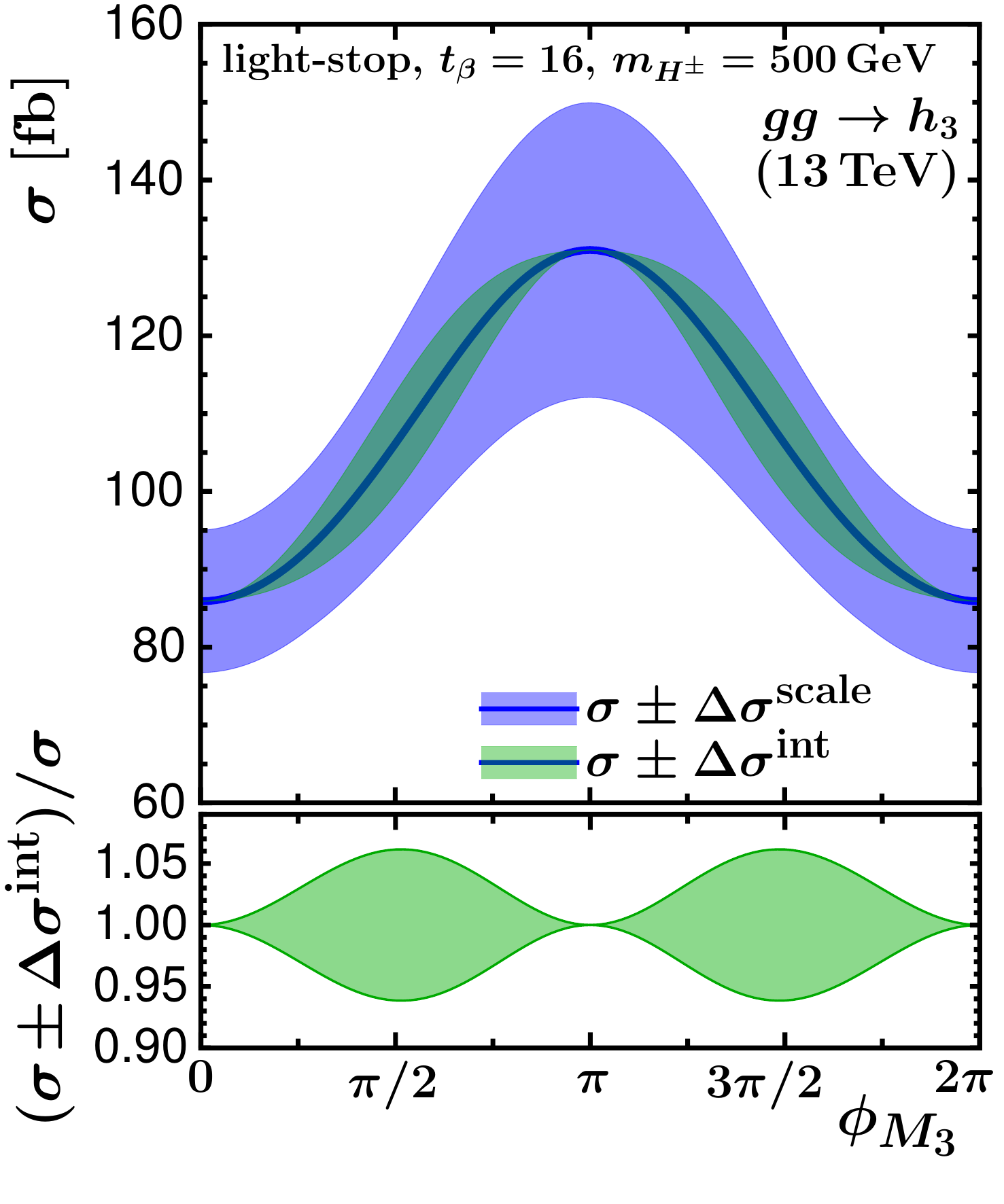}
\\[-.5cm]
 (a) & (b) \\
\end{tabular}
\end{center}
\vspace{-0.2cm}
\caption{Renormalisation and factorisation scale uncertainties $\Delta\sigma^{\mathrm{scale}}$ (blue) and interpolation
uncertainties $\Delta\sigma^{\mathrm{int}}$ (green) for the gluon-fusion cross section of (a) $h_2$ and (b) $h_3$
as a function of $\phi_{M_3}$ in the light-stop inspired scenario.
In the lower panel the upper and lower edge of the band of the cross
section prediction with the assigned interpolation uncertainty is normalised
to the cross section without this uncertainty.
}
\label{fig:lightstopint}
\end{figure}

\fig{fig:lightstopint} shows the renormalisation and factorisation scale uncertainties $\Delta\sigma^{\mathrm{scale}}$ as before
and in addition the above described interpolation uncertainty $\Delta\sigma^{\mathrm{int}}$, which in case of
the variation of $\phi_{M_3}$ can be substantial.
As can be seen in \fig{fig:lightstopint}, the 
interpolation uncertainty obtained from our conservative estimate can
in this scenario even exceed the scale uncertainty
for the gluon-fusion cross section of $h_2$. It should be noted that this is
an extreme case, while the interpolation uncertainty, which is an \nlo\
effect related to the squark and gluino loop contributions,
remains small for the other previously described scenarios 
(which we do not show here explicitly). This is simply
a consequence of the fact that the relative impact of the squark and
gluino contributions in the other scenarios is much smaller than in the
light-stop inspired scenario. The interpolation uncertainty 
for the gluon-fusion cross section of $h_3$ in \fig{fig:lightstopint}
is much less pronounced than for $h_2$, since as discussed above the 
squark loop corrections are significantly smaller in this case and would
vanish if $h_3$ were a pure \cp-odd state. 
The behaviour in the lower panels of 
\fig{fig:lightstopint} displays the fact that by construction the assigned
interpolation uncertainty vanishes for the phases 0 and $\pi$, where the
interpolated result in the MSSM with complex parameters merges the
known result of the MSSM with real parameters.
For the variation of $\phi_{A_t}$ the \lo{} cross section
incorporating squark contributions already includes the dominant
effect on the cross section, such that the uncertainty due to the interpolated \nlo{} contributions
is also less pronounced than in case of the variation of $\phi_{M_3}$.

\begin{figure}[t!]
\begin{center}
\begin{tabular}{cc}
\includegraphics[width=0.47\textwidth]{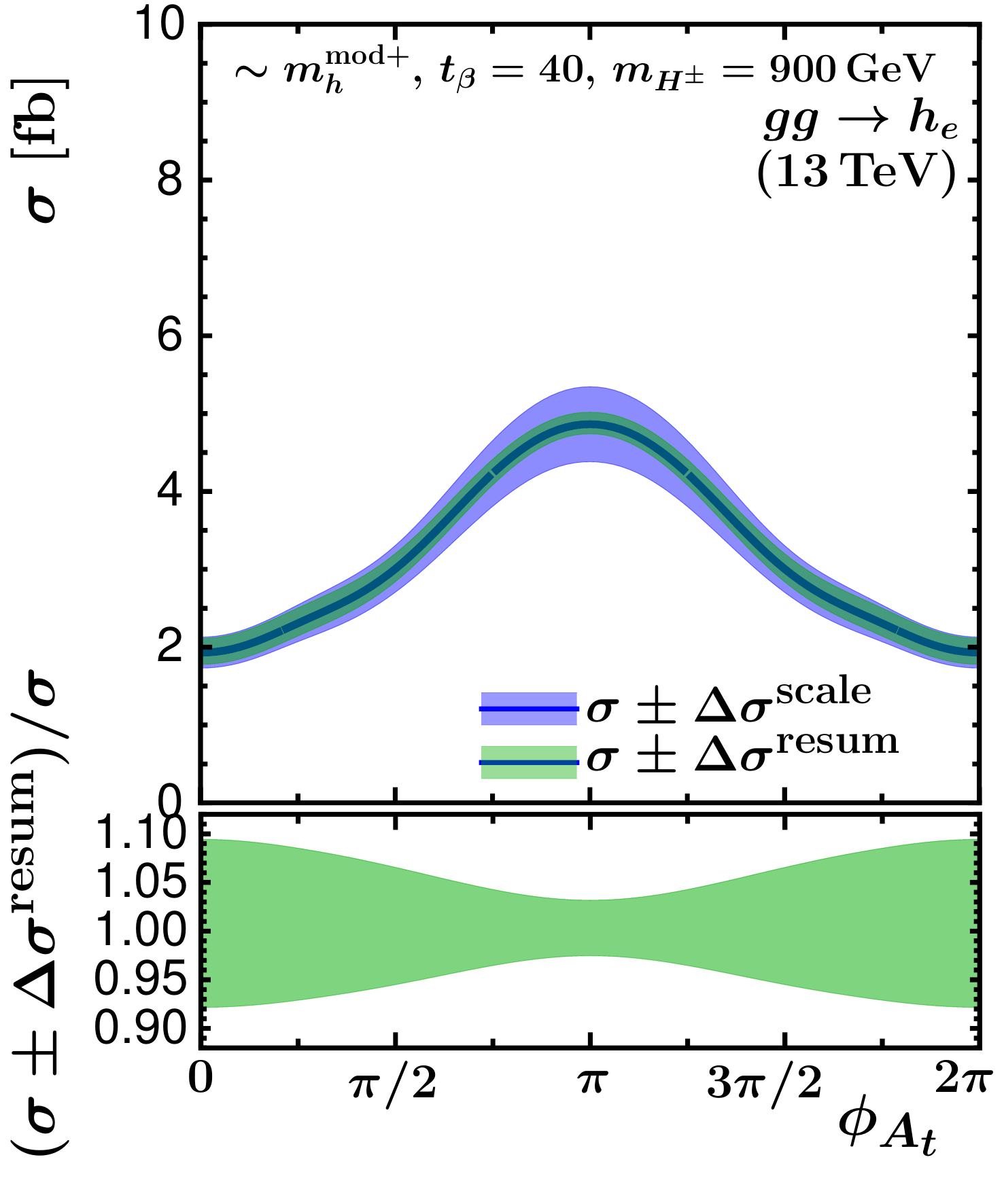} &
\includegraphics[width=0.47\textwidth]{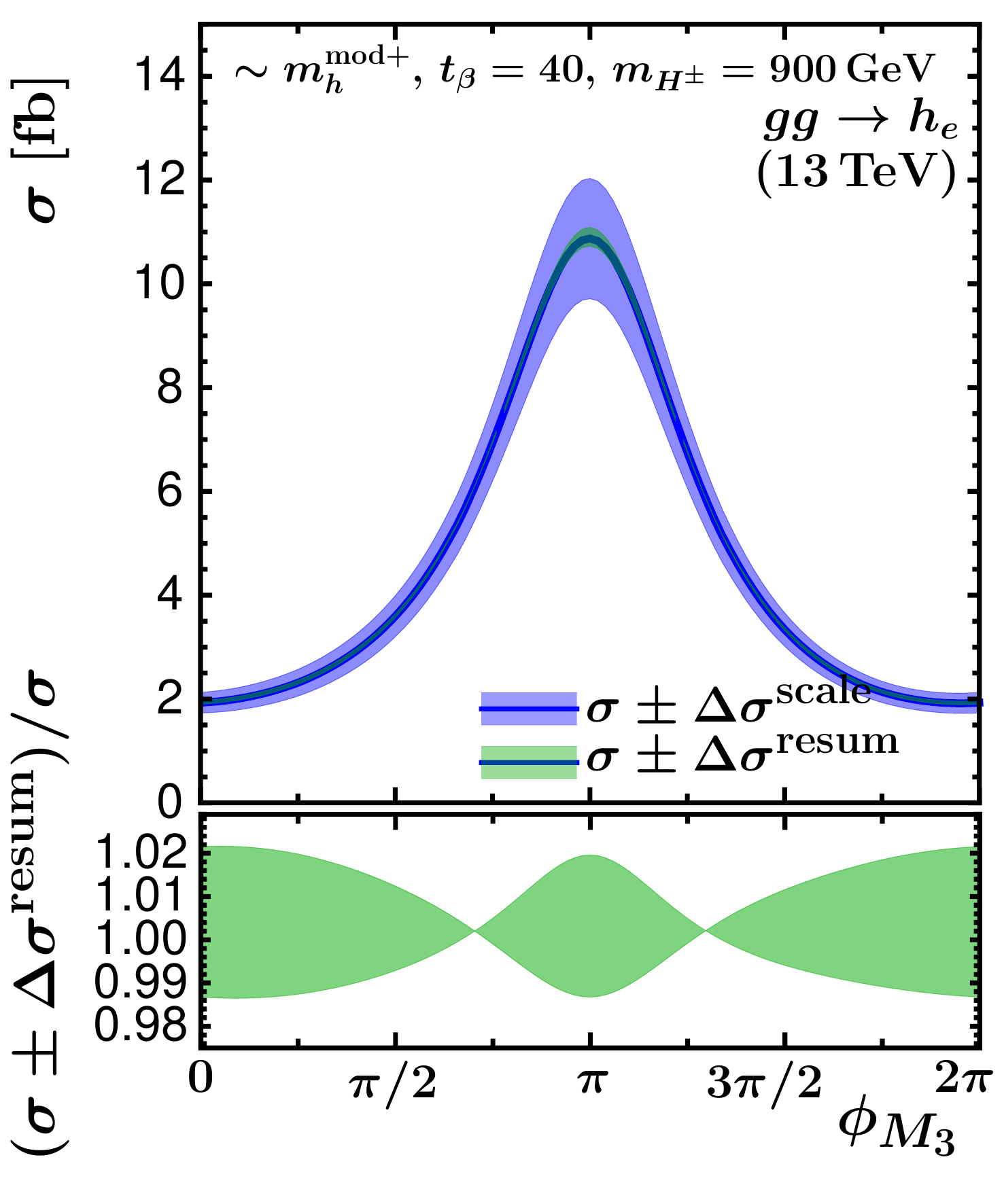}  \\[-.5cm]
 (a) & (b) \\
\end{tabular}
\end{center}
\vspace{-0.2cm}
\caption{Renormalisation and factorisation scale uncertainties $\Delta\sigma^{\mathrm{scale}}$ (blue) and resummation
uncertainties $\Delta\sigma^{\mathrm{resum}}$ (green) for the gluon-fusion cross section of $h_e$
as a function of (a) $\phi_{A_t}$ and (b) $\phi_{M_3}$ in the $\mhmod$-inspired scenario for $\tan\beta=40$.
In the lower panel the upper and lower edge of the band of the cross
section prediction with the assigned resummation uncertainty is normalised
to the cross section without this uncertainty.
}
\label{fig:mhmoddeltab}
\end{figure}

The described $\Delta_b$ uncertainties are depicted in
\fig{fig:mhmoddeltab}.
Since $\Delta_b$ crosses $0$ as a function of $\phi_{M_3}$ twice, the 
uncertainty that we have associated to it according to the prescription
discussed above
also vanishes there, as can be seen in the lower panel of 
\fig{fig:mhmoddeltab}(b). 
Even for the 
large value of $\tan\beta$ chosen here the assigned
$\Delta_b$ uncertainty of $\pm 10\%$ is much smaller than the scale
uncertainty of the displayed cross sections. Despite the different behaviour
with the phases $\phi_{A_t}$ and $\phi_{M_3}$ 
displayed in the lower panel of \fig{fig:mhmoddeltab} the qualitative effect
of the resummation uncertainties on 
the Higgs boson production cross sections 
is nevertheless rather similar. The latter is also true for the 
bottom-quark annihilation cross section, which is not depicted here.
The resummation uncertainties are of most relevance for large
values of $\tan\beta$, where the cross section of bottom-quark annihilation exceeds
the gluon-fusion cross section.

\section{Conclusions}
\label{sec:conclusions}

In this paper we have presented theoretical predictions for
inclusive cross sections for neutral Higgs boson production via gluon fusion
and bottom-quark annihilation in the \mssm{} with complex parameters, and
demonstrated the relevance of the \cp{}-violating phases on these cross sections.

The cross section predictions for the gluon-fusion process at leading-order
are based on an explicit calculation taking into account the dependence on all
complex parameters in the \mssm{}, and the complete form of the analytical formulae 
for the general \cp-violating case including Higgs mixing has been
presented in the literature for the first time.
The wave function normalisation
factors arising from the 
$(3\times 3)$-mixing of the lowest-order mass eigenstates of the 
Higgs bosons~$\lbrace h, H, A\rbrace$
into the loop-corrected
mass eigenstates $\lbrace h_1, h_2, h_3\rbrace$ have been described
with full propagator corrections using the
self-energies of the neutral Higgs bosons as provided by {\tt{FeynHiggs}}. Furthermore, the
\lo{} predictions for the gluon-fusion process in the \mssm{} with complex parameters
deviate from those of the \mssm{} with real parameters due to non-zero couplings
of the squarks to the pseudoscalar~$A$ and potentially different left- and right-handed
bottom-Yukawa couplings arising from the resummation of 
$\tan\beta$-enhanced sbottom contributions in $\D_b$.
We have supplemented the \lo{} computation of the cross section
by higher-order contributions:
using for the treatment of the higher-order corrections 
a simplified version of the 
$\D_b$ resummation we have included the full massive top- and bottom
quark contributions at \nlo{} \qcd{} and have interpolated the
\nlo{} \susy{} \qcd{} corrections
from the amplitudes in the \mssm{} with real parameters.
We have thoroughly discussed the uncertainties involved in using such an interpolation.
The interpolation uncertainty at \nlo, which is most relevant in
scenarios where the squarks and the gluino are relatively light in view of
the present limits from the LHC searches, could be avoided if an explicit
result for the squark-gluino contributions at 
\nlo{} \qcd{} in the \mssm{} becomes available for the general case of
complex parameters.
For the top-quark contribution in the effective theory of a heavy top-quark
we have added
\nnlo{} \qcd{} contributions for all Higgs bosons, and \nklo{3} \qcd{} contributions in
an expansion around the threshold of Higgs production
for the \cp{}-even component of the light Higgs boson~$h_1$ to match
the precision of the predictions for the \sm{} Higgs boson.
Electroweak effects, which include two-loop contributions with couplings of the
heavy gauge bosons to the \cp{}-even component of the Higgs bosons
mediated by light quarks, have been
added to the \cp{}-even component of the gluon-fusion cross section.

The results presented in this paper 
are currently the state of the art for neutral Higgs
production in the \mssm{} with complex parameters. Our calculations have been
implemented in an extension of the code \sushi{}
called \sushimi{}, which is 
linked to {\tt FeynHiggs}. \sushimi{} is available upon request.
Using \sushimi{}, we have investigated the phenomenological
effects of \cp{}-violating phases on the production
of Higgs bosons in the \mssm{} with complex parameters in two slightly modified benchmark
scenarios, light-stop and $m_h^{\text{mod+}}$. 
We have found in our analysis
of Higgs boson production through gluon fusion 
that  a proper description of squark and gluino loop contributions is essential. This
refers both to the loop contributions to the
gluon--gluon--Higgs vertex and to the corrections entering 
through $\Delta_b$.
Squark and gluino loop contributions furthermore enter the wave function
normalisation factors that are necessary to ensure the correct on-shell
properties of the produced Higgs boson. Where squark and gluino contributions
are sizeable the production cross sections show a significant dependence
on the \cp{}-violating phases. We have discussed
the remaining theoretical uncertainties in the
cross section predictions taking
into account renormalisation and factorisation scale uncertainties, a resummation
uncertainty for $\D_b$ and an uncertainty due to the performed interpolation of
\nlo{} \susy{} \qcd{} corrections. We have furthermore briefly
commented on other uncertainties that can directly 
be taken over from the case of the MSSM with real parameters.

A further important feature that occurs in the production processes
for the two heavy states $h_2$ and $h_3$ in the general case where
\cp-violating interactions are taken into account is the fact that there can
be a large mixing between these often nearly mass-degenerate states. Their
mixing effects are incorporated in the wave function normalisation factors
for the external Higgs bosons. 
For a proper interpretation of experimental exclusion limits
arising from \mssm{} Higgs searches, which so far have only been analysed in the
framework of the \cp{}-conserving \mssm{}, it will be important to take
into account interference effects in
the full process of Higgs production and decay. 
Our results for the cross sections for on-shell Higgs bosons
can be directly used in the context of a generalised narrow-width
approximation to incorporate these interference effects. This
topic will be addressed in a forthcoming publication.

\section*{Acknowledgements}

We thank Sebastian Pa\ss ehr and Pietro Slavich for discussions and Elina Fuchs
for discussions and comments on the manuscript.
The authors acknowledge support by Deutsche Forschungsgemeinschaft through the SFB~676 ``Particles, Strings and the Early Universe''
and by the European Commission through the ``HiggsTools'' Initial Training Network PITN-GA-2012-316704.

\appendix
\section{Formulas: Higgs-quark and Higgs-squark couplings}
\label{sec:appendix}

In \sushimi{} the Higgs--(s)quark couplings are expressed in terms of the \cp{}-even and \cp{}-odd
neutral gauge eigenstates $\phi_g\in\lbrace\phi_1^0,\phi_2^0\rbrace$ and
$\chi_g\in\lbrace\chi_1^0,\chi_2^0\rbrace$, respectively. In order to
obtain the couplings
of the squarks with the lowest-order mass eigenstates $\phi\in\lbrace
h,H,A,G\rbrace$ the gauge eigenstates are rotated
using the tree-level mixing matrix~$\mathcal{R}$ as depicted in the following Feynman diagrams
\begin{align}
 \parbox{27mm}{\includegraphics[width=0.2\textwidth]{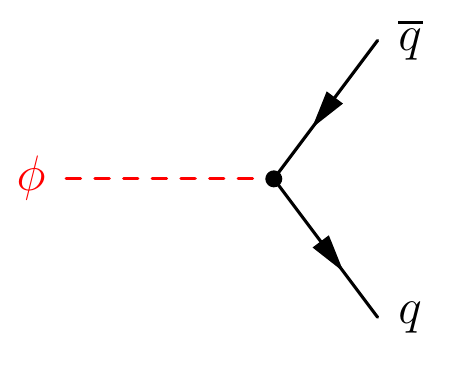}}=
 \left\lbrace\begin{matrix}i\frac{m_q}{v}\mathcal{R} (g_{q_L}^{\phi_g} P_L+g_{q_R}^{\phi_g}P_R)\\
              -\frac{m_q}{v}\mathcal{R} (g_{q_L}^{\chi_g} P_L-g_{q_R}^{\chi_g}P_R)\end{matrix}\right.\qquad\text{and}\qquad
 \parbox{27mm}{\includegraphics[width=0.2\textwidth]{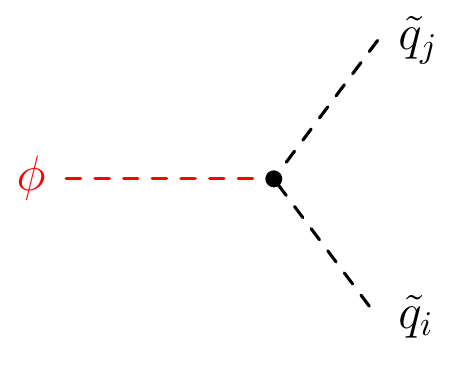}}=i\frac{1}{v}\mathcal{R} g_{\tilde{q},ij}^{\phi_g,\chi_g}
\end{align}
with $v = 2m_W/g = 1/ \sqrt{\sqrt{2}G_F}$ and the tree-level mixing matrix~$\mathcal{R}$ 
given in \eqn{eq:HiggsMixing_tree_neutral}.
At the amplitude level the results will then also be multiplied with the corresponding $\zhat$ factor.

The couplings between the gauge eigenstates and the third generation quarks are
$g_{q_L}=1/\cos\beta$ for $\phi_1^0$ and $\chi_1^0$ and 
$g_{q_L}=1/\sin\beta$ for $\phi_2^0$ and $\chi_2^0$.
For $\Delta_b$ corrections we refer to \eqn{eq:fulldeltab} and \eqn{eq:simpledeltab}.

The couplings between the gauge eigenstates and the third generation squarks
contain terms from the squark mass diagonalisation matrix $U_{\td{q}}$ (see \eqn{eq:sqark_mass_matrix_diag})
which is a unitary matrix with real diagonal elements and complex off-diagonal elements, i.e.\ it can be written as follows
\begin{align}
U_{\td{q}} = \begin{pmatrix}
U_{\td{q}11} & U_{\td{q}12} \\ -U_{\td{q}12}^* & U_{\td{q}22}
\end{pmatrix}\, .
\end{align}

They are obtained with {\tt MaCoR}~\cite{Liebler:2010bi,Liebler:2011qsa}.
For the \cp{}-even state $\phi_1^0$ we have the stop couplings (using
$s_{\bb} \equiv \sin {\bb}$, $c_{\bb} \equiv \cos {\bb}$, $t_{\bb} \equiv \tan {\bb}, s_W \equiv \sin \theta_{W}$ and $c_W \equiv \cos \theta_W$):
\begin{align}
g_{\td{t},11}^{\phi_1^0} &=  U_{\td{t}12}^* \left[ -\frac{U_{\td{t}11}m_t \mu}{\sinb} + \frac{4}{3} U_{\td{t}12} \cosb m_Z^2  s^2_W \right] - U_{\td{t}11}^* \left[\frac{U_{\td{t}12}m_t \mu^*}{\sinb} + \cosb m_Z^2 U_{\td{t}11} \left( \frac{s^2_W}{3} - c_W^2 \right) \right] \nn \\
g_{\td{t},12}^{\phi_1^0} &= U_{\td{t}12}^* \left[ -\frac{U_{\td{t}21}m_t \mu}{\sinb} + \frac{4}{3} U_{\td{t}22} \cosb m_Z^2  s^2_W \right] - U_{\td{t}11}^* \left[\frac{U_{\td{t}22}m_t \mu^*}{\sinb} + \cosb m_Z^2 U_{\td{t}21} \left( \frac{s^2_W}{3} - c_W^2 \right) \right] \nn \\
g_{\td{t},21}^{\phi_1^0} &= U_{\td{t}22}^* \left[ -\frac{U_{\td{t}11}m_t \mu}{\sinb} + \frac{4}{3} U_{\td{t}12} \cosb m_Z^2  s^2_W \right] - U_{\td{t}21}^* \left[\frac{U_{\td{t}12}m_t \mu^*}{\sinb} + \cosb m_Z^2 U_{\td{t}11} \left( \frac{s^2_W}{3} - c_W^2 \right) \right] \nn \\
g_{\td{t},22}^{\phi_1^0} &= U_{\td{t}22}^* \left[ -\frac{U_{\td{t}21}m_t
\mu}{\sinb} + \frac{4}{3} U_{\td{t}22} \cosb m_Z^2  s^2_W \right] -
U_{\td{t}21}^* \left[\frac{U_{\td{t}22}m_t \mu^*}{\sinb} + \cosb m_Z^2
U_{\td{t}21} \left( \frac{s^2_W}{3} - c_W^2 \right) \right] \, .
\end{align}
For the  \cp{}-even state $\phi_2^0$, the couplings are:
{\allowdisplaybreaks
\begin{align}
g_{\td{t},11}^{\phi_2^0} &= U_{\td{t}12}^* \left[ \frac{U_{\td{t}11}m_tA_t^*}{\sinb} + U_{\td{t}12} \left(\frac{2 m^2_t}{\sinb} -\frac{4}{3} \sinb m_z^2s^2_W  \right) \right]  \nn \\ &+ U_{\td{t}11}^* \left[ \frac{U_{\td{t}12}m_t A_t}{\sinb} + \frac{2U_{\td{t}11}m^2_t}{\sinb} + \sinb m_Z^2 U_{\td{t}11} \left(\frac{s^2_W}{3} - c^2_W \right)\right] \nn \\
g_{\td{t},12}^{\phi_2^0} &= U_{\td{t}12}^* \left[ \frac{U_{\td{t}21}m_tA_t^*}{\sinb} + U_{\td{t}22} \left(\frac{2 m^2_t}{\sinb} -\frac{4}{3} \sinb m_z^2s^2_W  \right) \right]\nn \\ &+ U_{\td{t}11}^* \left[ \frac{U_{\td{t}22}m_t A_t}{\sinb} + \frac{2U_{\td{t}21}m^2_t}{\sinb} + \sinb m_Z^2 U_{\td{t}21} \left(\frac{s^2_W}{3} - c^2_W \right)\right] \nn \\
g_{\td{t},21}^{\phi_2^0} &= U_{\td{t}22}^* \left[ \frac{U_{\td{t}11}m_tA_t^*}{\sinb} + U_{\td{t}12} \left(\frac{2 m^2_t}{\sinb} -\frac{4}{3} \sinb m_z^2s^2_W  \right) \right]\nn \\ &+ U_{\td{t}21}^* \left[ \frac{U_{\td{t}12}m_t A_t}{\sinb} + \frac{2U_{\td{t}11}m^2_t}{\sinb} + \sinb m_Z^2 U_{\td{t}11} \left(\frac{s^2_W}{3} - c^2_W \right)\right] \nn \\
g_{\td{t},22}^{\phi_2^0} &= U_{\td{t}22}^* \left[
\frac{U_{\td{t}21}m_tA_t^*}{\sinb} + U_{\td{t}22} \left(\frac{2
m^2_t}{\sinb} -\frac{4}{3} \sinb m_z^2s^2_W  \right) \right]\nn \\ &+
U_{\td{t}21}^* \left[ \frac{U_{\td{t}22}m_t A_t}{\sinb} +
\frac{2U_{\td{t}21}m^2_t}{\sinb} + \sinb m_Z^2 U_{\td{t}21}
\left(\frac{s^2_W}{3} - c^2_W \right)\right] \, .
\end{align}}
Similarly for the \cp{}-odd states $\chi_1^0$ and $\chi_2^0$ the stop couplings are given as:
\begin{align}
g_{\td{t},11}^{\chi_1^0} &= i\frac{m_t}{\sinb} \left[ -\mu U_{\td{t}12}^* U_{\td{t}11} + \mu^* U_{\td{t}11}^* U_{\td{t}12} \right] &&g_{\td{t},11}^{\chi_2^0} = i\frac{m_t}{\sinb} \left[ -A_t^* U_{\td{t}12}^* U_{\td{t}11} + A_t U_{\td{t}11}^* U_{\td{t}12} \right] \nn \\
g_{\td{t},12}^{\chi_1^0} &= i\frac{m_t}{\sinb} \left[ -\mu U_{\td{t}12}^* U_{\td{t}21} + \mu^* U_{\td{t}11}^* U_{\td{t}22} \right] &&g_{\td{t},12}^{\chi_2^0} =i\frac{m_t}{\sinb} \left[ -A_t^* U_{\td{t}12}^* U_{\td{t}21} + A_t U_{\td{t}11}^* U_{\td{t}22} \right] \nn \\
g_{\td{t},21}^{\chi_1^0} &= i\frac{m_t}{\sinb} \left[ -\mu U_{\td{t}22}^* U_{\td{t}11} + \mu^* U_{\td{t}21}^* U_{\td{t}12} \right] && g_{\td{t},21}^{\chi_2^0} = i\frac{m_t}{\sinb} \left[ -A_t^* U_{\td{t}22}^* U_{\td{t}11} + A_t U_{\td{t}21}^* U_{\td{t}12} \right] \nn \\
g_{\td{t},22}^{\chi_1^0} &= i\frac{m_t}{\sinb} \left[ -\mu U_{\td{t}22}^*
U_{\td{t}21} + \mu^* U_{\td{t}21}^* U_{\td{t}22} \right] &&
g_{\td{t},22}^{\chi_2^0} = i\frac{m_t}{\sinb} \left[ -A_t^* U_{\td{t}22}^*
U_{\td{t}21} + A_t U_{\td{t}21}^* U_{\td{t}22} \right] \, .
\end{align}
Analogously, the Higgs-sbottom couplings for the \cp{}-even state $\phi_1^0$ are:
\begin{align}
g_{\td{b},11}^{\phi_1^0} &= U_{\td{b}12}^* \left[ \frac{U_{\td{b}11}A_b^* m_b}{\cosb} + U_{\td{b}12} \left( \frac{2m_b^2}{\cosb} - \frac{2}{3} \cosb m_Z^2 s_W^2 \right) \right] \nn \\ &+ U_{\td{b}11}^* \left[ \frac{U_{\td{b}12}A_b m_b}{\cosb} + \frac{2U_{\td{b}11}m_b^2}{\cosb} - \cosb m_Z^2U_{\td{b}11} \left( \frac{s^2_W}{3}+ c_W^2   \right) \right] \nn \\
g_{\td{b},12}^{\phi_1^0} &= U_{\td{b}12}^* \left[ \frac{U_{\td{b}21}A_b^*
m_b}{\cosb} + U_{\td{b}22} \left( \frac{2m_b^2}{\cosb} - \frac{2}{3} \cosb
m_Z^2 s_W^2 \right) \right] \nn \\ &+ U_{\td{b}11}^* \left[
\frac{U_{\td{b}22}A_b m_b}{\cosb} + \frac{2U_{\td{b}21}m_b^2}{\cosb} - \cosb
m_Z^2U_{\td{b}21} \left(  \frac{s^2_W}{3} + c_W^2 \right) \right] \, , \nn 
\end{align}
\begin{align}
g_{\td{b},21}^{\phi_1^0} &= U_{\td{b}22}^* \left[ \frac{U_{\td{b}11}A_b^* m_b}{\cosb} + U_{\td{b}12} \left( \frac{2m_b^2}{\cosb} - \frac{2}{3} \cosb m_Z^2 s_W^2 \right) \right] \nn \\ &+ U_{\td{b}21}^* \left[ \frac{U_{\td{b}12}A_b m_b}{\cosb} + \frac{2U_{\td{b}11}m_b^2}{\cosb} - \cosb m_Z^2U_{\td{b}11} \left(  \frac{s^2_W}{3} +  c_W^2 \right) \right] \nn \\
g_{\td{b},22}^{\phi_1^0} &= U_{\td{b}22}^* \left[ \frac{U_{\td{b}21}A_b^*
m_b}{\cosb} + U_{\td{b}22} \left( \frac{2m_b^2}{\cosb} - \frac{2}{3} \cosb
m_Z^2 s_W^2 \right) \right] \nn \\ &+ U_{\td{b}21}^* \left[
\frac{U_{\td{b}22}A_b m_b}{\cosb} + \frac{2U_{\td{b}21}m_b^2}{\cosb} - \cosb
m_Z^2U_{\td{b}21} \left(  \frac{s^2_W}{3} + c_W^2  \right) \right] \, .
\end{align}
For the \cp{}-even state $\phi_0^2$ they are given as:
\begin{align}
g_{\td{b},11}^{\phi_2^0} &= U_{\td{b}12}^* \left[ - \frac{U_{\td{b}11} m_b \mu}{\cosb} + \frac{2}{3} U_{\td{b}12} \sinb m_z^2s_W^2 \right] - U_{\td{b}11}^* \left[ \frac{U_{\td{b}12}m_b \mu^*}{\cosb} - \sinb m_Z^2 U_{\td{b}11} \left( \frac{s^2_W}{3} + c_W^2 \right) \right] \nonumber \\
g_{\td{b},12}^{\phi_2^0} &= U_{\td{b}12}^* \left[ - \frac{U_{\td{b}21} m_b \mu}{\cosb} + \frac{2}{3} U_{\td{b}22} \sinb m_z^2s_W^2 \right] - U_{\td{b}11}^* \left[ \frac{U_{\td{b}22}m_b \mu^*}{\cosb} - \sinb m_Z^2 U_{\td{b}21} \left( \frac{s^2_W}{3} + c_W^2 \right) \right] \nonumber \\
g_{\td{b},21}^{\phi_2^0} &= U_{\td{b}22}^* \left[ - \frac{U_{\td{b}11} m_b \mu}{\cosb} + \frac{2}{3} U_{\td{b}12} \sinb m_z^2s_W^2 \right] - U_{\td{b}21}^* \left[ \frac{U_{\td{b}12}m_b \mu^*}{\cosb} - \sinb m_Z^2 U_{\td{b}11} \left( \frac{s^2_W}{3} + c_W^2 \right) \right] \nonumber \\
g_{\td{b},22}^{\phi_2^0} &= U_{\td{b}22}^* \left[ - \frac{U_{\td{b}21} m_b \mu}{\cosb} + \frac{2}{3} U_{\td{b}22} \sinb m_z^2s_W^2 \right] - U_{\td{b}21}^* \left[ \frac{U_{\td{b}22}m_b \mu^*}{\cosb} - \sinb m_Z^2 U_{\td{b}21} \left( \frac{s^2_W}{3} + c_W^2 \right) \right] .
\end{align}
Finally, for the \cp{}-odd states $\chi_1^0$ and $\chi_2^0$ the Higgs-sbottom couplings are:
\begin{align}
g_{\td{b},11}^{\chi_1^0} &= i\frac{m_b}{\cosb} \left[ -A_b^* U_{\td{b}12}^* U_{\td{b}11} + A_b U_{\td{b}11}^* U_{\td{b}12} \right] && g_{\td{b},11}^{\chi_2^0} = i\frac{m_b}{\cosb} \left[ -\mu U_{\td{b}12}^* U_{\td{b}11} + \mu^* U_{\td{b}11}^* U_{\td{b}12} \right] \nn \\
g_{\td{b},12}^{\chi_1^0} &=i\frac{m_b}{\cosb} \left[ -A_b^* U_{\td{b}12}^* U_{\td{b}21} + A_b U_{\td{b}11}^* U_{\td{b}22} \right] && g_{\td{b},12}^{\chi_2^0} = i\frac{m_b}{\cosb} \left[ -\mu U_{\td{b}12}^* U_{\td{b}21} + \mu^* U_{\td{b}11}^* U_{\td{b}22} \right] \nn \\
g_{\td{b},21}^{\chi_1^0} &= i\frac{m_b}{\cosb} \left[ -A_b^* U_{\td{b}22}^* U_{\td{b}11} + A_b U_{\td{b}21}^* U_{\td{b}12} \right] && g_{\td{b},21}^{\chi_2^0} = i\frac{m_b}{\cosb} \left[ -\mu U_{\td{b}22}^* U_{\td{b}11} + \mu^* U_{\td{b}21}^* U_{\td{b}12} \right]  \nn \\  
g_{\td{b},22}^{\chi_1^0} &= i\frac{m_b}{\cosb} \left[ -A_b^* U_{\td{b}22}^*
U_{\td{b}21} + A_b U_{\td{b}21}^* U_{\td{b}22} \right] &&
g_{\td{b},22}^{\chi_2^0} = i\frac{m_b}{\cosb} \left[ -\mu U_{\td{b}22}^*
U_{\td{b}21} + \mu^* U_{\td{b}21}^* U_{\td{b}22} \right] \, .
\end{align}

\end{document}